\documentclass[a4paper,11pt]{article}
\pdfoutput=1 
             
\usepackage{jheppub} 
                     \usepackage[T1]{fontenc} 
\usepackage[dvipsnames]{xcolor}
\usepackage{amsmath}
\usepackage{amssymb}
\usepackage{comment}
\usepackage{enumerate}
\usepackage{changes}
\usepackage{color}
\usepackage{tablefootnote}
\usepackage{slashed}
\usepackage{graphicx}

\usepackage[utf8]{inputenc}
\usepackage[utf8]{inputenc}

\def\be{\begin{equation}}
\def\ee{\end{equation}}
\def\ba{\begin{eqnarray}}
\def\ea{\end{eqnarray}}
\def\bi{\begin{itemize}}
\def\ei{\end{itemize}}	
\def\l{\left}
\def\r{\right}

\def\la{\label}
\def\d{\partial}

\def\mpl{M_{\rm pl}}

\def\eps{\epsilon}
\def\MD{{\mathcal{D}}}
\def\Ham{{\mathcal{H}}} 
\def\vk{\vec k}
\def\vx{\vec x}

\def\bea{\begin{eqnarray}}
\def\eea{\end{eqnarray}}

\newcommand\nn{\nonumber}

\def\la{\langle}
\def\ra{\rangle}
\def\r{\right}
\def\l{\left}

\def\op{{\mathcal O}}

\title{$\lambda \phi^4$ in dS}

\author[a,b]{Victor Gorbenko,}

\author[b,c]{Leonardo Senatore}

\affiliation[a]{Institute for Advanced Study, Princeton, NJ 08540, USA}

\affiliation[b]{Stanford Institute for Theoretical Physics, Stanford University, Stanford, CA 94305, USA}
\affiliation[c]{Kavli Institute for Particle Astrophysics and Cosmology, Physics Department and SLAC, Stanford University, Menlo Park, CA 94025, USA}

\emailAdd{gorbenko@ias.edu}

\emailAdd{senatore@stanford.edu}

\abstract{We resolve the issue of infrared divergences present in theories of light scalar fields on de Sitter space.}

\usepackage[utf8]{inputenc}

\begin{document}

\maketitle

\section{Introduction}
\subsection*{The problem}

The study of radiative corrections in spacetimes close to de Sitter space is extremely important for several reasons. First, since inflation is believed to be the first cosmological phase of the observable universe, we should understand its predictions at the quantum level, in order to show the full consistency of the theory and of its inferred observable consequences. In this sense, the study of radiative corrections is relevant to confidently connect inflation and current observations. Second, at a more conceptual, but not less important, level, there is the well-known difficulty of putting together gravity and quantum mechanics. While the inadequacy of field theory to describe gravity at energies of order the Planck scale is what has led to the development of String Theory, when we put together field theory and curved spacetime we often encounter surprises also in the infrared, finding challenges for our understandings of the laws of physics in general. Renowned example of these challenges are the black hole evaporation, from which the information problem arises, or the phenomenon of eternal inflation, which leads to the problem of the landscape and the measure problem, not to mention the interpretation of the de Sitter entropy. 

In this paper, we will take $M_{\rm pl}\to\infty$, $H={\rm const}$ limit, but, nevertheless, let us first focus on inflation. A satisfactory understanding of the radiative corrections has been reached in the case of single-field inflation or of multifield inflation where the additional light fields that might be present are derivatively coupled~\footnote{Depending on the topology of the moduli space, the case of derivatively coupled scalar fields might have some strong coupling issues that appear only at non-perturbative level, and whose onset is at very late times. This case was not covered in the former literature, {but can be studied with} the formalism we develop here.}. The main variable of interest, out of which we compute correlation functions of observable quantities, is $\zeta$. At quantum level, $\zeta$-correlation functions could receive many forms of corrections that could change its functional form:
\be\label{eq:potential-issues}
\langle\zeta_{k_1}(t)\ldots \zeta_{k_n}(t)\rangle\quad\supset\quad a(t)^n\,, \; (k\,L)^{n}\,,\; \log\left(\frac{k}{a(t)H}\right)\,,\; \log(k L)\,,\; \log(H(t_{\rm h.c.})/\mu)\ ,\ldots\ ,
\ee
where $a(t)$ is the scale factor, $H(t_{\rm h.c.})$ is the value of the Hubble rate at the time of horizon crossing for the modes of interests, $\mu$ is the renormalization scale and $L$ is  the comoving size of the universe ({\it i.e.} the space slices of the 3+1-dimensional cosmological spacetime). A first important paper by Steven Weinberg~\cite{Weinberg:2005vy} has shown that $a(t)$ could only appear inside a logarithm. This dispenses of the first term in~(\ref{eq:potential-issues}). A similar argument shows that no factors of $(k\,L)$ can appear but inside a logarithm. We give a simple derivation of this fact in section~\ref{sec:wavefunctionperturbativity}. We are left with the three logarithms at the end of~(\ref{eq:potential-issues}). The last logarithm is the IR-limit of the logarithmic divergences that we find in Minkowski~\cite{Senatore:2009cf}. In particular, its form, which does not violate gauge invariance, dispenses of some earlier results where the logarithmic divergence was claimed to take the form $\log(k/\mu)$, in violation of gauge invariance. Clearly, by choosing the renormalization scale close to the Hubble scale, this radiative correction is perturbatively small. Next, Ref.~\cite{Pimentel:2012tw,Senatore:2012ya} showed at all orders that $\log\left(\frac{k}{a(t)H}\right)$ is not present when the contribution of all the many diagrams is included, while it is present when looking only at a (unphysical) subset of diagrams. Lastly, factors of $\log(k L)$ do appear in correctly performed calculations~\cite{Giddings:2010nc}, but Ref.~\cite{Senatore:2012nq} showed that their form, which is constrained by diff. invariance~\cite{Giddings:2010nc}, is such that all the infrared dependence disappears once we compute physical observables. Therefore it appears that radiative corrections in the case of single field inflation or when additional light fields are present but are derivatively coupled, have been understood and shown to be satisfactorily very small, at least as long as the curvature perturbation $\zeta$ is perturbatively smaller than one.

It seems to us that the situation instead has not been settled for the case where during inflation there are additional scalar fields that are light but not just derivatively coupled. This case includes the one of spectators fields in de Sitter space. In this case, indeed, it has been known for a long time (see for example~\cite{Burgess:2009bs,Burgess:2010dd} and references therein) that in correlation functions of these scalar fields there appears factors of $\log\left(\frac{k}{a(t)H}\right)$ that lead, at late time, to a breakdown of perturbation theory:
\be\label{eq:potential-issues2}
\langle\phi_{k_1}(t)\ldots \phi_{k_n}(t)\rangle\quad\supset\quad\lambda  \log\left(\frac{k}{a(t)H}\right) \to \ \infty \ ,
\ee
where $\lambda$ is some small coupling constant associated to a potential term.
 These large logarithms are believed to be associated to a physical effect, representing the build up of non-linear effects at long times, which ultimately tell us that the asymptotic vacuum of these field theories is non-perturbatively far from the free-theory Bunch-Davies vacuum. Such a situation has lead to claims in the literature, most famously in~\cite{Polyakov:2012uc} (see also~\cite{Polyakov:2007mm,Polyakov:2009nq,Krotov:2010ma}) that de Sitter space might be unstable at quantum level, its symmetries being spontaneously broken, and ultimately the cosmological constant might be self-adjusted to zero. Clearly, the importance of the cosmological constant problem and of inflation demands this question to be answered.

\subsection*{Summary of the main results}

In this paper, we develop a rigorous formalism that allows us to compute correlation functions of scalar fields in a perturbative expansion in small parameters, therefore with the same accuracy as normal perturbative series in quantum field theory. We will work in the rigid de Sitter limit. Focusing for simplicity on a theory with potential $\lambda\phi^4$, we will develop an equation that allows us to find a non-perturbative solution to the vacuum correlation functions. The typical size of the fluctuations in this state is $\phi^2\sim 1/\sqrt{\lambda}$. We show how we can consistently compute order by order corrections to the correlation functions in the coupling $\sqrt{\lambda}$ and several {auxiliary parameters, $\epsilon$, $\delta$ and $\Delta$}, and explicitly compute some subleading in $\sqrt{\lambda}$ corrections. The fact that our expansion parameter is $\sqrt{\lambda}$ is another way to see that we are expanding around a non-perturbative solution. Instead, the presence of the auxilary parameters  is artificial. {The most important of them is $\eps$}. In fact, this parameter represents the value of an artificial cutoff that we insert at scales much longer than the de Sitter length, and that allows us to split the modes in `short' and `long'. The `short' modes are defined with a suitable infrared cutoff so that they do not experience large radiative corrections in the infrared: their contribution to correlation functions can be safely computed in ordinary perturbation theory. Instead, the `long' modes are strongly affected by the radiative corrections, and in order to solve for them, we develop a non-perturbative treatment. The presence of the cutoff $\epsilon$ that allows us to focus only on long modes is crucial for enabling us to develop such a non-perturbative equation. In fact, two major simplifications come to our help. We can perturbatively expand in the smallness of the gradients, that become irrelevant due to the presence of the de Sitter horizon, and also in the semi-classicality of the long modes (we will later review that in de Sitter long modes become more and more classical as they expand longer than the horizon). Both these expansions are controlled by powers of $\epsilon$. Clearly, the dependence on $\epsilon$ is artificial: it will cancel order by order in perturbation theory, but it nevertheless gives us control to solve our equations. Our non-perturbative equation  is an equation for the probability distribution of the long modes, and has a form similar, but with important differences, to a Fokker-Planck equation. The {``stochastic''} nature of this equation is due to the fact that, as time goes on, modes pass from being `short' to `long'. 

We stress that our formalism is rigorous: there is nothing that is dropped and that is not recoverable in perturbation theory. {In particular, nothing in the calculation is inherently stochastic, the probabilistic nature of correlation functions is the same as in the ordinary quantum field theory.} However, at zeroth order in all our expansion parameters, our formalism reduces to the one famously introduced by Starobinsky in the 1980's~\cite{Starobinsky:1984},\cite{Starobinsky:1986fx}. We comment in the next subsection on comparison with the former literature.

The main purpose of the paper is to develop such a systematic formalism. Endowed with this, we then compute several quantities of interest. Our formalism is such that it is workless to obtain our results for a growing number of fields, but it requires more and more labour to have a large number of spacetime points at which these need to be evaluated. Therefore, first, we compute  correlation functions at coincidence, $\langle\phi(\vec x,t)^n\rangle$, including the subleading correction in $\sqrt{\lambda}$. We show that the $\epsilon$-corrections consistently cancel at this order. Then, we compute several correlation functions at non-coincidence, such that  $\langle\phi(\vec x,t_1)^n\phi(\vec x,t_2)^m\rangle$, for $H (t_2-t_1)\gg 1$, $\langle\phi(\vec x_1,t)^n\phi(\vec x_2,t)^m\rangle$, for $H |\vx_2-\vx_1|\gg 1$, and, finally, $\langle\phi(\vec x_1,t_1)^n\phi(\vec x_2,t_2)^m\rangle$ in various kinematic regimes. This allows us to show that perturbation theory is well defined and that there is an equilibrium state, {\it i.e.} an attractor, where we can check that correlation functions are de Sitter invariant, {\it i.e. the de Sitter symmetry is non spontaneously broken} by radiative corrections. 

Though for cosmological observations we are interested in correlation functions among spacelike separated points that are many Hubble lengths far apart, it is also interesting, from a theoretical point of view, to consider correlation functions between points in the same Hubble patch. While in our calculations we use the so-called FRW slicing of de Sitter, such points can be described within a so-called static-patch set of coordinates. Upon rotation to Euclidean time, it has been known for quite some time~(see for example~\cite{Gibbons:1977mu}) that correlation functions should be thermal. In particular, they should obey the KMS condition~\cite{Kubo:1957mj, Martin:1959jp}. We explicitly verify that our correlation functions, in this kinematical regime, do indeed satisfy the KMS condition. This is a rather non-trivial check because, in our formalism, the KMS condition connects terms of the answer which are of different order in~$\sqrt{\lambda}$.

 We conclude by presenting another, apparently unrelated, reason for why the study we are going to develop is important.    
 In the case of single-field inflaton, all radiative corrections have been proven by the references above to be small as long as $\zeta\ll1$. There is one interesting radiative phenomenon that happens when $\zeta\sim 1$, even in the case of single-field inflation. Indeed, as described in~\cite{Senatore:2012nq}, the same coordinate transformation that removes the dependence on the comoving size of the universe, $L$, changes the mapping between comoving coordinates and physical distances in such a way that the physical volume covered by a certain comoving distance is {\it larger} than what would be obtained in the mapping given by the classical manifold (this is roughly the usual ${\rm Volume}_{\rm physical}=e^{3N_e} \Delta x^3_{\rm comoving}$, with $N_e$ being the number of $e$-foldings from horizon exit to the end of inflation). In practice, after a certain comoving mode has exited the horizon, the many shorter modes that subsequently exit the horizon before the end of inflation lead to an enhanced overall expansion of the universe, even though the local expansion rate is not changed. This {\it quantum} enhancement of the volume gives a relative correction of order $\zeta^2\ll1$ for $\zeta\ll1$. However, as we make the inflationary potential flatter and flatter, and $\zeta\sim H^2/\dot\phi$ becomes of order one, this enhancement of the volume become non-perturbatively under control. Ref.s~\cite{Creminelli:2008es,Dubovsky:2008rf,Dubovsky:2011uy,Lewandowski:2013aka} were able to study this non-perturbative regime and in particular to show that, as the inflationary potential is flattened and~$\zeta$ approaches one, the quantum enhancement of the volume at the end of inflation grows and there is critical value beyond which the probability to generate an infinite inflationary volume passes from being zero to non-zero. This highly non-perturbative phenomenon where quantum effects completely change the asymptotics of the spacetime and make the spacetime manifold stochastic on the largest distances is what is called slow-roll eternal inflation (see for example~\cite{Goncharov:1987ir}). Ref.s~\cite{Creminelli:2008es,Dubovsky:2008rf,Dubovsky:2011uy,Lewandowski:2013aka} provided a first gauge-invariant, quantitative understanding of this phenomenon. For example, they also found that, whenever the volume is finite, there is a sharp, universal, bound to the maximum finite volume that inflation can produce, and that this bound is related to the de Sitter entropy. However, all of these results were obtained by using Starobinsky's stochastic approach to treat quantum corrections in inflationary spacetimes which was simply assumed to be correct.  Since, as we mentioned, we will derive a systematic method that, at leading order in the coupling constants and $\epsilon$, $\delta$ and $\Delta$, reduces to the one of Starobinsky used in~\cite{Creminelli:2008es,Dubovsky:2008rf,Dubovsky:2011uy,Lewandowski:2013aka}, our results will also automatically provide the missing link to obtain a rigorous establishment of slow-roll eternal inflation.

\subsection*{Comparison with former literature}

Obviously, our work strongly relies on the former work by Starobinsky and collaborators~\cite{Starobinsky:1984,Starobinsky:1986fx,Starobinsky:1994bd} (see also~\cite{Goncharov:1987ir}), which deserves a lot of credit. However, it is very unclear if the problem of the IR-divergences can be declared to have been solved until we can actually compute {any corrections to the leading answers}~\footnote{We thank an anonymous appreciated colleague for words of this meaning.}.
In a sense, we show in which sense Starobinsky's original intuition was ultimately correct: Starobinsky's equations for the classical probability distribution represent the zeroth order truncation in the parameters and $\sqrt\lambda$, $\epsilon$, $\delta$ and $\Delta$, of some more general equation (that we derive for the first time) for the probability distribution of the fields.  We establish that by iterating in $\sqrt\lambda$ and other parameters the answer reaches arbitrary precision (at least in the sense of asymptotic series). The fact that the rigorously-established formalism that we develop here has, as we will see in detail, multiple new ingredients with respect to the `Starobinsky' equation provided in~\cite{Starobinsky:1986fx,Starobinsky:1994bd} shows that the formalism provided in~\cite{Starobinsky:1986fx,Starobinsky:1994bd} was not complete. 

For the same reasons, before our work, it was unclear, at least to us, if the stochastic approach could provide a formalism that was not just order-one correct, but could provide an answer to arbitrary precision. Indeed, for this to be the case, as we show in this paper, it is necessary to include the {\it short}-wavelength component of the fields, a component which was never not-even mentioned in the literature, as well as a careful treatment of the momentum of the field. 

However, several ingredients that appear in our formalism had already appeared in several earlier works, see, for example,~\cite{bondprivate,Tokuda:2017fdh,Riotto:2011sf,Moss:2016uix}. Ref.s~\cite{Burgess:2014eoa,Burgess:2015ajz,Collins:2017haz} embark into a rigorous derivation of the Stochastic approach.  Though the general philosophies are similar, the resulting formalism does not appear  to us to have been developed in any level of detail beyond the one necessary for reproducing  the leading order equation that reduces to Starobinsky's. Wherever it would be easy to compare, that is with the expressions of the subleading corrections to correlation functions of $\phi$, or with the explicit expressions of the equations to solve for,  in~\cite{Burgess:2014eoa,Burgess:2015ajz,Collins:2017haz} these have not been computed, with the exception of the case of the free massive field.

We also mention a few other approaches and results. Ref.~\cite{Gautier:2013aoa,Gautier:2015pca} considers an $O(N)$ symmetric model of $N$ scalar fields, and is able to solve for the two-point function at leading order in $\sqrt{\lambda}$ and subleading in $1/N$. Ref.~\cite{Burgess:2009bs,Burgess:2010dd} develop and employ dynamical-RG equations, which however only partially ameliorates the problem, as the authors themselves recognize. Ref.~\cite{Tsamis:2005hd} explicitly checks that the Stochastic equations of~\cite{Starobinsky:1986fx,Starobinsky:1994bd} agree with the leading-in-$\lambda H t$ IR-divergences, with $t$ being time and $H$ being the Hubble rate, that are found in calculations from standard quantum field theory methods. Ref.~\cite{PhysRevD.99.086009,Akhmedov:2013xka,Akhmedov:2017ooy} argues that these IR-divergences are present even for heavy fields, once we go to high-enough order in perturbation theory.

{\bf Euclidean-space approach:} We conclude this section by offering a comparison with the literature that attacks the problem of the IR-divergences in de Sitter space by computing correlation function directly in the Euclidean rotation of de Sitter space, which is a four-sphere. We find that none of the attempts that we are going to list below address or solve the problem. Let us go in order. Ref.~\cite{Marolf:2010zp,Marolf:2010nz,Higuchi:2010xt} shows that, if the field has a finite mass, then the loop diagrams are IR-finite: they do not grow with time. In our opinion, this does not address the main issue: even though the loop diagrams are IR-finite, they scale as powers of $\lambda H^4/m^4$ (see for example~\cite{Burgess:2010dd,Garbrecht:2013coa}), so, for sufficiently small masses ($m^2\lesssim \sqrt{\lambda} H^2$), perturbation theory is ill defined. Indeed, our results show that for masses in this range
 the actual solution is non-perturbative.

An attempt to address the relevant issue of the case where the mass is small enough so that perturbation theory breaks down in Euclidean signature has been started by~\cite{Rajaraman:2010xd}, which correctly emphasized that the zero-mode on the Euclidean four-sphere is strongly coupled. Indeed, as further developed in~\cite{Beneke:2012kn}, the non-perturbative treatment of the zero mode makes Euclidean space correlation functions well behaved. However, as somewhat anticipated already at the end of~\cite{Beneke:2012kn}, and very clearly and explicitly explained in~\cite{Nacir:2016fzi,LopezNacir:2016gfj,LopezNacir:2018xto}, this treatment is not sufficient to make the correlation functions well behaved once they are rotated back to Lorentzian space. In other words, upon rotation back to Lorentzian space of the correlation functions  obtained in Euclidean space in this way, one still has IR (specifically secular) divergences for the same small enough masses. In order to address these issues, Ref.s~\cite{Nacir:2016fzi,LopezNacir:2016gfj,LopezNacir:2018xto} realize that they need to resum infinite serieses of diagrams. To proceed, they focus on an $O(N)$ symmetric model of $N$ scalar fields, and develop a double expansion in $\sqrt{\lambda}$ and $1/N$. At each order, a new infinite series of diagrams needs to be resummed. 
Though this research direction is interesting and well founded, it still needs to be fully developed to reach results beyond the leading order. {To enable comparison, in section \ref{sec:largeN} we apply our formalism to the large-$N$ theory and obtain explicit results for the two-point functions at leading order in $1/N$ and to  subleading order in $\sqrt{\lambda}$.}

\section{Strategy\label{sec:strategy}}

\subsection{Origin of the IR-divergences}
It is useful to start by stating the problem and explaining its origin, so that we will be able to highlight the main strategy with which we are going to address it.
Consider massless, $\lambda\phi^4$ in dS, and restrict to the Poincar\'e patch, which we describe in FRW slicing~\footnote{Of course setting the mass exactly to zero in an interacting QFT without a shift symmetry is not a well-defined procedure. What we mean is that we set the mass to be relatively small~($m\ll H$) at energy scales of order Hubble. Since the theory is weakly coupled at this scale this is well-defined, moreover all our logic generalizes trivially to arbitrary potentials, in particular to those with larger masses. }. If one computes the in-in two-point functions with the usual quantum field theory formalism (see for example~\cite{Senatore:2009cf}), one finds that at one-loop the result is both secular and IR-divergent:
\be\label{eq:exactperturbationIR}
\langle\phi_{\vec k}(t)\phi_{\vec k'}(t) \rangle'\sim\frac{H^2}{k^3}  \left(1+\lambda {\log\left(\frac{k}{H e^{H t}}\right) \log\left(k H e^{H t} L^2\right)} \right)\ ,
\ee
where $\langle\ldots\rangle'$ means that we dropped the momentum conserving $\delta$-function and where we dropped the numerical coefficients. Here $H$ is the de Sitter Hubble rate, $t$ is the time of the FRW slicing, 
 and $L$ is the {comoving} IR-cutoff. Clearly, at late enough times, or large enough $L$'s, the perturbative calculation breaks down. {This result has been obtained for example by~\cite{Burgess:2009bs,Burgess:2010dd}, where however we replaced the IR cutoff there, which is fixed in physical coordinates, with one that is fixed in comoving coordinates~\footnote{The result is not qualitatively changed if one chooses an IR cutoff which is fixed in physical coordinates, as in~\cite{Burgess:2009bs,Burgess:2010dd}: one still finds logarithmic IR and secular divergences.}.}

It is particularly enlightening to see the emergence of the secular divergence in a familiar context. It will turn out that the IR divergence ultimately has the same physical origin. Consider, to start, a free massive field. Its two-point function can be found exactly, and gives: 
\be\label{eq:massive2pt}
\langle\phi_{\vec k}(t)\phi_{\vec k'}(t) \rangle'= {\frac{\pi}{4}}\frac{e^{-3 H t}}{H} \left|H^{(1)}_{\nu=\sqrt{9/4-m^2/H^2}}\left(\frac{k}{aH}\right)\right|^2\ .
\ee
This expression is not secular divergent, and it actually goes to zero at late times. However, this limit is very different from the same limit for a massless field, which goes to a constant: the effect of the mass is large at late times. In fact, if we Taylor expand the above expression in $m^2/H^2\ll1$, we obtain
\be\label{eq:massiveTaylored}
\langle\phi_{\vec k}(t)\phi_{\vec k'}(t) \rangle'\quad \to\quad \frac{H^2}{k^3}\left(1+\frac{m^2}{H^2}\log\left(\frac{k}{a(t) H}\right)\right)\ ,
\ee
where we dropped order one numbers. This expansion in $m^2/H^2\ll1$ loses perturbative control at late enough times that $\frac{k}{a(t) H}\lesssim e^{- \frac{H^2}{m^2}}$. 

Expression~(\ref{eq:massiveTaylored}) is the same one that we would have obtained if we had treated the mass as a small perturbation. Solving perturbatively the equation of motion for a massive scalar field, and denoting by $\phi^{(0)}$ and $\phi^{(1)}$ respectively the zeroth and first order solution, we would have obtained at late times when the mode is outside the horizon ({\it {\it i.e.}}~$\frac{k}{a H}\ll1$):
\bea
&&\ddot \phi_{\vec k}(t)+3 H\dot \phi_{\vec k}(t)+\frac{k^2}{a^2}\phi_{\vec k}(t)=m^2 \phi_{\vec k}(t) \\ \nn
&&\Rightarrow \quad \phi^{(1)}_{\vec k}(t)\supset \int_{\frac{k}{a(t) H}\sim 1}^t dt'\; \frac{1}{3H} m^2 \phi^{(0)}_{\vec k}(t')\sim  \phi^{(0)}_{\vec k}(t)\, \frac{m^2}{H^2}  \log\left(\frac{k}{a(t)H}\right)\ ,
\eea
where in the last passage we have focussed on the contribution from when the mode is outside the horizon, and used the fact that a massless field is time-independent in this regime. We see that we obtain the same secular divergence as in~(\ref{eq:exactperturbationIR}). The divergence is associated to the fact that the leading order wave function is time-independent outside of the Horizon, but the mass perturbation does not shut down in this limit. 

\begin{figure}[htbp] 
	\centering
		\includegraphics[width=.8\linewidth,angle=0]{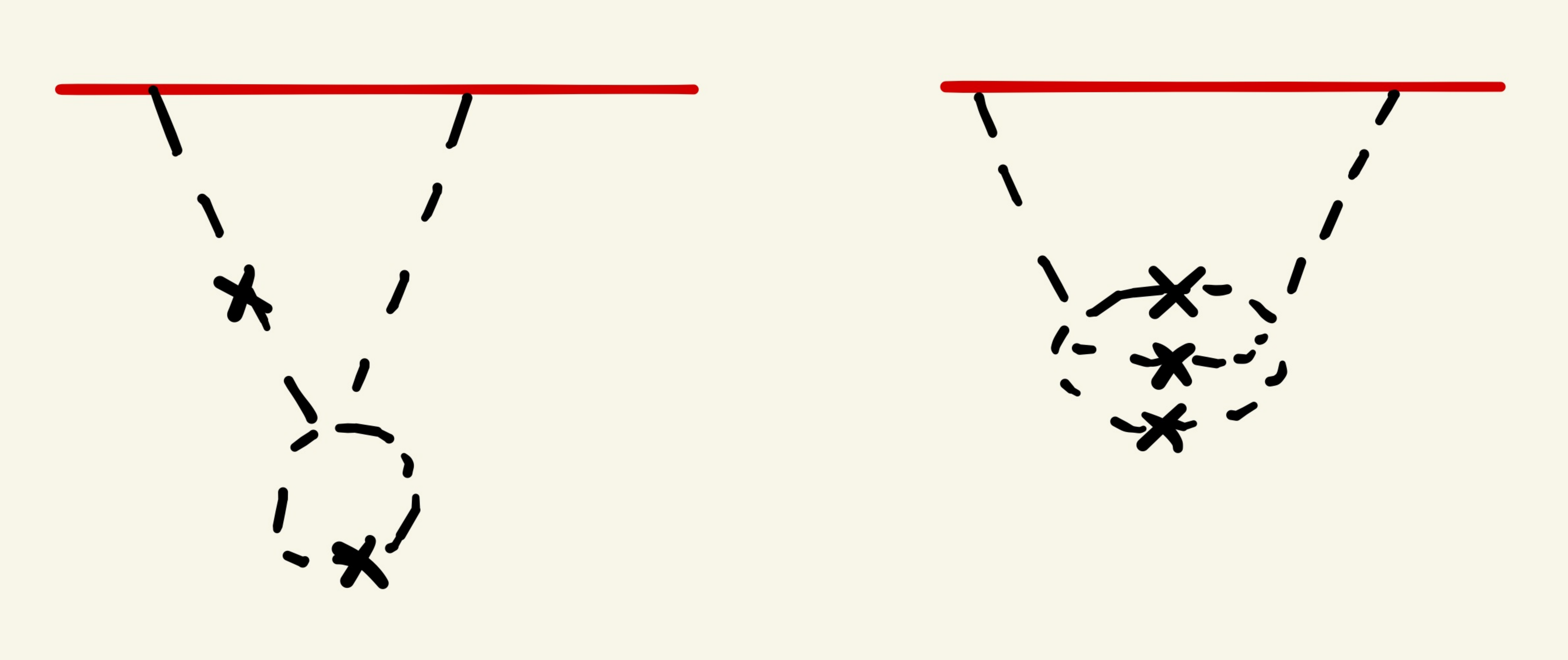}
	\caption{\small One-loop ({\it left}) and two-loop ({\it right}) diagrams for the two-point function of $\phi$. Lines with crosses represent correlation functions, without crosses represent Green's functions.}
	\label{fig:olddiagrams}
\end{figure}

Indeed, we can check that this is the same phenomenon that happens in massless $\lambda\phi^4$. Solving perturbatively in $\lambda$, and again focussing on the contribution from times when the mode is outside of the Horizon, we have
\bea\label{eq:phi1schamatic}
&&\ddot \phi_{\vec k}(t)+3 H\dot \phi_{\vec k}(t)+\frac{k^2}{a^2}\phi_{\vec k}(t)=\lambda \phi^3(\vec k,t) \\ \nn
&& \Rightarrow \quad \phi^{(1)}_{\vec k}(t)\sim\int_{\frac{k}{a(t') H}\sim 1}^t dt'\;\frac{1}{3H} \lambda \phi^{(0)}{}^3(\vec k,t') \ .
\eea
When we compute expectation values, we have (see Fig.~\ref{fig:olddiagrams} on the left)
\bea\label{eq:simplediv1}\nn
&&\langle\phi_{\vec k}(t)\phi_{\vec k'}(t) \rangle'\quad\supset\quad \langle\phi^{(1)}_{\vec k}(t)\phi^{(0)}_{\vec k'}(t) \rangle'\sim\int_{\frac{k}{a(t') H}\sim 1}^t dt'\;\frac{1}{3H} \l\langle\lambda\phi^{(0)}{}^3(\vec k,t') \phi^{(0)}_{\vec k'}(t')\r\rangle'\\ \nn
&&\qquad \sim\int_{\frac{k}{a(t') H}\sim 1}^t dt'\;\frac{1}{3H} \lambda \l\langle\phi^{(0)}{}^2({\vec k=0},t') \r\rangle' \ \l\langle\phi^{(0)}_{\vec k}(t')\phi^{(0)}_{\vec k'}(t)\r\rangle'\\ 
&&\qquad\sim \frac{H^2}{k^3}\ \lambda {\log\left(\frac{k}{H e^{H t}}\right) \log\left(k H e^{H t} L^2\right)}  ,
\eea
where we cutoff the integral in internal momentum in $ \l\langle\phi^{(0)}{}^2({\vec k=0},t') \r\rangle' $ to include only modes outside of the Horizon. We obtain the result of~(\ref{eq:exactperturbationIR}), and we learn the following lesson. The origin of the secular and the IR divergences is due to the fact that there is an accumulation of the effect of the interaction because the wave function of the massless field is independent of time and scale invariant outside of the Horizon, and the potential interactions do not shut down as the gradient of the modes become negligible. 

Very naively, one could interpret the result of~(\ref{eq:simplediv1}) as due to a simple mean-field effect ({\it {\it i.e.}} as an effective mass). However, it is not so. At two loops, one gets the following contribution, which does not have, at least to us, a mean field interpretation (see Fig.~\ref{fig:olddiagrams} on the right):
\bea\label{eq:simplediv1}
&&\langle\phi_{\vec k}(t)\phi_{\vec k'}(t) \rangle'\quad\supset\quad \langle\phi^{(1)}_{\vec k}(t)\phi^{(1)}_{\vec k'}(t) \rangle'\\ \nn
&&\quad\quad\quad\quad\sim\int_{\frac{k}{a(t_1) H}\sim 1}^t dt_1 \int_{\frac{k}{a(t_2) H}\sim 1}^t dt_2\;\left(\frac{1}{3H}\right)^2 \lambda^2 \langle\phi^{(0)}{}^3(\vec k,t_1)\phi^{(0)}{}^3(\vec k',t_2) \rangle'\\ \nn
&&\quad\quad\quad\quad \sim\int_{\frac{k}{a(t_1) H}\sim 1}^t dt_1 \int_{\frac{k}{a(t_2) H}\sim 1}^t dt_2\;\int^{q_1/a(t_1)\sim H}_{1/L} d^3q_1\int^{q_2/a(t_2)\sim H}_{1/L} d^3 q_2\left(\frac{1}{3H}\right)^2 \lambda^2 \ \times \\ \nn
&&\quad\qquad\quad\times \ \langle\phi_{\vec q_1}(t_1) \phi_{\vec q_1}(t_2)\rangle'\langle\phi_{\vec q_2}(t_1) \phi_{\vec q_2}(t_2)\rangle'\langle\phi_{\vec k-\vec q_1-\vec q_2}(t_1) \phi_{\vec k-\vec q_1-\vec q_2}(t_2)\rangle' \\ \nn
&&\quad\quad\quad\quad\sim   \frac{H^2}{k^3}\ \lambda^2 \left[{\log\left(\frac{k}{H e^{H t}}\right) \log\left(k H e^{H t} L^2\right)}\right]^2\ .
\eea

Clearly, given that the potential of $\lambda \phi^4$ is unbounded at high values of $\phi$, we expect a saturation effect. Given that the energy scale of the problem is $H$, we expect that an equilibrium state will form where all values of $\phi$ are populated such that $V(\phi)\sim H^4$. This means that $\phi\sim H/\lambda^{1/4}$, which is a non-perturbative configuration. We will indeed see that this is what happens, { through a controlled derivation}.

\subsection{Strategy: mode splitting, locality, classicality and the  wave function}

{\bf Parametrics:} It is evident from the former expressions that the secular and IR divergences arise when the modes are outside of the horizon. When the modes are still inside the horizon, in fact, their wave function oscillates and is not scale invariant, and there is no room for secular or IR divergences. What is physically happening is that as the modes are outside of the horizon, there is an accumulation of the field value at a given location that makes the effect of the potential term larger and larger. Clearly, there should be a saturation effect that is invisible in perturbation theory. A non-perturbative treatment is required.

However, the situation is not so hopeless.  In fact, modes outside the horizon follow a simplified dynamics, where the gradients are small, so the evolution is approximately local, and they become semiclassical, as we review next. On the contrary, modes shorter than the horizon are quantum and have an evolution where gradients are non-negligible, but for them, perturbation theory works very well. It therefore emerges the following natural strategy. We are going to split the mode into `short' and `long', according to wether their physical wavenumber $k/a(t)$ is respectively higher or lower than a fictitious, but useful, energy cutoff. In comoving coordinates, therefore, this takes the form of a time-dependent cutoff
\be\label{eq:physical_cutoff}
\Lambda(t)=\epsilon H\,a(t) \ , 
\ee
and treat the long and short modes differently. Notice that, in FRW slicing, modes start short and then become long after some time. For the short modes, we are going to use perturbation theory. For the long mode, we instead are going to find a non-perturbative formalism, which can however be handled efficiently.
First the evolution is {``ultra-local''}, with perturbatively small corrections of order $\epsilon^2$ (by analyticity and rotation invariance). {By ultra-local evolution we mean that the long fields at different spacial points, to leading order, behave as independent degrees of freedom. Since we are interested in correlators of finitely many fields, in practice we will need to deal with a system of finitely many variables (as opposed to a usual field theory), which can be solved non-perturbatively.  } The second simplifying feature is that the evolution is quasi-classical, with perturbative corrections scaling as~$\epsilon^3$. {The parameters $\delta$ and $\Delta$ mentioned in the introduction will be introduced below. Their physical meaning is not as important as that of $\epsilon$ so we will not mention them until they become necessary}.

Though the semiclassical approximation will play almost a marginal role, compared to the role of ultra-locality, we here  review why it holds. The fact that the evolution of modes outside the Horizon becomes semiclassical is beautifully explained in the original paper by Guth and Pi~\cite{Guth:1985ya}. Here we mention the key facts from this reference. Let us consider a free scalar field in de Sitter space with a small mass, as this will be the case of interest in our study. Each Fourier mode $\phi_{\vk}$ evolves independently and its wave function $\Psi(\phi_{\vec k})$ is given by, when the mode is outside of the Horizon,
\be
\Psi\left(\phi_{\vk},\frac{k}{a(t) H}\ll 1\right)\simeq e^{-\left( \frac{k^3}{2H^2}+i\left(\frac{1}{6} \frac{m^2}{H}e^{3 H t}+\frac{k^2}{H}e^{H t}\right)\right)\phi_{\vec k}^2} \ ,
\ee
where we dropped subleading terms in $m^2/H^2$.
The semiclassicality of a wave function is controlled by the {change in the phase of the wave function}
associated to the shift in $\phi_k$ that it takes to change the modulus by order one. This is given by
\be\label{eq:semicl}
\l({\rm Semiclassicality}\r)^{-1}\sim{\rm Min}\left[\frac{k^3}{m^2 H a(t)^3},\; \frac{k}{H a(t)}\right]\ .
\ee
While this is a well known criterion, Ref.~\cite{Guth:1985ya} explicitly checks that in this regime, the correlation functions of $\phi$ and its conjugate momentum, $\pi$, can be computed using a classical distribution that satisfies the classical Boltzmann equation up to mistakes of order the above correction in~(\ref{eq:semicl}).  
Therefore, the quantum corrections for the long modes are peaked at the first moment when they become `long', and they scale as ${\rm Min}\left[\frac{H^2}{m^2}\epsilon^3,\;\epsilon\right]$.

By choosing $\epsilon\ll 1$, we can control the quasi-local and semiclassical approximations. However, we cannot take $\epsilon$ too small, as otherwise perturbation theory for the short modes breaks down. In fact, the time $\Delta t$ a given short mode spends outside the horizon is $\sim \frac{1}{H}\log\epsilon$, and we need to ensure the perturbative corrections, that scale as $\lambda H \Delta t$ are under control.  In reality, as we will show later, perturbative corrections around our non-perturbative solution scale as  $\sqrt{\lambda} H \Delta t$. We therefore can choose $\epsilon$ to lie in the interval
\be
e^{-\frac{1}{\sqrt{\lambda}}}\ll \epsilon\ll1 \ . 
\ee
This interval is parametrically large for $\lambda\ll 1$. Given this interval, it is natural to choose $\epsilon\ll {\lambda}^{1/4}$, so that
\be\label{eq:epsbounds}
e^{-\frac{1}{\sqrt{\lambda}}}\ll \epsilon\ll{\lambda}^{1/4} \ ,
\ee
which is also parametrically large for $\lambda\ll 1$. In fact, in this regime, the corrections  from the selfinteractions, which, as we will see, scale as $\sqrt{\lambda}\ll1$, are the leading ones. Furthermore, 
{we will see that the potential term gives a phase to the wave function so that semiclassical corrections scale as those for a massive field with}
 $m^2\sim \sqrt{\lambda}H^2$~(\footnote{As we will see in section~\ref{sec:quantum}, strictly speaking the quantum corrections scale as $\epsilon^3/\lambda^{1/4}$. Since, from~(\ref{eq:epsbounds}), we can naturally take $\epsilon$ exponentially smaller than $\lambda$, we neglect to mention these power-law corrections to the expansion parameter, as they have a small influence for small $\lambda$'s. Similarly, as we will see next, if we take $\epsilon\sim e^{-\frac{1}{\lambda^{1/n}}}$, with $n$ large but sufficiently small that~(\ref{eq:epsbounds}) is not violated, we will find that the self-interactions of the short modes scale as corrections $\sqrt{\lambda}\log\epsilon\sim\lambda^{\frac{1}{2}-\frac{1}{n}}$. Notice that as $\lambda\to 0$, $n$ can be made arbitrarily large. We will simply mention these corrections  as $\sqrt{\lambda}$.}).
In summary, using this strategy, we will find an equilibrium non-perturbative solution for correlation functions, around which, both for the long and the short modes, we will be able to develop a perturbative expansion in powers of $\epsilon\ll1 $ and $\sqrt{\lambda}\ll1$.

We add three more-technical comments. First: Of course, $\Lambda=\epsilon Ha(t)$ is an artificial cutoff that separates where different approximations are valid. The theory does not change as a mode moves across $\Lambda$. Therefore, as we make approximations that corresponds to expansions in $\epsilon\ll1$ and $\sqrt{\lambda}\log \eps\ll1$, the corrections in $\epsilon$ should cancel order by order in perturbation theory. This cancellation will be an important consistency check of our formalism and an algebraic one for our computations. Needless to say, this is very similar to the role played by the renormalization scale in very-familiar flat-space quantum field theory calculations. 

 Second: if we were to split the modes into longs and shorts by using a sharp cutoff in Fourier space, the resulting long-wavelength theory would be non-local in space. While this would result in a theory that technically is not harder to deal with than a local-in-space one, we find it would be a less intuitive one. For this reason, we split the modes into long and short by using a smooth cutoff in Fourier space, centered around $k=\Lambda$, but with width of order $\Delta k\sim \delta \Lambda$. As we will show later by choosing $e^{-\frac{1}{\sqrt{\lambda}}} \ll \delta\ll \sqrt{\lambda}$, the resulting long-wavelength theory will be local in space, with small, local, corrections scaling as $\sqrt{\lambda} \log\delta\ll 1$. As for the dependence in $\epsilon$, any dependence on $\delta$ will automatically cancel in a correct calculation. However, we will content ourself to check the cancellation of the factors of $\epsilon$, and we will assume that the dependence on $\delta$ to cancel similarly, leaving an explicit check of this fact to future work.

Third and last comment: our radiative corrections will include, naturally, also the ones from the very short modes. This leads to the requirement of renormalizing our Lagrangian, {which in particular leads to the generation of a mass term}. For this reason, it is convenient to assume that the field has a mass of order $m^2\lesssim {\lambda} H^2$. In this regime, perturbation theory is still IR-divergent, and a non-perturbative treatment is still required. Indeed, we will treat the {physical} mass as a small perturbation, {assuming the bare mass has absorbed} the UV divergences.\\

{\bf Strategy: }Having stated the general ideas and simplifications that allow us to control this non perturbative system, we now move on to give some additional information on the steps by which we will address this problem. In this way one can get the general picture of the derivation of the  results without having to follow the details. 

We will start by studying the wave functional of the theory in a particular state, the Bunch-Davies (BD) vacuum, according to which each mode is in the Minkowski vacuum in the infinite past once it is deep inside the horizon~\footnote{ We are using the BD vacuum with a slight abuse of notation. In the original reference, the BD vacuum is referred to a free theory. Here we refer to it for an interacting theory in the sense in which we just described.}.  We will prove that such a wave function is not affected by IR-divergence and can be reliably computed in perturbation theory as a formal expansion in powers of $\phi$. We stress that this fact does not imply that correlation functions can be extracted from the wave function in any straightforward manner since the functional integral which has to be evaluated in order to do so is strongly coupled in the IR. In this sense the wave function is strongly non-gaussian. In particular, in order to compute the leading long-distance contribution to a correlator, one needs to take into account infinitely many terms in the wave function. This is why we need to develop a non-perturbative formalism.

Nevertheless, the perturbative expression for the wave function can be reliably used for the three operations that we will actually need: first, to express the action of the momentum operator $\hat\Pi$ on the wave function $\Psi_{BD}[\phi,t]$~(\footnote{We will use square brackets for objects that are functionals of three- or four-dimensional field configurations to distinguish them from functions of a finite number of variables. We will also keep explicit the time dependence of the operators.}); second: to evaluate expectation values of the short modes with the long modes kept fixed; third: to compute the wave function at one time with the insertion at an earlier time of an operator (this will be mostly needed for the quantum corrections). The reliability of such perturbative calculations can only be proven {\it a-posteriori} after one is able to estimate the typical size of the $\phi$ fluctuations, {\it i.e.}  after computing $\phi$-correlation functions.

We will then consider the probability distribution of the field values,  
\be\label{eq:Pdefinition}
P[\phi,t]=\Psi_{BD}^*[\phi,t]\Psi_{BD}[\phi,t]\ ,
\ee 
and its evolution equation, which is controlled by the quantum-mechanical functional continuity equation. From this, we will derive an equation for the analogous quantity but restricted to the long modes. Such an equation will have a `diffusion' term that originates from  the contribution of short modes becoming long, and a `drift' term, associated  to the term that was already present in the equation for the probability distribution for the full field. 

The evolution equation for the probability distribution, either for the full field or for the long field, is not a closed  equation: it in fact involves directly the  wave function, since there is a term where the momentum of the field acts on the wave function, in such a way that one cannot combine the two wave functions into a probability distribution. Schematically, there is a term of the form $\Psi^*[\phi]\,\hat \Pi\, \Psi[\phi]$. The continuity equation therefore cannot be solved directly for $P[\phi]$ unless one is able to express the momentum operator in terms of fields, allowing one to combine $\Psi^*[\phi]$ and $\Psi[\phi]$ back into $P[\phi]$. At this point we use the knowledge of the wave function.
The evaluation of various terms in the equation for the long distribution also requires the calculation of the short modes expectation values for which again the wave function is needed.

At this point, the continuity equation for the the long fields probability distribution is a closed equation, and, using the simplifications we mentioned at the beginning of this section, can be solved non-perturbatively. The solution allows us to compute correlation functions of $\phi$'s, determining the typical size of the field fluctuations and, as mentioned, proving {\it a-posteriori} the consistency of the whole calculation. We stress that we do not neglect finite, although small, momenta of the long modes, as effects associated with it give corrections that are perturbative in $\sqrt{\lambda}$. Since we will work mostly in position space for the long modes, keeping track of this momentum leads to several technical complications, which we overcome.

We end this section by addressing a possible limitation of our approach. To develop our formalism we need to specify a wave function, {\it i.e.} a particular state. Though the state we use, the BD vacuum, is a particularly-well physically motivated one, our approach is more general than that. Our procedure, in particular, allows us to compute correlation functions of fields inserted at different times. Then by computing correlation functions with an arbitrary large number  of fields inserted in the past, one can compute equal time correlation functions in a state prepared by these insertions. This already gives a rather large set of states.  In order to further  shed light on the state  independence of our results, in section \ref{sec:stability} we show that small deformations of the long-modes-part of the wave function decay on time scales of order Hubble. Additionally, in App.~\ref{app:Wigner}, we will develop a formalism that is naturally more state-independent, valid in the vicinity of any state whose correlation functions are close to the one of the BD vacuum. These results convincingly show that the BD state is an attractor, that the correlation functions computed in the main part of the paper are universal and that the state-dependent corrections decay at late times.

\subsection{Reader's guide}
We must admit that the resolution of the problem of IR {{and}} or secular divergences in de Sitter turned out rather complicated, and, consequently, our paper is somewhat technical. To guide the reader, we briefly mention where the benchmarks of the implementation of the above strategy, as well as the main results, appear. Throughout the paper we work with the expanding part of dS. In section~\ref{sec:wavefunctionperturbativity}, we discuss the calculation of the wave function in perturbation theory, presenting the general form it will takes in eq.~(\ref{eq:wavefunction_schematic}). In section \ref{sec:EFT} we introduce the probability distribution of the long modes and derive the functional equation which governs its time evolution, eq.~\eqref{eq:full}.
In section~\ref{sec:Singlepoint} we study the distribution marginalized over the fields at all but one space-time point. Making several controlled approximations in both $\lambda$ and $\eps$ we derive the partial differential equation which this ditribution satisfies to leading order, Eq.~\eqref{1pteqtime}. {This is, in fact, the only equation that needs to be solved non-perturbatively.} In section \ref{sec:two-point}  we study the leading equation for the distribution of the long fields at two space-time points. It requires the analysis of the time dependence of the single-point distribution in states that differ from BD, and the application of ``sudden'' perturbation theory. The solution for the equal-time case is given in \eqref{2ptfinal} in coordinate space, and in   \eqref{2ptfourier} in momentum space. {In subsection \ref{nlocation} we compute the three-point function at leading order and sketch the calculation of higher-point functions. } In section \ref{sec:subleadP1} we study subleading, {\it i.e.} order $O(\sqrt{\lambda})$, corrections to the single-point distribution. The equation governing the field distribution at subleading order is given in \eqref{1pteqsublead} and the associated subleading correlation functions are computed in Eq.~\eqref{corrsublead}. {In section \ref{sec:Eigenvalues} we compute the subleading corrections to the exponents controlling the two-point distribution at large separations.} In section~\ref{sec:generic-2-point} we study the probability distribution for two generic spacetime points, checking in Eq.~\eqref{eq:generictwospacetimes} that the two-point function is de-Sitter invariant. These results allow us to compute correlators of arbitrary bi-local operators to subleading order in $\sqrt{\lambda}$. In section~\ref{sec:thermality} we focus on the static patch and check that the two-point function satisfies the KMS condition as expected from thermality. In section~\ref{sec:largeN} we study the system at large number of fields.   In section \ref{sec:quantum} we study the corrections due to non-classical behavior of the long modes and, in particular, confirm that they are suppressed by powers of $\eps$. Last, in section~\ref{sec:gradients}, we show how to compute corrections from gradients, showing they are of order $\epsilon^2$. {The most technical parts are relegated to the Appendices referred to in the main text.}

\section{Perturbative calculation of the wave function\label{sec:wavefunctionperturbativity}}

We are now going to compute the wave function using perturbation theory. What we will find is that the wave function is not affected by IR divergences, however, it contains particular growing secular terms that are potentially large for low momenta. We will bound these secular terms from above and show that given our scaling of the fields at low momenta the secular terms do not invalidate our expansion. 
To streamline the presentation we focus on the $\lambda \phi^4$ potential, restoring general $V(\phi)$ where it is obvious to do so.

The wave function, $\Psi[\phi,t]$, satisfies a functional Schroedinger equation
\be\label{eq:shrodinger}
i \frac{\d}{\d t}\Psi[\phi,t]={\cal H}\left[-i\frac{\delta}{\delta \phi},\phi,t\right] \Psi[\phi,t]\ ,
\ee
where the momentum operator is represented as $\hat\Pi(\vec x)=-i\frac{\delta}{\delta\phi(\vec x)}$ and where ${\cal{H}}$ is the Hamiltonian. As familiar from non-relativistic quantum mechanics, the solution to this equation is offered by a path integral, where the choice of the boundary conditions specifies the specific state we are dealing with. In the context of de Sitter space, the formalism to compute the wafecuntion has been well developed originally and generically by~\cite{Maldacena:2002vr}, and by~\cite{Anninos:2014lwa} for massless $\lambda\phi^4$. In this work, we will be mainly interested in the BD vacuum, which, as we specified above, is the vacuum that corresponds to imposing that, at early enough times, when a given mode is well inside the horizon, it is in the Minkowski vacuum. For this calculation let us switch to conformal time $\eta$. Then we have
\be\label{eq:path integral}
\Psi_{BD}[\phi,\eta]=\int {\cal{D}}\varphi\; e^{i S[\varphi]}\ ,
\ee 
in which we integrate over field configurations that, at the final time, satisfy the  boundary condition $\varphi(\vec k,\eta'=\eta)=\phi(\vec k)$, and, in the past, go as $\varphi(\vec k,\eta')\sim e^{i k\eta' }$, with a $\eta'$-contour deformed in such a way that, at early times, the oscillation is damped. This choice of the~$i\varepsilon$ prescription guarantees that we are describing the state corresponding to the interacting vacuum.
For example, the wave function for the BD vacuum of a free massless field is the following Gaussian~\cite{Guth:1985ya}
\be
\Psi_{BD}[\phi,\eta]=\prod_{\vec k} \left(\frac{2 k^3}{\pi}\right)^{1/4} {\rm Exp}\left( \frac{i}{2 H^2}\left(\frac{k^2}{\eta\left(1-i k\eta\right)} \phi(\vec k)\phi(-\vec k)\right)\right) \frac{e^{-ik\eta/2}}{\sqrt{1-ik\eta}}\ .
\ee
We notice, in passing, that the imaginary part is much larger than the real part once the modes are outside of the horizon, a well known fact that leads to the modes being semiclassical once outside of the horizon~\cite{Guth:1985ya}.

With interactions, the perturbative evaluation of~(\ref{eq:path integral}) can be organized in terms of Feynman diagrams that involve two propagators (see~\cite{Anninos:2014lwa} for a review of these aspects): the bulk-to-bulk propagator $G(\eta_1,\eta_2,k;\eta)$, which connects vertices at time $\eta_1$ and $\eta_2$, while the wave function is evaluated at time $\eta$, and whose expression is given by
\bea
\label{Bulk2Bulk}
&&G(\eta_1,\eta_2,k;\eta)=\Theta(\eta_2-\eta_1)\times \\ \nn
&&\qquad\times\ \frac{H^2}{2 k^3} (1-i k\eta_2 ) \left(-\frac{(1+i k\eta) (1-i\eta_1 k)}{(1-i k \eta) }e^{-i k (2 \eta   -\eta_1-\eta_2)}+(1+i k\eta_1 ) e^{-i k (\eta_1-\eta_2)}\right)+\\ \nn 
&&\qquad+\{\eta_1\leftrightarrow\eta_2\}\ ;
\eea
and the bulk-to-boundary propagator, $K(\eta_1, k;\eta)$, which connects a vertex at time $\eta_1$ to the final time $\eta$:
\be
\label{eq:Bulk2Boundary}
K(\eta_1, k;\eta)=\frac{(1-i k\eta_1) }{(1-ik\eta)} e^{-i k (\eta-\eta_1)}\ .
\ee
For a wave function at time $\eta$, the term in the exponent of order $\phi^n$ is obtained by computing all {\it connected} diagrams in which $n$ points at time $\eta$ are connected with vertices by bulk-to-boundary propagators, and vertices among themselves with bulk-to-bulk propagators. The $i\varepsilon$ prescription is implemented by writing $\eta_{i}=\tilde\eta_i +e^{-i\varepsilon} \Delta\eta_{i}$, with $\Delta\eta_{i}\in[-\infty,0]$,  $\pi>\varepsilon>0$, and $\tilde \eta_i$ being the end-point of the integration in $\eta_i$ (typically the final time or the time of insertion of another vertex). Notice also that each vertex carries a factor of $i$. 

Readers familiar with AdS/CFT will recognize in these Feynman rules something very similar to the Witten diagrams that are used to compute the partition function in AdS. Indeed,  $\Psi_{BD}[\phi,\eta]=Z_{EAdS}[\phi,z=-i\eta]$ (with $L=-i/H$), where $Z_{EAdS}$ is the partition function in Euclidean AdS, $z$ is the usual radial coordinate in  Poincare slicing, and  $L$ is the AdS length \cite{Maldacena:2002vr},
\cite{Harlow:2011ke}. In fact, Ref.~\cite{Anninos:2014lwa} explicitly shows that the two calculations are the same for the $\lambda \phi^4$ theory we study here.

\begin{figure}[htbp] 
	\centering
		\includegraphics[width=.4\linewidth,angle=270]{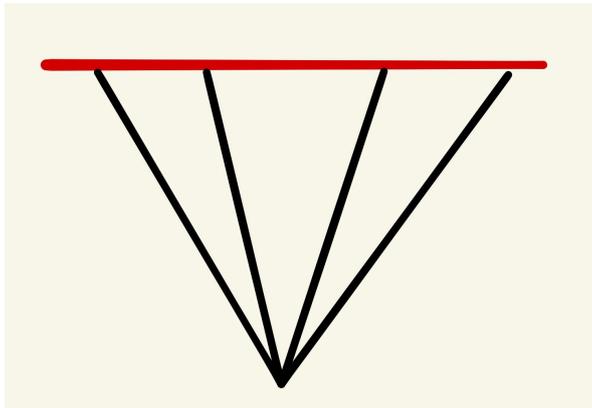}
	\caption{\small Tree-level Witten diagram contributing to the $\lambda\phi^4$ term in the exponent of the wave function.}
	\label{fig:quartic_tree_level}
\end{figure}

The first few contributions to  the wave function were computed in~\cite{Anninos:2014lwa}\cite{Collins:2017haz}. We can organize them in terms of number of fields appearing in the exponent of the wave function. With four-$\phi$'s, the tree-level diagram of Fig.~\ref{fig:quartic_tree_level} gives the following contribution in the regime $k_i\eta\ll1$ we are mainly interested in
\bea\label{eq:wavefuncquartic}
&&\log\left(\Psi_{BD}[\phi,\eta]\right)\supset \frac{\lambda}{12}\int \left(\prod_{i=1,\ldots,4}\frac{d^3k_i}{(2\pi)^3}\right) (2\pi)^3\delta^{(3)}\left(\sum_{i=1,\ldots,4}\vk_i\right)\ \times\\ \nn
&&\qquad\qquad\qquad\qquad \times\ \left(\frac{i}{H^4\eta^3}+\frac{k_{\Sigma^3}}{H^4}\log(-k_\Sigma\eta)+\ldots\right) \phi(\vk_1) \phi(\vk_2) \phi(\vk_3) \phi(\vk_4)\,,
\eea
where
\be
k_\Sigma=\sum_{i=1,\ldots,4}k_i\ , \quad k_{\Sigma^3}= \sum_{i=1,\ldots,4}k_i^3\ ,
   \ee
and the $\ldots$ represent subleading terms in the $k_i\eta\ll1$ limit. More explicitly, the imaginary part contains a subleading term in two powers in $k_i\eta$, while the real part has a subleading part suppressed by $1/\log(k_i\eta)$.

There are several observations to make about te result in (\ref{eq:wavefuncquartic}). First, we notice that the  momentum operator applied to the wave function gives:
\be\label{eq:momentum_leading}
\hat\Pi[\phi,\vec x]\,\Psi_{BD}[\phi,\eta]=\left(-a^3\frac{\lambda\phi(\vec x)^3}{3H}+O\left(k\eta,\sqrt{\lambda}\r)\right) \Psi_{BD}[\phi,\eta] \ ,
\ee
where $O(k\eta)$ means subleading corrections when all the modes are long. This result is nothing but the familiar, classic and classical, value of the momentum of $\phi$ on the slow-roll solution with potential $\lambda\phi^4$.

Next, notice that the real part of the wave function contains logarithmic secular terms. This is important, because, once we compute correlation functions of $\phi$, the imaginary part of the wave function does not contribute, and since the real quadratic part of the wave function goes to a constant, then the interacting part becomes the leading contribution, and perturbation theory for correlation functions breaks down. This is a manifestation in the wave function formalism of how the secular divergences of the more familiar perturbative calculation we described earlier shows up when we compute the  same quantities using the wave function. We will see later how the infrared divergences manifest themselves. 

{\bf   Leading late-time behavior of the wave function:} We can derive some general properties of the form of the wave function by studying the degree of secular growth of the diagrams. The argument is slightly lengthy, however the result is quite intuitive. The uninterested reader can skip directly to eq.~(\ref{eq:wavefunction_schematic}), the justification of which is what we are going to prove in the following. Arguments similar to what we present below were used in \cite{Weinberg:2006ac} and \cite{Harlow:2011ke}, however they do not prove exactly the statement we need here. The latter reference also emphasized the qualitative difference between the path integrals computing the wave function and correlation functions in dS.

 \begin{figure}[htbp] 
	\centering
		\includegraphics[width=.4\linewidth,angle=270]{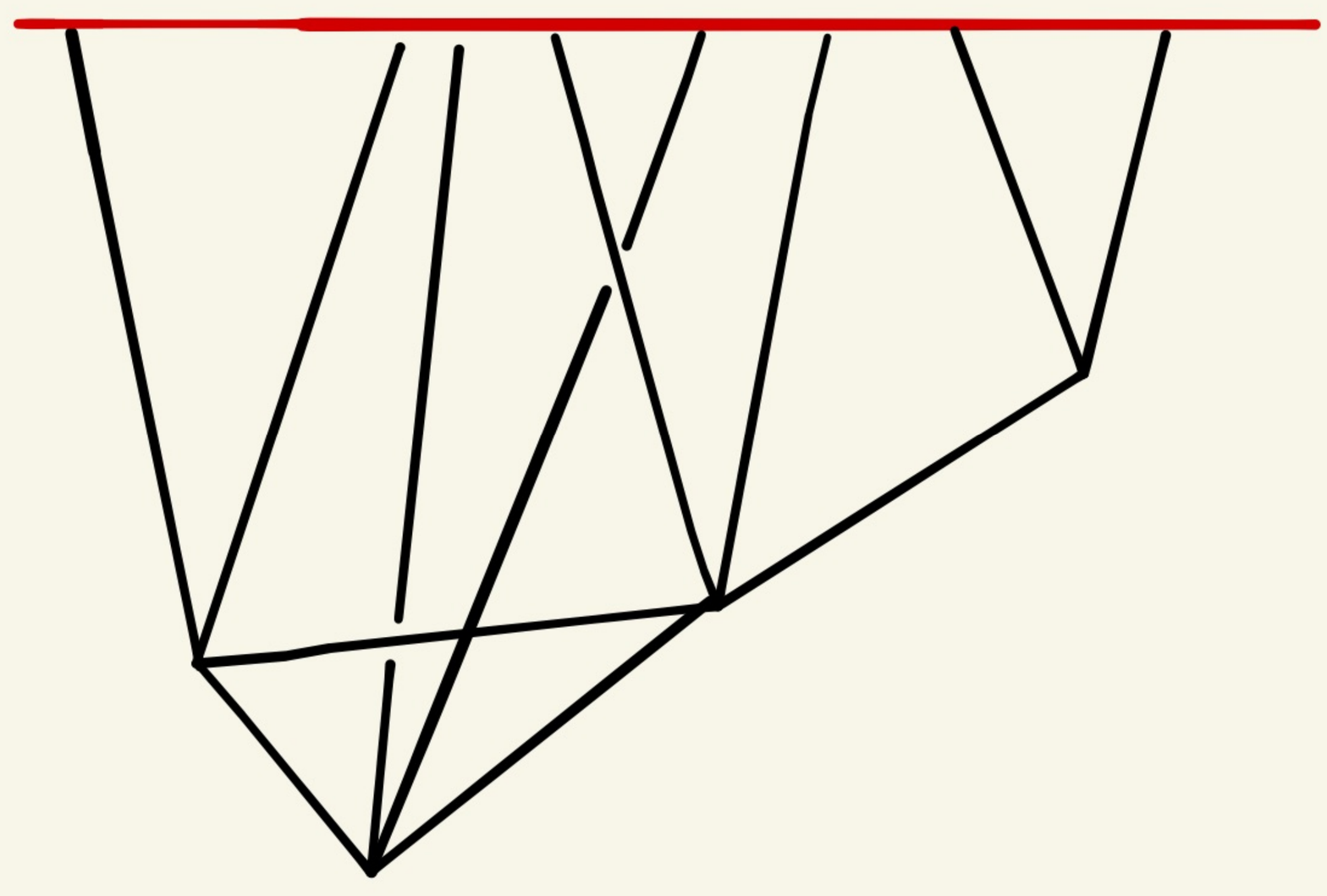}
	\caption{\small Generic Witten diagram contributing to the wave function.}
	\label{fig:generic_diagram}
\end{figure}

 First, it is possible to show that the exponent of the wave function grows at most as $1/\eta^3$. To do this, let us consider a generic connected diagram with $V$ vertices, such as the one represented in Fig.~\ref{fig:generic_diagram}.  We are interested in the behavior of this diagram as the final time is taken to be late. Generically, the diagram will have UV divergences that need to be regulated, and ultimately renormalized by adding other diagrams with just less vertices that have a similar  form to the one we consider here.  The regulator  must preserve diff. invariance, and one can therefore, for example, use dimensional regularization, which, in dS, must  be done with care~\cite{Senatore:2009cf}. One can take this path, but the same result is obtained perhaps more intuitively if  one cuts off the loop integrals at a fixed physical momentum $q_j/a(\eta_V)=\Lambda^{\rm UV}_{\rm ph}$, where $q_j$'s are the loop momenta (see again~\cite{Senatore:2009cf}). Since different vertices are inserted at different times, with this regulating scheme there is the peculiarity that for a vertex that occurs at late time, modes that were above the cutoff at some earlier time are  now below the cutoff. Clearly, in this diagram these modes cannot contribute, therefore one should regulate the theory by defining the physical momentum using the scale factor of the earliest-occurring vertex, that we label by $\eta_V$: $q_{{\rm ph},i}=q_i/a(\eta_V)=-H q_i\eta_V$. 
 
 Each vertex in our diagram carries a factor of $1/\eta_i^4$, and an integral in $\int^\eta d\eta_i$.
 Let us consider a contribution with some particular time ordering of vertices. We then label the time integration variables so that
 \be
 \eta_V\leq \eta_{V-1}\leq\ldots\leq \eta_1<0\ .
 \ee
 Now, by changing variables of integration from $q_i$ to $q_{{\rm ph},\,i}$,  we have that each {loop  carries a factor of $ \int^{{\Lambda_{\rm ph}^{\rm UV}}}_{\eta_V/L} d^3q_{{\rm ph},\,i}/\eta_V^3$}, where $L$ is some comoving IR cutoff that  we have introduced to regulate potentially-present  IR divergences. 

Let us first quickly establish that there are no IR-divergences. {The internal momenta integrals are only associated to the bulk-to-bulk propagators \eqref{Bulk2Bulk}, which, in the limit $q_{\rm ph}\to 0$, go to a constant in $q_{\rm ph}$~(\footnote{{Here for simplicity we assumed that $q_{\rm ph}$ is the total momenta flowing through the leg.} This is the same behavior in the usual in-in  formalism for correlation functions of the retarded Green's function, it is however  not the  same behavior of the free-field correlation functions, that scales as $1/q_{\rm ph}^3$ and also enters in the in-in diagrams for correlation functions.}). {This is the key technical difference between the perturbative calculations of the wave function and of correlation functions}. Therefore, since each integral carries a factor of $ \int_{\eta_V/L} d^3q_{{\rm ph},\,i}$,} there are no IR-divergences in the calculation of the wave function, and in particular we can safely take the limit $L\to \infty$.
 
 Now, all integrals  in $q_{{\rm ph},\,i}$ are integrated from zero up to the same, time-independent cutoff $\Lambda^{\rm UV}_{\rm ph}$. This means that we can evaluate all  the time integrals prior than doing the momentum integral. In this case,  the late time behavior  of the  diagram is determined by taking the late time limit of all the integrals,  {since, as we will see momentarily, all diagrams are late-time-dominated. Moreover, doing momenta integrals will not introduce any new secular divergences, since those integrals are both UV and IR finite}. 

Let us  explore the time integrals. The external fields are connected to vertices by the bulk-to-boundary propagators \eqref{eq:Bulk2Boundary}. { We can bound the bulk-to-boundary propagators by one, $K(k_j,\eta_i)\leq 1$, while we can take into account of the exponential suppression induced by these terms at early times by restricting the integration region of the time integrals to be $\eta_i<1/k_j$}~(\footnote{{At this point it is important that the mass, if any, is treated as a perturbation. Of course a finite mass included in the propagators would only make the late-time behavior of the wave function better, but it would alter some of the intermediate steps of our argument.}}). { Now, it is useful to notice that the bulk-to-bulk propagator, $G$, is bounded by 
\be
 |G(\eta_m,\eta_n,q_{\rm ph,\,j},k_i;\eta)| \leq - c \,H^2 \,\eta_m^3\,,
\ee
where $\eta_m$ is the latest of the two times $\eta_m$ and $\eta_n$, $q_{\rm ph,\,j}$ and $k_i$ are some loop and external momenta flowing through the propagator, and $c$ is a positive constant. The proof of this bound follows by simply, though tediously, checking that the function $G(\eta_m,\eta_n,q_{\rm ph,\,j},k_i;\eta)/\eta_m^3$ is bounded in the whole region of integration, and so it attains a maximum in the region.
Let us now go back to the generic diagram, and take the integrals in  chronological order. Dropping factors of $H$,  the time integral are bounded by the following expression:
 \be\label{eq:timeintegral}
\lambda^V c^{\sum_i I_i}\int^{\eta}_{-\infty} \frac{d\eta_1}{\eta_1^4}\eta_1^{3 I_1}\ldots \int^{\eta_{V-2}}_{-\infty} \frac{d\eta_{V-1}}{\eta_{V-1}^4} \eta_{V-1}^{3I_{V-1}}\int^{\eta_{V-1}}_{-\infty} \frac{d\eta_V}{\eta_V^{4+3 L}}\ ,
\ee
where $L$ is the number of loops and $I_i$ is the number of internal lines for which $\eta_i$ is the later of their endpoint vertices. Consider the first $V-j$ integrals, associated to the earliest $V-j$ {vertices}.  This defines a subdiagram with $\bar I_{V-j}$ total internal lines, {\it i.e.} internal lines for which both endpoints are attached to  any of the vertices $V_{V-j}\ldots V_V$, and $L_{V-j}$ associated loops. Because of the topological constraint  $(V-j)+L_{V-j}- \bar I_{V-j}=1$, it is straightforward to establish that the $j$-th integral behaves as $1/\eta_j^{4+3(L-L_{V-j})}$. Since $L\geq L_{V-j}$, we therefore learn that each integral is therefore dominated by the latest time, and the full integral goes at most as $1/\eta^3$. Furthermore, never at each step we encounter a logarithmic divergence. The smallest negative power an integral can take is $d\eta_i/\eta_i^{4}$. So, there is no logarithmic enhancement on top of the $1/\eta^3$ factor. 

We notice explicitly that the bound on the growth of the diagram as $1/\eta^3$ applies in particular to the limiting case $ k_i\eta\to0$, which is therefore a non-singular limit. In this limit, the contribution to the wave function is local~\footnote{ This is consistent from scaling invariance $\eta\to \Delta\cdot \eta,\, x\to \Delta\cdot x$. In the local limit, these diagrams give rise to a term that in the wave function goes as $\int d^3x\; \phi(\vec x)^E f(\eta)$, with $E$ being the number of external legs of the diagram, and $f(\eta)$ a function of $\eta$. Scaling invariance tells us that $f(\eta)\sim 1/\eta^3$. 
}. Of course this is a well-known result in the AdS/CFT context. A conceptual difference though is that we do not disregard this ``divergent'' piece, instead it will play a major role in our computations. By keeping track of the powers of $i$ one also easily sees that the leading piece is purely imaginary.}

Indeed it is now possible to show that the maximum degree of divergence  of the real part is $(\lambda\log(-k\eta))^V$, {\it i.e.} there is no $1/\eta^3$ factor. This comes from unitarity, that imposes that the wave function satisfies the following equation, which directly derives from the Schroedinger equation and which will be of great use for us later:
\bea \label{eq:firstprobabilityconservation}
\frac{\d}{\d t} \left(\Psi^\star[\phi,t]\Psi[\phi,t]\right)= -\frac{i}{2 a^3} \int  d^3 x \frac{\delta }{\delta\phi(\vec x)}\left(\Psi[\phi,t]^* \frac{\delta}{\delta\phi(\vec x)}\Psi[\phi,t]-\Psi[\phi,t]\frac{\delta}{\delta\phi(\vx)} \Psi^*[\phi,t]\right)\,.\nn \\
\eea
Writing $\Psi[\phi]$ as an exponential of a real part  and the imaginary part of the exponent, it very simply follows that for this equation to be true, if the imaginary part goes as $1/\eta^3$, the real part can go at most logarithmically in~$\log(-k\eta)$, without any power of $1/\eta$. In our exemplifying theory with an interaction of the form $\lambda\phi^4$, it holds the following topological relation between the number of external legs, $E$, the number of vertices, $V$, and the one of internal legs, $I$: $4V=E+2I$. Since, in the calculation of the wave function, each external leg corresponds to a factor of $\phi$, and since the maximum number of logarithmic factors and $\lambda$'s is $(\lambda\log(-k\eta))^V $, we conclude that the real part of the exponent of the wave function goes at most as 
\be
{\rm Re}\left[\log(\Psi)\right]\sim  \lambda^{\frac{E-2}{2}} \phi^Ek^3 \left(\log(-k\eta)\right)^{\frac{E-2}{2}} \left(\lambda\log(-k\eta)\right)^L \ ,
\ee
where $k$ represents a general combination of momenta. This expression teaches us also that each loop insertion carries potentially a factor of $\lambda\log(-k\eta)$. {The factor of $k^3$ can be derived by scaling invariance $\eta\to \Delta\cdot \eta,\, k\to k/\Delta$.}

{ We have established that the leading term in the Taylor expansion of the wave function in the external momenta is analytic. We already know that subleading terms in this Taylor expansion in $k_i\eta\ll1$ can contain logarithms.} For example such log is present in eq. \eqref{eq:wavefuncquartic}. Our next goal is to understand the structure of these logarithmic terms. From \eqref{eq:timeintegral} we see that to get a log-divergence one needs to bring three powers of $\eta$'s in the numerator. Then the integral stops being IR (late time) dominated and is divergent if all momenta are formally set to zero. Said in other words, if we differentiate our integral with respect to $k$'s three times, we can no longer simply take the $k_i\eta\to0$ limit because the time integral will diverge. In practice, for finite $k_i$, some time integral $d\eta_j$, sensitive to the momentum~$k_i$, will be cutoff { by the exponential factors in the propagators} at $\eta_j=-1/k_i$ leading to a term that goes as $\log \l(k_i\eta_j \r)$. We then learn an important lesson: for any long momentum $k_i$, each log of the form $\log \l( k_i\eta\r)$ in the final expression for the wave function is suppressed at least by the third power of the {{\it same}} momentum $k_i$. Indeed, if it were multiplied by powers of some larger momentum $k_l\gg k_i$, the corresponding time integrals would be cutoff at times $\eta_j=-1/k_l$, thus replacing $\log \l( k_i\eta\r)$ with $\log \l( k_l\eta\r)$.

This concludes the derivation of the general properties of the wave function that we wished to establish. It is now possible to see that, in computing correlation function of $\phi$, there will be IR-divergences proportional at most to powers $\log[k_i L]$. Indeed, if we use the wave function $\Psi_{BD}[\phi,\eta]$ to compute correlation functions of $\phi$ in a perturbative manner, we end up Taylor expanding the wave function around the Gaussian part, using the additional terms in $\phi$ as `vertices', and perform Wick contractions using a `final-propagator' that goes as $\delta^{(3)}(\vec k+\vec k')/k^3$. Then, consider a generic connected diagram and cut out of it a generic subdiagram, which, in general, will have some external legs, some loops, vertices and internal legs. We are interested in the case where some momenta become soft, while others are hard. Each loop carries a factor of $\int d^3k$, and each vertex carries a factor of the hard momenta cubed, unless all momenta are soft, in which case it is a factor of the soft momenta cubed. For a generic configuration of the external momenta, only one internal leg for each loop will be soft, in which case we obtain at most a logarithmic IR-divergence. Additional internal legs can become soft, but at the cost of forcing particular configurations for external momenta, and so of imposing to be soft the phase space either of the other loops in the subdiagram, or of the external legs of the sub-diagram, that are going to be, in-their turn, integrated in $\int d^3k$ as they are either part of another loop, or, if they are connected to the external points of the $n$-point function, because physically relevant correlation functions are computed for fields in real space. Therefore, we introduce a term cubic in the soft momenta from $\int d^3k$ for each additional internal line becoming soft, so that, again, we obtain at most logarithmic divergences. As we will see in a moment, such logarithmic divergences are indeed present if one attempts to compute the correlators from the wave function directly.

{\bf Perturbativity of the wave function:} We are now in the position to discuss the sense in which the perturbative expression for the wave function can be used. From the  discussion above, it emerges that the perturbatively-computed wave function takes the following schematic form:
\bea\label{eq:wavefunction_schematic}
&&\log \Psi_{BD}[\phi,\eta]\sim \sum_L \int \left\{ \frac{\phi(\vec k)^2}{H^2} \left[\frac{i \tilde m^2}{\eta^3}+\frac{i k^2}{\eta}+ \sum k_i^3 \left(\lambda \log\left(- k_j\eta\right)\right)^L\right]\right.\\ \nn
&&\quad\left.+\lambda \,\frac{\phi(\vec k)^4}{H^4} \left[\frac{i}{\eta^3}+ (1+i)\sum k_i^3 \log\left(- k_i\eta\right)\left(\lambda \log\left(- k_j\eta\right)\right)^L\right]\right.\\\nn
&&\quad\left.+\lambda^2 \,\frac{\phi(\vec k)^6}{H^6} \left[\frac{i}{\eta^3}+(1+i) \sum k_i^3 \left(\log\left(- k_i\eta\right)\right)^2\left(\lambda \log\left(- k_j\eta\right)\right)^L\right]+\right.\\  \nn
&&\quad\left.+\ldots+\lambda^{\frac{E-2}{2}} \,\frac{\phi(\vec k)^{E}}{H^E} \left[\frac{i}{\eta^3}+(1+i) \sum k_i^3 \left(\log\left(- k_i\eta\right)\right)^{\frac{E-2}{2}}\left(\lambda \log\left(- k_j\eta\right)\right)^L\right]+\ldots\right\}\ ,
\eea
where we dropped all numerical coefficients and did not keep track of the  various wavenumbers. We focused on the long-wavenumbers contribution and just highlighted in this expression several important features: the leading part goes as $1/\eta^3$ and is purely imaginary; the largest number of fields for a given number of factors of the coupling constant comes from the tree-level contribution and goes as $\lambda^n\phi^{2n+2}$; the real part starts with terms that are not larger than $k_i^3\log(-\eta k_i)^V$ and we stress again that each log is protected by the third power of the same momentum; $\tilde m^2$ is the ``mass'' term appearing in the wave function. It is neither the bare mass in the Lagrangian, nor the ``physical'' mass that we denote $\bar m^2$ and that we assume is small ($\bar m^2\ll\sqrt{\lambda} H^2$) in order to simplify the counting.\footnote{We would like to point out that the wave function potentially contains pieces that explicitly depend on positive powers of the UV cutoff. Indeed, the renormalization procedure should be carried out in a way that keeps the correlation functions finite. The path integral that computes correlators from the wave function may have some UV divergences and hence the wave function should explicitly contain the counter terms to render those finite. This observation will not play any role in what follows. }

In order to compare the various terms in \eqref{eq:wavefunction_schematic},  and to determine which ones one needs in order to compute correlation functions of $\phi$ to a given order, which are affected by IR and secular divergences, one needs to know the typical value  of $\phi$. In order to address this apparent problem, we will next establish a non-perturbative formalism that will allow us  to compute 
correlation functions of $\phi$ upon assuming that the wave function above is controlled by a perturbative expansion. Within such an assumption, we will find 
that the  typical value  of $\phi$ in position space is $\lambda^{-1/4}$ and that each momentum mode $\phi_{\vk}$ for small $k$ decays in time as $(k\eta)^{c\sqrt{\lambda}}$, where $c$ is some 
constant. This implies, in particular, that  the secular logarithms saturate at most to an $O(1/\sqrt{\lambda})$ number. Now, by inspection of (\ref{eq:wavefunction_schematic}), we find that the imaginary part of the exponent is dominated  by the quartic term, with some other  terms contributing by corrections of  order $\sqrt{\lambda}$ or $(k\eta)^2$.  The real terms, as well as the additional subleading imaginary terms, are instead more complicated to estimate, as they involve logarithms. Let us take the leading logarithmic terms that scale as $k_i^3 \left(\lambda \phi^2\log(k_i\eta)\right)^n$. It turns out that weather they are hierarchically organized or not  depends on the actual calculation one uses the wave function for. If one were to compute correlation functions of $\phi$'s in a perturbative manner, only the real part would matter, {and IR divergences would reappear. Moreover, if IR divergences are regulated by a cutoff or by a small mass, one would also see that the factors of $k_i^3$ protecting the { secular} log's are cancelled by the propagators, and { these} log's become IR-dominated.} This recovers the secular divergences observed in  in-in perturbation theory for correlators. One can also see that even if one uses the physical distribution of the long modes which we will prove to be true and which has a faster decay at low $k$ than the free theory one, {so that each log divergence is effectively replaced by $\lambda^{-1/2}$, infinitely many terms in the wave function contribute at the same order and} ordinary perturbation theory is not saved.

Even though, in our formalism,
we still need to compute various { contributions from terms involving}  logarithms~\footnote{In fact, this only becomes necessary when one computes corrections of higher order in $\lambda$ than what we do in this paper, however, nothing precludes us from computing these corrections in principle.},  the factor of $k_i^3$ always remains protecting the logs, hence the corresponding $d^3 k$ integrals are UV dominated and the logs only give ${O(\log\eps)}$ not $O(1/\sqrt{\lambda})$ contributions. Thus for us, in order to compute correlators to any finite order in $\lambda$, as well as in our other parameters such as $\eps$, we only need finitely many terms in the wave function. We { show this more explicitly} in section \ref{sec:subleading-tadopoles}, once we have our non-perturbative probability distributions and correlators computed.

\section{EFT for long modes}
\label{sec:EFT}
\subsection{Continuity equation}

As explained above, we expect it to be possible to derive a closed set of equations that describes the dynamics of the long modes. Even though these modes are strongly coupled, their evolution is ultra-local which leads to a major simplification. Using the Shroedinger equation as well as knowledge of the wave function of the BD state  we are going to derive the equation the probability distributions for field values in this state satisfy. We first focus on distributions of fields on the same time slice, generalization to the case of several times will be considered in section~\ref{sec:nonequil1pt} and \ref{sec:generic-2-point}. We would also like to stress that it is possible to develop a formalism which is state-independent (within some class of states) however, at the cost of a significant increase in complexity. Some outline of this formalism is provided in appendix \ref{app:Wigner}. {We also discuss the universality of our result with respect to variations of the state in section \ref{sec:stability}}. 

As we already pointed out several times, knowledge of the wave function still does not allow us to compute the $n$-point probability distributions directly. Instead, we are going to go back to the Shroedinger equation and use the wave function only to compute  some of the terms in the equation.
For simplicity let us specify to the theories with the Hamiltonian of the form
\be
\Ham[\pi,\phi,t]=\int d^3 x\; \left(-\frac{1}{2 a(t)^3}\l(\frac{\delta}{\delta\phi(\vx)}\r)^2+\frac{a(t)^3}{2 }V\l(\phi(\vx)\r)+\frac{a(t)^3}{2}\frac{(\nabla\phi)^2}{a(t)^2}\right)\;,
\ee
{where for now we will keep a general $V(\phi)$, assuming that it has some perturbative parameter like $\lambda$ in $\lambda \phi^4$.}
Theories with higher powers of momentum in the Hamiltonian can be treated identically, as long as these higher terms are suppressed by some high UV scale. Then the full functional probability distribution $P[\phi,t]$ satisfies the continuity equation which was already used above in~\eqref{eq:firstprobabilityconservation} to bound the real part of the wave function. We repeat this equation here:
\bea
\label{continuityeq}
&&\frac{\d P[\phi,t]}{\d t}= -\frac{i}{2 a^3} \int  d^3 x \frac{\delta }{\delta\phi(\vec x)}\left(\Psi[\phi,t]^* \frac{\delta}{\delta\phi(\vec x)}\Psi[\phi,t]-\Psi[\phi,t]\frac{\delta}{\delta\phi(\vx)} \Psi^*[\phi,t]\right)\,.
\eea
This is not yet a closed equation for $P[\phi,t]$, however, for a given $\Psi[\phi,t]$ we can define and compute~\footnote{Notice that we have included a factor of $1/a^3$ in this definition, for later convenience}: 
\bea
\label{Pidef}
&&\Pi[\phi,\vx,t]=-\frac{1}{a^3}\frac{\delta}{\delta\phi(\vx) }{\rm{Im}}\log\Psi[\phi,t]\,.
\eea
For a semiclassical wave function $a^3 \Pi[\phi,\vx,t]$ is just the value of the momentum as a function of the field value on the classical equations of motion. We will often call $\Pi$ and related quantities just ``momentum'' for shortness.  For a general wave function, up to a factor of $a^{3}$, it is the real part of the momentum operator, $\hat \Pi$, evaluated on the wave function in question. What is important for us is that for a given state this is some known-to-us functional of fields. This is the main application of the fact that the wave function can be computed in perturbation theory, and consequently in practice is known to us. Now we can write a closed form equation for the distribution:
\bea
&&\frac{\d P[\phi,t]}{\d t}= - \int  d^3 x \frac{\delta }{\delta\phi(\vec x)}\l(\Pi[\phi,\vx,t]P[\phi,t]\r)\,.
\eea


\subsection{Functional equation for long modes\label{sec:functionequation}}
Our next step is to separate the field into `long' and `short' components by introducing a splitting in Fourier space  at a fixed physical wavenumber. We therefore define 
\bea
\phi(\vx)=\int_0^{\Lambda(t)}\frac{d^3 k}{(2\pi)^3}\;  \;e^{i \vk \cdot \vx}\phi({\vk})+\int_{\Lambda(t)}^\infty  \frac{d^3 k}{(2\pi)^3} \;e^{i \vk \cdot \vx}\phi(\vk)\equiv\phi_\ell(\vx)+\phi_s(\vx)\;,
\eea
where the integration limits are implied for the absolute value of the momentum and 
\be
\Lambda(t)=\eps a(t) H\;.
\ee
In what follows we will often work in the coordinate representation for the long modes and in momentum space for the short modes. In this case, by $\phi_s$, where `$s$' stays for `short', we will mean the collection of all modes $\phi_k$ with $|k|\geq\Lambda(t)$.

\begin{figure}[htbp] 
	\centering
		\includegraphics[width=.5\linewidth]{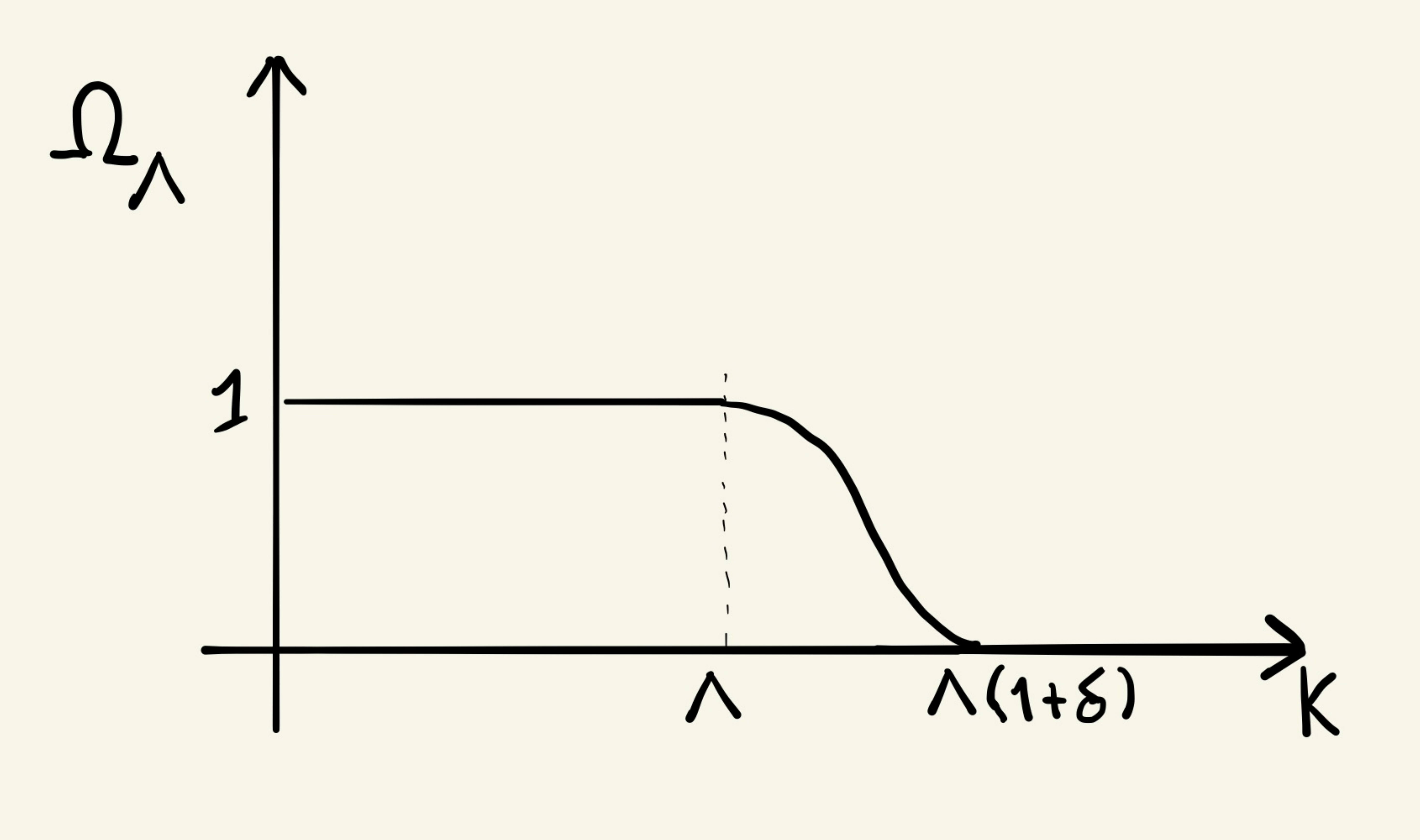}
	\caption{\small Pictorial representation of the window function that splits the long and the short modes.}
	\label{fig:window}
\end{figure}

Not so surprisingly,  the sharp momentum cutoff introduced above can lead to violation of locality in coordinate space. To circumvent related complications (see Appendix \ref{app:locality-space} for discussion) 
we also define a smooth window function $\Omega_{\Lambda(t)}(k)$ with a smooth transition around the time-dependent comoving momentum $\Lambda(t)$,
and with a width of order $\delta \Lambda(t)$, with $\frac{e^{-\frac{1}{\sqrt{\lambda}}}}{\epsilon}\ll \delta\ll \sqrt{\lambda}$,  such that (see Fig.~\ref{fig:window})
\begin{equation}
\Omega_{\Lambda(t)}(k) = \left\{
\begin{array}{rl}
1 & \text{for}\; k \leq \Lambda(t),\\
0 & \text{for }\; k \geq (1+\delta) \Lambda(t)\,.
\end{array} \right.
\end{equation}
and smoothly transitions between $1$ and $0$ for $\Lambda(t)\leq k \leq  (1+\delta) \Lambda(t)$. In the limit $\delta\to 0$, the window function must converge to a unit-step function. An example of such a function is the one defined at eq.~(29) of~\cite{Polchinski:1983gv}, though its explicit form will not be needed here. We then define the smoothly cutoff long field as
\bea\nn
\phi_\Omega(\vx)=\int\frac{d^3 k}{(2\pi)^3}\; \Omega_{\Lambda(t)}(k)\; \;e^{i \vk \cdot \vx}\phi({\vk})\,.
\eea

\begin{figure}[htbp] 
	\centering
		\includegraphics[width=.5\linewidth]{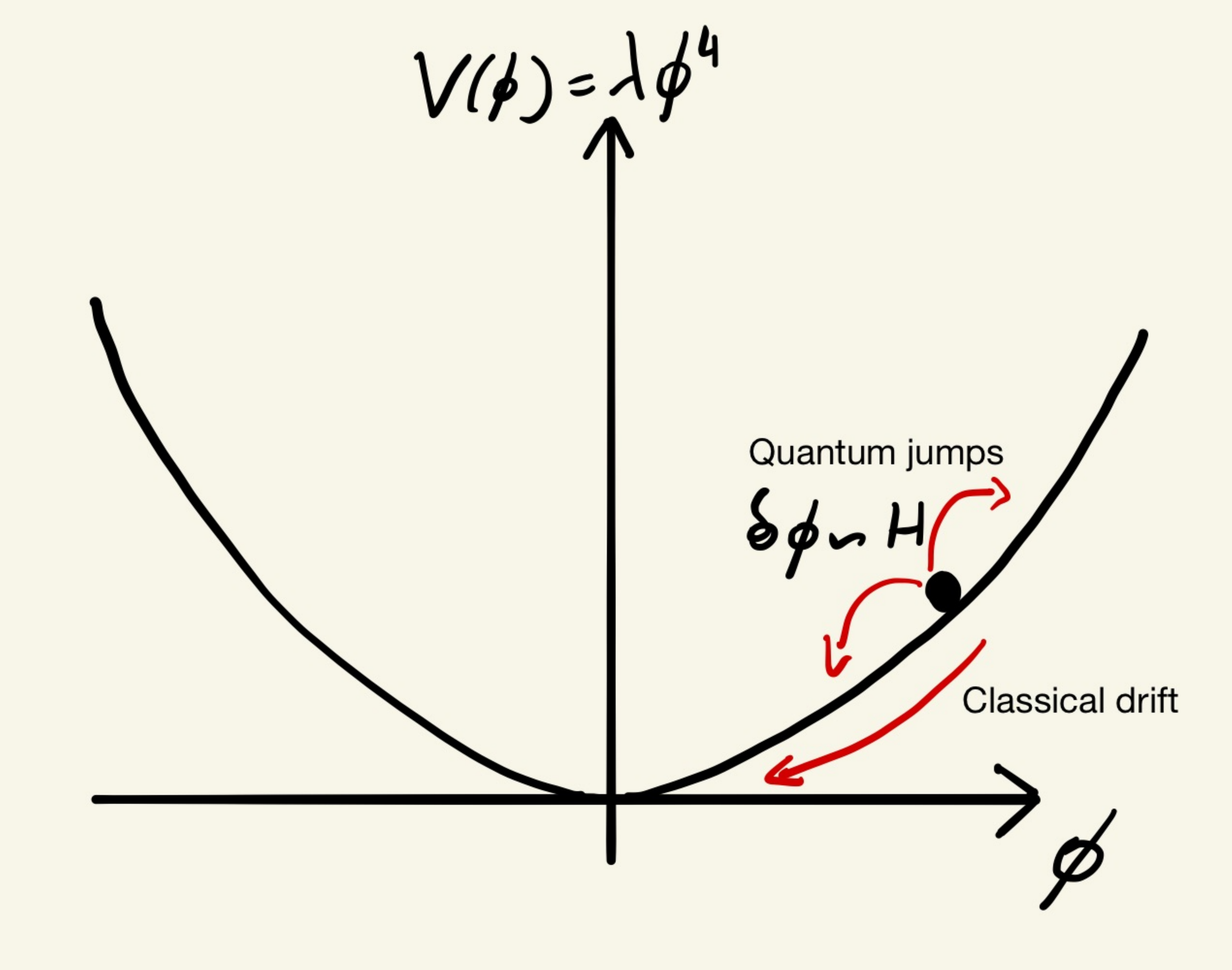}
	\caption{\small Pictorial representation to the two contributions to the long-wavelength dynamics. The diffusion due to the short modes becoming long, and the drift from the evolution  of the long modes. It is expected that an equilibrium value with the typical values, for $V=\lambda\phi^4$, of order $\phi\sim H/\lambda^{1/4}$ is reached.  We will make this picture rigorous. }
	\label{fig:schematicFokker}
\end{figure}

Now we can define the probability distribution of these smoothly cutoff long modes:
\be
P_\Omega[\phi_\Omega,t]=\int {\cal{D}}\phi \;\delta\left[\phi_{\Omega }(\vec x)-\int\frac{d^3 k}{(2\pi)^3}\; \Omega_{\Lambda(t)}(k)\; e^{i\vec k\cdot\vec x}\; \phi(\vec k)\right]\;  P[\phi,t]\ .
\ee
Notice the square brackets indicating the functional $\delta$-function. 
Let us derive an equation which describes the time evolution of the long distribution. Clearly, when we take the time derivative of  $P_\Omega[\phi_\Omega,t]$, there will be two terms: one, which we will call the `Drift', is when the time derivative acts on the probability distribution inside the integral and the other, the `Diffusion' term, when it acts on the $\delta$-functions (see Fig.~\ref{fig:schematicFokker}). Thus, schematically, we can write:
\be\label{eq:schematic}
\frac{\d}{\d t} P_\Omega [\phi_\Omega,t]= {\rm Diffusion} + {\rm Drift}\; .
\ee

Let us derive explicitly these terms, and let us start with the drift term:
\bea\label{eq:drift_general}
&&{\rm Drift}=\int {\cal{D}}\phi \; \delta\left[\phi_\Omega(\vec x)-\int \frac{d^3 k}{(2\pi)^3}\; \Omega_{\Lambda(t)}(k)\;  e^{i\vec k\cdot\vec x}\,  \phi(\vec k)\right] \frac{\d }{\d t}P[\phi,t] =\\ \nn
&&\qquad \ =-\int {\cal{D}}\phi \; \delta\left[\phi_{\Omega }(\vec x)-\int \frac{d^3 k}{(2\pi)^3}\; \Omega_{\Lambda(t)}(k)\; e^{i\vec k\cdot\vec x}\, \phi(\vec k)\right] \int  d^3 x \frac{\delta }{\delta\phi(\vec x)}\l(\Pi[\phi,\vx,t]P[\phi,t]\r)\;,
\eea
where $\Pi[\phi,\vx,t]$ is given in \eqref{Pidef}. At this point, it is convenient to approximate the window function as a unit step function at $k=\Lambda(t)$, which is correct up to $O(\delta)$ corrections. To do this we define the long distribution of the sharply cutoff long modes
\be
P_\ell[\phi_\ell,t]=\int {\cal{D}}\phi \;\delta\left[\phi_{\ell }(\vec x)-\int^{\Lambda(t)}\frac{d^3 k}{(2\pi)^3}\; e^{i\vec k\cdot\vec x}\; \phi(\vec k)\right]\;  P[\phi,t]\ .
\ee
In fact, up to corrections of order $\delta$, smooth and sharp distributions coincide:
\be
\label{deltacor}
P_\ell[\phi_\ell,t]=P_\Omega[\phi_\ell,t]\l(1+O(\delta)\r)\;.
\ee
 We will discuss such corrections in $\delta$ later on in Appendix \ref{app:delta_corrections}. Then we can explicitly use the momentum representation and separate the variational derivatives into those with respect to short and long modes. 
Derivatives with respect to short modes are total derivatives and get integrated to zero. Derivatives with respect to long modes can be integrated by parts to act on the $\delta$-function and then pulled out of the integral. Consequently, we get
\bea
&&\!\!\ {\rm Drift}=-\int d^3 x \;\int_0^{\Lambda(t)}\frac{d^3 k'}{(2\pi)^3}\; e^{i \vk'\cdot \vx}\; \times\\ \nn 
&&\times\frac{\delta}{\delta \phi_\ell(\vx)}\int {\cal{D}}\phi \; \delta\left[\phi_{\ell }(\vec y)-\int_0^{\Lambda(t)} \frac{d^3  k}{(2\pi)^3}\, e^{i\vec k\cdot\vec y} \phi(\vec k)\right]  \Pi[\phi,\vk',t]P[\phi,t ]\l(1+{{O}}\l(\delta\r)\r)\ .
\eea
Now we can take integrals over $\vk'$. We are doing it so pedantically because there is actually a subtlety:
\bea
\int_0^{\Lambda(t)}\frac{d^3  k'}{(2\pi)^3}\; e^{i \vk'\cdot \vx}\Pi[\phi,\vk',t]=\int_0^{\Lambda(t)}\frac{d^3  k'}{(2\pi)^3}\; e^{i \vk'\cdot \vx}\; \int d^3\vx'\;e^{-i \vk'\cdot \vx'}\Pi[\phi,\vx',t]\equiv\Big[ \Pi[\phi,\vx,t]\Big]_{\Lambda(t)}\;.\nn\\
\eea
Indeed, the real-space expressions appearing in the equation for the long modes has to be transformed to momentum space and transformed back in real space by using only the modes up to $|k|=\Lambda(t)$. This is what we mean by the definition $\left[\dots \right]_{\Lambda(t)}$. This is one of the effects related to the fact that long modes have small but non-vanishing momentum: several long modes can source a short mode in the equation. Keeping this in mind, the drift term reads:
\bea\nn\label{eq:drift-general}
&&{\rm Drift}=-\int d^3 x \frac{\delta}{\delta \phi_\ell(\vx)}\l( \left\langle\Big[ \Pi[\phi,\vx,t]\Big]_{\Lambda(t)}\right\rangle_{\phi_\ell}P_\ell[\phi_\ell,t]\r)\l(1+{{O}}\l(\delta\r)\r)
\eea
Here we defined the `expectation values in the long background' $\left\langle\op[\phi]\right\rangle_{\phi_\ell}$ in the following way
\be\label{eq:expvalue}
 \left\langle\op[\phi]\right\rangle_{\phi_\ell} P_\ell[\phi_\ell,t] \equiv \int \MD \phi_s \; \op[\phi] \,P[\phi,t]\;.
\ee 
It is intuitively clear that this object can be computed in perturbation theory in a way that does not suffer from IR divergences since the long modes are kept fixed and not integrated over.  Showing this is a simple application of the proof on the maximum degree of IR divergence that we discussed in  section~\ref{sec:wavefunctionperturbativity}. We will check that it is indeed the case by explicit calculations.

It is tempting to connect the drift term in~(\ref{eq:drift-general})  with the Coleman-Weinberg effective potential that accounts for the renormalization of the potential. Indeed, it is true that the corrections in~(\ref{eq:expvalue}) contain the renormalization to the Hamiltonian as due to the short modes. In practice, though, these are different in form with respect to the ones in Minkowski space  because the propagators for modes longer than the Horizon are different. Maybe more importantly, while the Coleman-Weinberg effective potential indeed controls part of the dynamics, this does not exhausts it. In fact, as we will see, the long modes move in this effective potential, but are also influenced by what we call the `diffusion' term, which also receives radiative corrections. In a sense, the dynamics controlled in Minkowski by the Coleman-Weinberg potential is here generalized to a dynamics controlled by a non-Hamiltonian equation, a Fokker-Planck-like equation, that we are constructing, and for which one can include radiative corrections, as we will see.

Next, we turn to analyzing the diffusion term in \eqref{eq:schematic},
\bea
&&{\rm Diffus.}=\int {\cal{D}}\phi \; \frac{\d}{\d t}\left( \delta\left[\phi_{\ell }(\vec x)-\int \frac{d^3 k}{(2\pi)^3}\; \Omega_{\Lambda(t)}(k)\;  e^{i\vec k\cdot\vec x}\,  \phi(\vec k)\right]\; \right)\; P[\phi,t]\;.\\ \nn
\eea
Here it is so far important to keep the smooth window function. The time derivative acts on $\Omega_{\Lambda(t)}(k)$, and therefore affects only the intermediate modes in the momentum shell of modulus in $[\Lambda(t),(1+\delta)\Lambda(t)]$. By carefully Taylor expanding the $\delta$-function for these modes, we show in App.~\ref{app:smooth-diffusion} that the diffusion term takes the form:
\bea
&&{\rm Diffus.}=\left(\int {\cal{D}}\phi \; \left(\int d^3 x\; \frac{\delta}{\delta\phi_{\ell }(\vx)} \,\delta\left[\phi_{\ell }(\vec y)-\int_0^{\Lambda(t)} \frac{d^3  k}{(2\pi)^3}\; e^{i\vec k\cdot\vec y} \phi(\vec k)\right]\ \times \right.\right.\\ \nn
&&\left.\qquad\qquad\qquad\qquad\qquad\qquad\qquad\qquad\qquad\qquad\qquad\qquad \;\times\ (-\dot\Delta \phi(\vx))\; P[\phi,t]\right.+\\\nn
&&\left.+ \int d^3 x \;\int d^3 x'\;\frac{\delta^2}{\delta\phi_{\ell }(\vx)\delta\phi_{\ell }(\vx')} \,\delta\left[\phi_{\ell }(\vec y)-\int_0^{\Lambda(t)} \frac{d^3  k}{(2\pi)^3}\; e^{i\vec k\cdot\vec y} \phi(\vec k)\right]\ \times\right.\\ \nn
&&\qquad\qquad\qquad\qquad\left. \left. \;\times\ \left(\dot\Delta \phi(\vx)\Delta \phi(\vx')\right)\; P[\phi,t]\right) \right) \ \times (1+O(\delta))\;,
\eea
where
\bea\label{dgamma}
&&\Delta\phi(\vx)={\int_{\Lambda(t)}^{(1+\delta)\Lambda(t)}\frac{d^3  k}{(2\pi)^3}\Omega_{\Lambda(t)}(k)}\; e^{i\vec k\cdot \vec x} \phi(\vec k)\ , \\ \nn
&&\dot{\Delta}\phi(\vx)={\int_{\Lambda(t)}^{(1+\delta)\Lambda(t)}\frac{d^3  k}{(2\pi)^3}\;\frac{\d}{\d t}\left(\Omega_{\Lambda(t)}(k)\right)\; e^{i\vec k\cdot \vec x} \phi(\vec k)}\ .
\eea
We show in App.~\ref{app:smooth-diffusion} in explicit computations in perturbation theory that both the terms that are linear and quadratic in $\Delta\phi$ produce finite contributions in the limit $\delta\to 0$, while terms of order  $(\Delta\phi)^3$ and higher give vanishing contributions in the limit $\delta\to 0$, due to more than one phase space integrations over the momentum shell with thickness $\sim\delta$. Consequently the Taylor expansion of the $\delta$-functions leads to a perturbative structure in~$\delta$.

Now we can perform manipulations similar to those we did for the drift term. We pull the derivatives with respect to $\phi_\ell$ out of the integral, and evaluate the expectation values of $\Delta\phi$'s in perturbation theory, {using the perturbative wave function and keeping the long modes fixed in the path integral for correlation functions.} Perturbation theory is still valid for modes comprising ~$\Delta\phi$'s, and, even though they belong to the support of the window function $\Omega_{\Lambda(t)}$, we will treat them as short modes since the $\delta$-functions were now Taylor expanded. After these manipulations, the diffusion term reads:
\bea\label{eq:diffstep}\nn
&&{\rm Diffus.}= \l( \int d^3 x\;\frac{\delta}{\delta\phi_{\ell }(\vx)}\l(\left\langle-\dot\Delta \phi(\vx)\right\rangle_{\phi_\ell}\; P_\ell[\phi_\ell,t]\r)\;+\r.\\ \nn
&&\l.\ +\;\int d^3 x \;\int d^3 x'\;\frac{\delta^2}{\delta\phi_{\ell }(\vx)\delta\phi_{\ell }(\vx')} 
\l( \left\langle\dot\Delta \phi(\vx) \Delta \phi(\vx')\right\rangle_{\phi_\ell}\; P_\ell[\phi_\ell,t]\r)\r)\ \times\\ 
&&\ \times\ \left(1+O(\delta)\right)\;.
\eea

Let us finally put together the drift \eqref{eq:drift-general} and diffusion terms to get the full equation for the long-modes functional probability distribution:
\bea\label{eq:full}
&&\frac{\d}{\d t} P_\ell[\phi_\ell,t]=\Bigg[\l.-\int d^3 x \frac{\delta}{\delta \phi_\ell(\vx)}\l( \left\langle\Big[ \Pi[\phi,\vx,t]\Big]_{\Lambda(t)}\right\rangle_{\phi_\ell}P_\ell[\phi_\ell,t]\r)+\nn\r.\\
&&+ \int d^3 x\;\frac{\delta}{\delta\phi_{\ell }(\vx)}\l(\left\langle-\dot\Delta \phi(\vx)\right\rangle_{\phi_\ell}\; P_\ell[\phi_\ell,t]\r)\;+\\ \nn
&& +\;\int d^3 x \;\int d^3 x'\;\frac{\delta^2}{\delta\phi_{\ell }(\vx)\delta\phi_{\ell }(\vx')}\l(  \left\langle\dot\Delta \phi(\vx) \Delta \phi(\vx')\right\rangle_{\phi_\ell}\; P_\ell[\phi_\ell,t]\r) \Bigg]\l(1+O\l(\delta\r)\r)\,,
\eea
where on the left hand  side we also used \eqref{deltacor}.


\subsection{Structure of the perturbative expansion: $\lambda$, $\hbar$, $\nabla$, and $\delta$\label{sec:perturbatie-structure}}

Let us summarize the  structure of the functional equation we just derived. We would like to stress that so far we neglected only the  corrections from the finite width of the window function, which are of order $\delta\ll{\sqrt{\lambda}}$. So \eqref{eq:full} is formally all orders in $\lambda$ and gradients or, equivalently, $\epsilon$. This is a functional equation with a similar structure to the renowned Fokker-Planck equation, with a drift and a diffusion term. However, there are a few important differences. For example the diffusion terms contains a term which is first order in field-derivatives of the long modes and the expectation values are, in principle, complicated functionals of the long field.

So far, equation  \eqref{eq:full} is still too complicated to be solved even to leading order in $\lambda$, because of its functional nature. Most importantly, even if we were  able  to solve it, in order to compute correlation functions of $\phi$'s, we would still need to compute a strongly non-Gaussian path integral over {$\phi_\ell$, which would be practically impossible.} At this point we can drop the gradients of long modes in the Hamiltonian, keeping in mind that they give corrections of order $\eps^2$ only. After this is done, as we will see, the long modes equation becomes ultra-local, {\it i.e.} different spatial points affect each other in a perturbative expansion in powers of $\sqrt{\lambda}$, and we can consider distributions for the values of long fields localized at a finite number of spacial points. In fact the theory admits an expansion in the number of locations, which is in some sense opposite to the derivative expansion employed in usual effective field theories. Physically, this originates from the presence of the horizon in de Sitter space.\footnote{It is an interesting question what are the conditions under which such an expansion becomes useful.  } More concretely, we define the distribution for fields in $n$ spacial locations, the ``$n$-locations'' distribution: \footnote{{To avoid proliferation of lower indices we denote long modes localized at the spacial location $\vx_i$ by $\phi_i$ with the index $\ell$ dropped.}} 
\bea\label{eq:n-location-definition}
&&P_n\l(\phi_\ell(\vx_1)=\phi_1,t;\phi_\ell(\vx_2) = \phi_2,t;\ldots;\phi_\ell(\vx_n)=\phi_n,t\r)\;=\\ \nn
&&\qquad\qquad\qquad\qquad\qquad\qquad\qquad\qquad\;=\int\MD\phi_\ell  \prod_{i=1}^n \delta\l(\phi_{\ell }(\vx_i)-\phi_{i}\r)P_\ell[\phi_\ell,t].
\eea
These objects are functions of finitely many variables and satisfy partial differential equations, as opposed to functional ones. We will see below that for low $n$'s, these equations can be easily solved perturbatively in $\lambda$. In fact, at a given order in $\lambda$, the evaluation of some $n$-point distribution requires knowledge of the distributions with different number of points, but only finitely many of them, so that one always gets a closed system of equations.

To compute non-equal time correlators we need an object more general than \eqref{eq:n-location-definition}. For short modes non-equal time correlators can be computed, as usual, with the help of perturbation theory, for example, by inserting extra operators in the path integral that defines the wave-functional. If we treat long modes as classical, non-equal time correlators are also computed straight-forwardly.  In particular, we can define a joint-probability of the form $P_\ell^{[n]}[\phi_\ell,t;\phi_{\ell, 2},t_2;\ldots;\phi_{\ell , n},t_{n}]$, which is the probability to have $\phi_{\ell , n}$ at time $t_{n}$, $\phi_{\ell , n-1}$ at $t_{n-1},\, \ldots$ .
 Its $t$-derivative satisfies the same equation as $P_\ell[\phi_\ell,t]$ and the initial conditions are given by
\be
\label{Pnfunct}
P_\ell^{[n]}[\phi_\ell,t;\phi_{\ell, 2},t_2;\ldots;\phi_{\ell , n},t_{n}]\Big|_{t\to t_{2}}=\delta\l[\phi_\ell(\vx)-\phi_{\ell, 2}(\vx)\right]P_\ell^{[n-1]}[\phi_{\ell 2},t_2;\ldots;\phi_{\ell , n},t_{n}]
\;.
\ee
{From the functional $P^{[n]}$ one can proceed to define arbitrary multiple-point distributions which keep track of fields at several space-time points. Then partial differential equations that these objects satisfy can be derived following our methodology. Examples are given in sections \ref{sec:two-point} and \ref{sec:generic-2-point}.
If we would like to include quantum mechanical corrections for long modes, suppressed at least by $\hbar \eps^3$, the definition of $P_\ell^{[n]}$ requires some extra care. Corresponding subtleties are discussed in section~\ref{sec:quantum}.}

At last, there are the corrections in $\delta$, which are conceptually straightforward. They simply come from taking into account of the form of window function $\Omega_{\Lambda(t)}(k)$ in the diffusion and drift terms. We discuss these in App.~\ref{app:delta_corrections}. In summary, we have anticipated how to express the exact solution in a perturbative expansion in three small parameters, $\epsilon$, $\sqrt{\lambda}$ and $\delta$. In the rest of the paper, we are going to make this construction more explicit, {and, in such a process, we will include an additional expansion parameter, $\Delta$,{ which controls time-dependent perturbation theory used for solving the differential equation for the $n\geq2$ $n$-point distributions.}}.

\section{Single-point distribution}
\label{sec:Singlepoint}
\subsection{Equation for $P_1(\phi_1,t)$}
Let us start with studying the one-point distribution $P_1(\phi_\ell(\vx_1)=\phi_1,t)$, or, for shortness $P_1(\phi_1,t)$. 
The equation for $P_1$ can be obtained by taking the time derivative of the definition~\eqref{eq:n-location-definition} and using the equation for $P_\ell$, \eqref{eq:full}. We can integrate by parts the functional derivatives and trade them for derivatives with respect to $\phi_1$, after which we get:
\bea
\label{1ptfull}
&&\frac{\d}{\d t} P_1(\phi_1,t)=\Bigg [-\frac{\d}{\d \phi_1} \int {\cal {D}}\phi_\ell\; \delta\l(\phi_{\ell }(\vx_1)-\phi_{1}\r)\left\langle\Big[ \Pi[\phi,\vx_1,t]\Big]_{\Lambda(t)}\right\rangle_{\phi_\ell}P_\ell [\phi_\ell,t]+\nn\\
&&+\l(\frac{\d}{\d\phi_{1}}\int {\cal{D}}\phi_\ell  \; \delta\l(\phi_{\ell}(\vx_1)-\phi_{1}\r)\left\langle-\frac{\d}{\d t}\Delta \phi(\vx_1)\right\rangle_{\phi_\ell}\; P_\ell[\phi_\ell,t]\;+\r.\\ \nn
&&\l. +\;\frac{\d^2}{\d\phi_{1}^2} \int {\cal{D}}\phi_\ell  \; \delta\l(\phi_{\ell}(\vx_1)-\phi_{1}\r)\left\langle\dot\Delta \phi(\vx_1) \Delta \phi(\vx_1)\right\rangle_{\phi_\ell}\, %
\ P_\ell[\phi_\ell,t]\r)\Bigg]\times
\nn\\
&& 
\times\;\l(1+O\l(\delta\r)\r)\;.\nn
\eea
Notice that, with respect to \eqref{eq:full}, the functional $\delta$-functions have been replaced by ordinary ones. Note also the reason why we did not get a closed equation for $P_1$ yet: the expectation values present in the equation depend on the long field at different spacial locations. For example, as we will see, the truncation in momentum space present in the drift term in the first line of the equation requires knowledge of the long field at least at two locations, or, equivalently, of the distribution $P_2\l(\phi_\ell(\vx_1)=\phi_1,t;\phi_\ell(\vx_2) = \phi_2,t\r)$. Similarly, the tadpole terms in the third line are in general non-vanishing due to the fact that the long-modes background is not exactly spatially homogeneous and knowledge of the equal-time two-location distribution is also necessary to carefully account for this effect. We will see in section~\ref{sec:subleading-tadopoles}, however, that all these effects are suppressed at least by a full power of $\lambda$, and consequently even at the subleading order in $\sqrt{\lambda}$ can be dropped~\footnote{We note that if we did not use a smooth window function, but rather took $\delta\to0$, the tadpole term would only be suppressed by $\sqrt{\lambda}$. We explain this in Appendix \ref{app:sharp-diffusion}.},   while at arbitrary order  in $\lambda$ one can use perturbation theory in $\sqrt{\lambda}$ to obtain a solvable closed set of equations. 
  
 After these simplifications,~\eqref{1ptfull} reads 
\bea
\label{1pt}
&&\frac{\d}{\d t} P_1(\phi_1,t)=\Bigg [-\frac{\d}{\d \phi_1}\l(\left\langle\Pi_1(\phi_1,\phi_s(\vx_1),t)\right\rangle_{\phi_1}P_1 (\phi_1,t)\r)+\nn\\
&&+\frac{1}{2}\;\frac{\d^2}{\d\phi_{1}^2} \l(\left\langle\dot\Delta \phi(\vx_1) \Delta \phi(\vx_1)\right\rangle_{\phi_1}\, 
 P_1(\phi_1,t)\r)\Bigg]\times\l(1+O\l(\lambda,\eps^2,\delta\r)\r)\,,
\eea
where by $\Pi_1$ we denoted the long part of the momentum at $x_1$, which only depends on the fields at $x_1$.
We stress that in the above equation we have, for the first time, dropped terms that provide a relative correction of order $\lambda$, such as, for example, the tadpole term for~$\phi$, see section~\ref{sec:subleading-tadopoles} for a explicit estimates. At this order it only matters the dependence on the field at one spacial location.


\subsection{Leading equation} 
\label{sec:leadeq}
It is now time to evaluate the expectation values of the short modes. In order to simplify the study of our equation, we will have to make some assumptions about the leading solution around which we are going to expand the equation. We will assume that the potential is such that there exists an equilibrium configuration and that the state of the field is close to this equilibrium. In particular, if one is interested in a state that is far from the equilibrium one, the contribution of various terms in the equation may be different than the one we will assume next. 

Let us do the counting for $V\sim\lambda\phi^4$ for simplicity, however, a similar counting can be performed for any potential. We expect, as anticipated earlier on, $\phi_\ell\sim H\lambda^{-1/4}$ and localized at very long wavelengths. On the other hand, for long modes with momenta $k\sim\Lambda(t)/\text{few}$, we do not expect any enhancement by inverse powers of $\lambda$, as the mode has not been outside of the horizon for long enough time, so that the amplitude of those modes is just of order $H$. Equivalently short modes all have amplitudes of order $H$. In these estimates we ignore powers of $\log\eps$ due to the hierarchy discussed in section \ref{sec:strategy}. 

This counting implies that, up to corrections of order $\sqrt{\lambda}$ the fields with momentum of order $k\sim\Lambda(t)$ can be treated as free massless fields and that $\Pi_1(\phi_1,\phi_s(\vx_1),t)$ depends only on the long part of the field. The leading expression for $\Pi_1$ can be easily inferred from~(\ref{eq:momentum_leading}), giving $\Pi_1(\phi_1)=-\lambda \phi_1^3/(3 H)$. 

This gives  us the drift term, and now we move to compute the diffusion term. The two-point function of a free massless scalar in momentum space is given by
\be
\label{phicorr}
\langle\phi(\vk,t)\phi(\vk',t)\rangle =(2\pi)^3\delta(\vk+\vk')\frac{H^2}{2k^3}\l(1+O\l(\sqrt{\lambda}\r)\r),
\ee
We thus get
\bea
\label{derivediff}
&& \left\langle\dot\Delta \phi(\vx_1) \Delta \phi(\vx_1)\right\rangle=\frac{1}{2}\int \frac{d^3k}{(2 \pi)^3}\int d^3k' \frac{d}{d t}\l(\Omega_{\Lambda(t)}(k)\Omega_{\Lambda(t)}(k')\r)\delta(\vk+\vk')\frac{H^2}{2k^3}=\nn\\
&& =\frac{1}{2}\int \frac{d^3k}{(2 \pi)^3} \frac{d}{d t}\l(\Omega_{\Lambda(t)}(k)\r)^2\frac{H^2}{2k^3}=\\
&&=\frac{1}{2}\l(\int \frac{d^3k}{(2 \pi)^3}\,\delta(k-\Lambda(t)) \frac{H^2}{2\Lambda(t)^3}\r) \l(-\int dk\; \dot \Lambda(t) \frac{d}{dk} \l(\Omega_{\Lambda(t)}(k)\r)^2\r)\l(1+O(\delta)\r)=\nn\\
&&=\frac{H^3}{8 \pi^2}\l(1+O(\delta)\r)\nn\,,
 \eea
 where in going from the second to the third lines in the equation we used the fact that $1/k^3$ is a smooth function of $k$, as compared to the window function, and consequently it can be replaced with
 $1/\Lambda(t)^3$ ignoring corrections of order $\delta$.
Substituting this into \eqref{1pt} gives
\be
\label{1pteqtime}
\frac{\d}{\d t}  P_1(\phi_1,t)= \Gamma_{\phi_1} P_1(\phi_1,t)\l(1+O(\eps^2,\delta,\lambda^{\frac{1}{2}})\r)\;,
\ee
where the differential operator $\Gamma_\phi$ is defined as 
\be
\label{Gammaphi}
\Gamma_{\phi}= \frac{\d}{\d\phi}\frac{V'(\phi)}{3H}+\frac{H^3}{8\pi^2}\frac{\d^2 }{\d\phi^2} \;.
\ee
This equation agrees with the one obtained by Starobinsky~\cite{Starobinsky:1986fx,Starobinsky:1994bd}. 
However, now we have derived this equation rigorously with control over all the approximations that were made.

It is intuitively clear that the state we study at the moment corresponds to the equilibrium solution of this equation, although, we would like to stress that formally we have not proven it yet. We have only shown that the one-point distribution of our state evolves in time according to equation \eqref{1pteqtime}. One way to see that the BD  wave functional indeed corresponds to a $P_1$ independent of time is to note that time dependence in it enters only through $k/a(t)$, or equivalently through $a(t) x$. But since $P_1$ is $x-$independent, it should also be time-independent. 

\begin{figure}[htbp] 
	\centering
		\includegraphics[width=.5\linewidth]{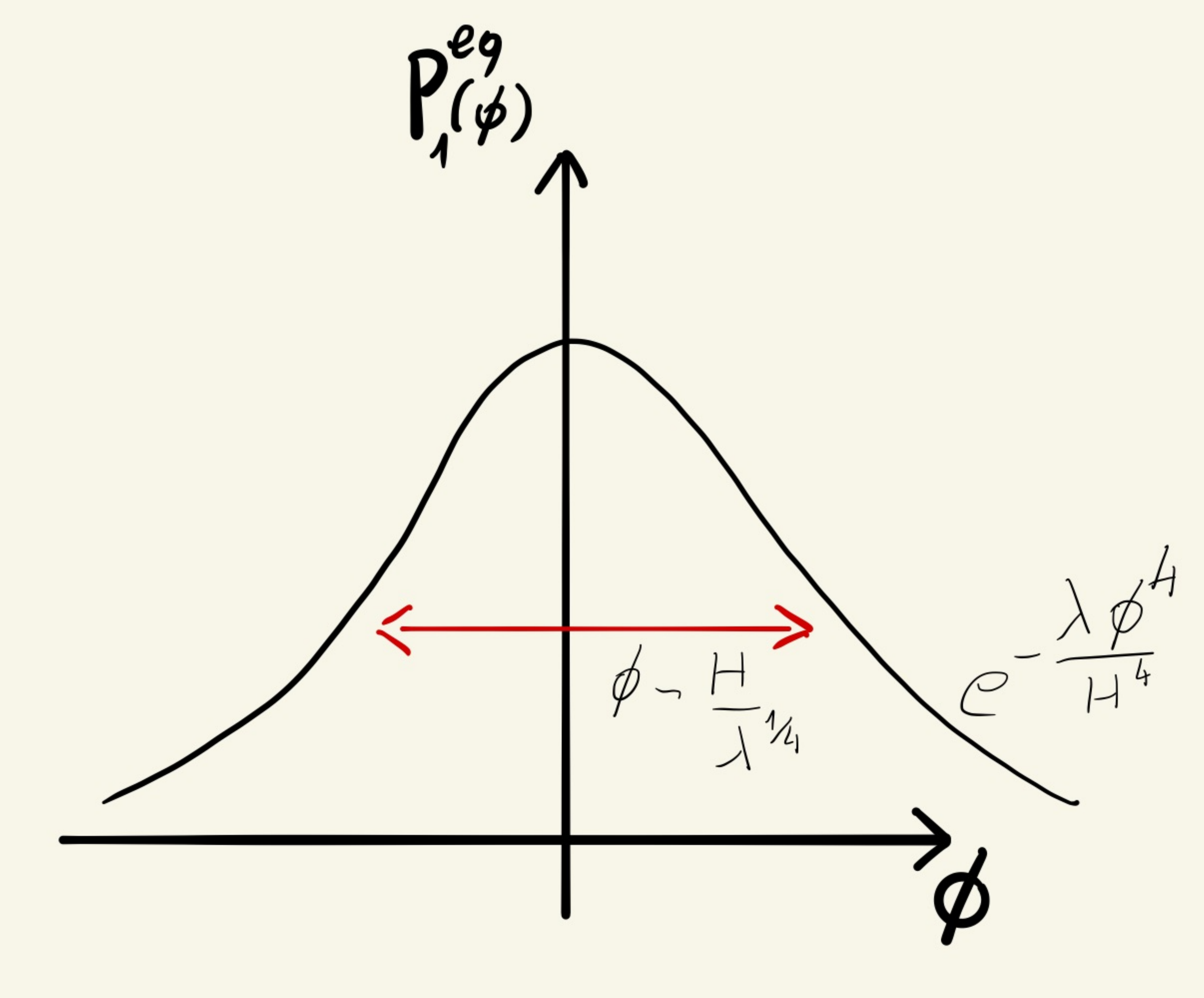}
	\caption{\small Pictorial representation of the distribution the long-wavelength field at a single location for the BD state. The typical values of the field, for $V=\lambda\phi^4$, are of order $\phi\sim H/\lambda^{1/4}$.}
	\label{fig:equilibrium}
\end{figure}

Hence, let us first study the equilibrium solution which satisfies 
\be
\label{1pteqphi}
\Gamma_{\phi_1} P^{eq}_1(\phi_1)=0\;.
\ee
 To fix it uniquely
we pick the boundary conditions for which all the moments of the distribution are finite, so that $ P^{eq}_1(\phi_{1})$ decays at infinity faster than any power. 
It is straightforward to check that the corresponding solution is 
\be
\label{1pteqsol}
 P^{eq}_1(\phi_1)=N e^{-{\frac{8 \pi^2 V(\phi_1)}{3 H^4}}}\;,
\ee
where $N$ is a constant normalization factor. 
Now that we obtained the solution, we can confirm the validity of our assumption, {\it i.e.} the characteristic value of $\phi_1$, and more generally of the long field $\phi_\ell(\vx)$, is of order $\lambda^{-1/4}$ (see Fig.~\ref{fig:equilibrium}). In the next section  we will see that the equilibrium solution we just found is also an attractor.  

The one-location equilibrium distribution allows us to compute the leading order expectation values of the operators $\hat\phi^{2 n}(\vx)$, to which the long modes give the leading contribution.\footnote{{We sometimes use the $\hat\phi$ notation if we want to stress that we are computing expectation values of operators.}} For the the $\lambda\phi^4/4$ potential  they are given by 
\bea
\label{1ptphinlead}
\langle \hat\phi^{2 n}(\vx_1) \rangle=\langle\phi_1^{2 n}\rangle \l(1+O(\sqrt{\lambda})\r)=\l(\frac{3}{2\pi^2}\r)^{\frac{n}{2}}\lambda^{-\frac{n}{2}}H^{2 n} 
\frac{\Gamma\l(\frac{n}{2}+\frac{1}{4}\r)}{\Gamma\l(\frac{1}{4}\r)}\,.
\eea

\subsection{Non-equilibrium one-point distribution}
\label{sec:nonequil1pt}

 Eq.~(\ref{1pteqtime}) has been derived using the BD wave function to express the action of the momentum, $\hat \Pi$, on the wave function, $\Psi[\phi]$, in terms of a multiplication by a function of $\phi$ times $\Psi$: $\Hat\Pi \Psi[\phi]\sim \lambda \phi^3 \Psi[\phi]$. Therefore, this equation will hold unchanged for all the states for which the momentum is the same function of $\phi$ as for the BD vacuum. This is expected to be a large class of states. In fact, since modes are long, we can use our classical intuition to notice that the expression of the momentum in terms of $\phi$ that we are using is nothing but the functional form that the momentum has on the attractor solution that exist in de Sitter for slow rolling scalar fields. {To support this intuition in section \ref{sec:stability} we will show directly that small variations in $\Pi$ as a function of $\phi$ decay in times of order $H^{-1}$ and that $\Pi$ approaches the BD value.} With this in mind, let us now turn to analyzing the time-dependence of the one-point distribution of the field {in states where the probability distribution of the long field is different.} In particular, we will see that the equilibrium distribution is an attractor of this time evolution, which means that there is a class of states in dS space that after some time all have the same one-point distribution as the state we study. It will also allow us to introduce some objects that will be useful for studying the higher-point correlation functions.

It is convenient to decompose the probability distribution into { a basis of Eigenfunctions, $P_1(\phi_1,t)=\sum_n c_n(t)\Phi_n(\phi_1)$, where $\Phi_n(\phi_1)$ are the Eigenfunctions} of the differential operator $\Gamma_{\phi_1}$, defined by:
\be
\label{Phin}
\Gamma_{\phi_1}\Phi_n(\phi_1)=-\lambda_n\Phi_n(\phi_1) \;.
\ee
If the potential is smooth and grows fast enough at infinity the spectrum will be discrete, the Eigenfunctions will decay exponentially at infinity and form a complete orthonormal set of functions with the inner product given by
\be
\int d\phi\, \Phi_n(\phi)\Phi_m(\phi)\mu(\phi)=\delta_{mn},
\ee
where the measure $\mu$ is~\footnote{The normalization constant $N$ is included in the measure for some future convenience.}
\be
\mu(\phi)=N^{-1}e^{\frac{8 \pi^2 V(\phi)}{3 H^4}}\;.
\ee
 These conditions are satisfied, in particular, by the $\lambda \phi^4$ potential. More general potentials, for example those that do not have a ground state, can be studied in a similar way but with corresponding modifications.
 
 For most potentials it is impossible to find all the Eigenvalues and Eigenfunctions analytically, however, it is relatively simple to do this numerically. Few general results still hold due to the special form of the operator $\Gamma_\phi$~\cite{Starobinsky:1994bd}. First, all the Eigenvalues are non-negative. Second, the smallest Eigenvalue, $\lambda_0$, is equal to zero and the corresponding Eigenfunction corresponds to the equilibrium solution given in \eqref{1pteqsol}: 
\be
\Phi_0(\phi)=P^{eq}_1(\phi)\;,
\ee
consequently, at late times the probability distribution of $\phi_1$ will tend to the equilibrium one which means that the latter is an attractor. At late times the time-dependence is controlled by the leading non-zero Eigenvalue $\lambda_1$, which for the $\lambda\phi^4/4$ potential is approximately equal to~\cite{Starobinsky:1994bd} 
\be
\lambda_1\approx1.37 \sqrt\frac{\lambda}{24 \pi^2}H\;.
\ee
In general the $n$-th Eigenvalue scales as $\lambda_n\approx n^{\alpha}\sqrt{\lambda} H$ at large $n$, {where $\alpha$ is some positive power, {\it e.g.} $\alpha=3/2$ for the quartic potential.} This shows that the characteristic timescale of the evolution of the distribution is of order $\lambda^{-1/2} H^{-1}$.

\section{Two-point distribution}
\label{sec:two-point}
Our aim in this section is to compute the correlation functions of bi-local operators of the form 
\be
\l\langle\hat\phi(x_1^\mu)^n \hat\phi(x_2^\mu)^m\r\rangle\;.
\ee
According to our general logic, each operator is split into long and short parts $\hat\phi(x^\mu)=\hat\phi_\ell (x^\mu)+\hat\phi_s (x^\mu)$ and the contribution of long modes is enhanced by the powers of $\lambda$. 
To compute the correlators of long modes, we need the two-point probability distribution. So we start by computing that.
In this section we treat equal $x$ (two-times) and equal $t$ (two-locations) cases, postponing the general case to section \ref{sec:generic-2-point}.

\subsection{Two-times distribution}
\label{sec:twotimes}
Our goal will be to compute the joint probability distribution of two long fields at the same spacial position $\vx_1$ but at different times, $ P_2(\phi_1,t;\phi_1',t')$. Here we treat the long modes at classical level.
We already encountered a similar, but more general, functional distribution in section \ref{sec:perturbatie-structure} and noted that its derivative with respect to the latest time satisfies the same equation as the single-time distribution, while the boundary conditions are given by \eqref{Pnfunct}. Consequently, assuming $t>t'$,
\be
\label{twoteom}
\frac{\d}{\d t}  P_2(\phi_1,t;\phi_1',t')=\Gamma_{\phi_1} P_2(\phi_1,t;\phi_1',t')\;,
\ee
where $\Gamma_\phi$ is given in \eqref{Gammaphi}, while the boundary condition reads
\be
\label{twotbc}
P_2(\phi_1,t';\phi_1',t')=\delta(\phi_1-\phi_1') P_1(\phi_1',t') \;.
\ee
The $\delta$-function  can be decomposed into the Eigenfunctions of $\Gamma_{\phi_1}$, defined in~(\ref{Phin}), as
\be
\label{completeness}
\delta(\phi-\phi')=\sum_{n=0}^\infty\Phi_n(\phi)\Phi_n(\phi')\mu(\phi')\;,
\ee
which allows us to find the two-point distribution immediately:
\be
P_2(\phi_1,t;\phi_1',t')=P_1\l(\phi_1',t'\r)\sum_{n=0}^\infty e^{-\lambda_n(t-t')}\Phi_n(\phi)\Phi_n(\phi')\mu(\phi')\;.
\ee

As an application of this result let us compute the connected two-point function of the field operator at large time-separation in the equilibrium state.  We have
\bea
\label{2ptfeqtime}
&&\l\langle\hat\phi(\vx,t)\hat\phi(\vx,t')\r\rangle=\l\langle\hat\phi_\ell(\vx,t)\hat\phi_\ell(\vx,t')\r\rangle\l(1+O\l(\sqrt{\lambda}\r)\r)=\nn\\
=&&\int d\phi_1 d\phi'_1\,P_2^{eq} (\phi_1,t;\phi_1',t')\phi_1\phi_1'\l(1+O\l(\sqrt{\lambda},\eps^3\hbar\r)\r)=\\
&&=C_1 e^{-\lambda_1 (t-t')} \int d\phi'_1\,P^{eq}_1 (\phi_1')\mu(\phi')\Phi_1(\phi')\phi_1'\l(1+O\l(\sqrt{\lambda},\eps^3\hbar\r)\r)=\nn\\
&&=C_1^2 e^{-\lambda_1 (t-t')}\l(1+O\l(\sqrt{\lambda},e^{(\lambda_1-\lambda_2)(t-t')},\eps^3\hbar\r)\r)\;,\nn
\eea
where $C_1=\int d\phi \Phi_1(\phi)\phi\sim\lambda^{-1/4}$. In the first equality in \eqref{2ptfeqtime} 
we dropped the short modes contribution, and in the second equality we expressed the correlator of long modes with the help of the two-times distribution and  took the least-decaying contribution. Since we focus on the connected part, we dropped the contribution from the $\lambda_0=0$ Eigenvalue (for symmetric potentials, this contribution automatically vanish). We added $O\l(\eps^3\hbar\r)$ at this stage to stress that we did not yet show how to compute non-equal-time correlators of long modes at quantum level (see section \ref{sec:quantum}). In the last transition we used that $\mu(\phi) P^{eq}_1(\phi)=1$. Note also that if the potential is symmetric, the even-$n$ Eigenfunctions will be even, consequently, the subleading contribution to this correlator will come from $\lambda_3$, not~$\lambda_2$. 

{This is maybe a good place to show, again, that the non-perturbative solution for massless $\lambda\phi^4$ in de Sitter that we are finding cannot be interpreted as the dynamical generation of a mass term, such that, after taking that into account, the resulting massive theory is weakly coupled. For example, this is what happens in massless $\lambda\phi^4$ at finite temperatures in Minkowski space, see for example~\cite{Laine:2016hma}. If the same situation where to happen here, the long time behavior of the same-location $n$-point functions would be the same, upon a suitable mapping between $\lambda$ and $m^2/H^2$. The fact that this is not the case can be easily checked by inspection of the Eigenvalues of the operators $\Gamma_{{\phi}_1}$. For massless $\lambda\phi^4$ theory, we can straightforwardly find the leading Eigenvalues: $\sqrt{\lambda}H\cdot\{0,0.089,0.289,0.537,\ldots\}$. Instead, for a free massive theory, the Eigenvalues are $\frac{1}{3}\frac{m^2}{H^2} H\{0,1,2,\dots\}$. We clearly see that the set of Eigenvalues cannot be made compatible with any choice of $m$.}

\subsection{Two-locations distribution, sudden perturbation theory \label{sec:twolocations}}
Let us now turn to the more complicated and interesting case of the equal time two-point distribution for the fields located at two different spacial locations $\vx_1$ and $\vx_2$, $P_2(\phi_\ell(\vx_1)=\phi_1,t;\phi_\ell(\vx_2)=\phi_2,t)$, that we denote by $ P_2(\phi_1,t;\phi_2,t;\vx_{12})$ for shortness. The equation for this distribution can be derived analogously to the single point case just with the insertion of two $\delta$-functions instead of one, see equations \eqref{1ptfull} and \eqref{1pt}. The only new term that appears in the case of two points is related to the mixed correlators
\be
\l\langle\dot\Delta\phi(\vx_1)\Delta\phi(\vx_2)\r\rangle_{\phi_\ell}
\ee
between the modes at different locations. All the simplifications arising at the leading order that appeared in the single point case continue to apply, and these correlators can be easily computed. Using \eqref{phicorr} and the same manipulations as in \eqref{derivediff}, we get
\bea
&&\l\langle\frac{\d}{\d t}\l(\Delta\phi(\vx_1)\Delta\phi(\vx_2)\r)\r\rangle=\int \frac{d^3  k}{(2\pi)^3}\; \frac{d\Lambda(t)}{dt}\delta(k-\Lambda(t))e^{i \vk\cdot\vx_{12}}\frac{H^2}{2k^3}\l(1+O\l(\sqrt{\lambda},\delta\r)\r)=\nn\\
&&\qquad \qquad \qquad= \frac{H^3}{4\pi^2}j_0(\eps a H x_{12})\l(1+O(\sqrt{\lambda},\delta)\r)\;,
\eea
where $j_0(z)=\sin(z)/z$. We thus arrive to the following equation
\bea
\label{2pt}
&&\frac{\d}{\d t} P_2(\phi_1,t;\phi_2,t;\vx_{12})= \\ \nn
&&=\l(\Gamma_{\phi_1}+\Gamma_{\phi_2}+\frac{H^3}{4\pi^2}j_0(\eps a H x_{12})\frac{\d^2}{\d{\phi_1}\d{\phi_2}}\r) P_2(\phi_1,t;\phi_2,t;\vx_{12})\l(1+O(\sqrt{\lambda}, \epsilon^2,\delta)\r)\;.
\eea

This agrees with the equation for the two-location distribution given in~\cite{Starobinsky:1986fx,Starobinsky:1994bd}. We presented here a more careful derivation of this result so that we can analyze this equation systematically, as well as compute perturbative corrections. 

To develop some intuition about this equation it is useful to consider two limits, $\eps a x_{12}H\to0$ and $\eps a x_{12}H\to\infty$. {In the first case $j_0(\eps a x_{12}H)$ approaches one 
 and the diffusion part of the operator in the right hand side of \eqref{2pt} becomes identical to the diffusion part of the one-location operator but with respect to the variable $(\phi_1+\phi_2)/2$.
  Additionally, as $\eps a x_{12} H\to 0$, $\phi_1$ must be equal to $\phi_2$, as, in this limit, they represent the field at the same point {so the distribution should be proportional to $\delta(\phi_1-\phi_2)$ in this limit. It is easy to see that} , as $\eps a x_{12} H\to 0$, it exists a solution that can be written as $ P_1((\phi_1+\phi_2)/2,t)\, \delta(\phi_1-\phi_2)$, where  $ P_1$ satisfies the equation for the one-point distribution. 
   On the other hand, for large $\eps a x_{12}H$, the mixed term goes to zero and the differential operator becomes equal to $\Gamma_{\phi_1}+\Gamma_{\phi_2}$. Therefore the variables $\phi_1$ and $\phi_2$ separate, and the distribution factorizes into the product of two one-point distributions $P_2(\phi_1,t;\phi_2,t;\vx_{12})\to P_1(\phi_1,t)P_1(\phi_2,t)$, each satisfying \eqref{1pteqtime}. The behavior of the two-location distribution that we just described is of course just the manifestation of the fact that fields are perfectly correlated while they are separated by distances much shorter that the averaging scale $1/(a\eps H)$, and are completely uncorrelated when they are separated by distances that are much longer.

The exact solution of equation \eqref{2pt} seems to be rather complicated. However, the timescale during which $j_0$ changes form from being very close to 1 to being very close to 0 is of order $H^{-1}$, which is much shorter than the characteristic timescale of the evolution of the distribution which we estimated to be of order $H^{-1}\lambda^{-1/2}$. This allows us to approach the problem with techniques very similar to the sudden approximation in Quantum Mechanics. To stress the analogy let us formulate the problem in the corresponding language. For a moment we keep $x_{12}$ fixed, and think about the time evolution of $P_2$ given by \eqref{2pt}, which we will rewrite in the following form, inspired by Quantum Mechanics:
\be
\frac{\d}{\d t}P_2(\phi_1,t;\phi_2,t;\vx_{12})=-(H_1+j_0(\eps a H x_{12}) H') P_2(\phi_1,t;\phi_2,t;\vx_{12})\;,
\label{Shroed}
\ee
where
\be\label{eq:Hprdef}
H_1=-\Gamma_{\phi_1}-\Gamma_{\phi_2}\ ,\qquad H'=-\frac{H^3}{4\pi^2}\frac{\d^2}{\d{\phi_1}\d{\phi_2}} \ .
\ee
It is convenient to decompose $\bar P_2$ into the Eigenfunctions of the late-time ``Hamiltonian'' $H_1$:
\be\label{latetime}
 P_2(\phi_1,t;\phi_2,t;\vx_{12})=\sum_{n,m}A_{nm}(t)\Phi_n(\phi_1)\Phi_m(\phi_2)\ .
\ee
Here $\Phi_n(\phi)$ are the Eigenfunctions of the operator $\Gamma_\phi$ introduced in \eqref{Phin}.

Let us also assume that the field is in the equilibrium state, and call the associated two-location distribution as $\bar P_2^{eq}(\phi_1,t;\phi_2,t;\vx_{12})$. The initial conditions for $ P_2$ are given by 
\be
 P_2^{eq}(\phi_1,t;\phi_2,t;\vx_{12})\Big|_{t=-\infty}=\delta(\phi_1-\phi_2) P^{eq}_1\l(\frac{\phi_1+\phi_2}{2}\r)\;.
\ee
Consequently, at $t=-\infty$, the coefficients $A_{nm}(-\infty)\equiv A_{nm}^{(0)}$ are given by
\bea
\label{Anm0}
&& A_{nm}^{(0)}=\int d \phi_1 d\phi_2 \; \mu(\phi_1)\mu(\phi_2)\;\Phi_n(\phi_1)\Phi_m(\phi_2) \Phi_0\l(\frac{\phi_1+\phi_2}{2}\r)\delta(\phi_1-\phi_2)=\\ \nn
&&\qquad=\int d\phi_1\; \mu(\phi_1)^2\;\Phi_n(\phi_1)\Phi_m(\phi_1)\Phi_0(\phi_1)=\delta_{nm}\;,
\eea
where in the last transition we used orthonormality of the wave functions and the fact that $\mu(\phi)=\Phi_0(\phi)^{-1}$ and $\Phi_0(\phi)= P^{eq}_1(\phi)$.
At this point the leading sudden approximation is straightforward to obtain. One simply determines the coefficients $A_{nm}$ from the initial conditions and finds that they stay constant until the instant of time 
\be
\label{t0}
t_0=-H^{-1}\log\l(\eps H x_{12}\r)\;,
\ee
when the "interaction" $H'$ is "suddenly switched off". Then, in the leading approximation, the distribution does not have time to change in the process of switching off of  the interaction,
and after that the evolution is given by \eqref{latetime} with $A_{nm}(t)=A_{nm}^{(0)} e^{-(\lambda_n+\lambda_m)(t-t_0)}$. This approximation is enough if one is interested, for example, only in the large distance behavior of the distribution at leading order in $\lambda$. The $x_{12}$-dependence  comes uniquely through  $t_0$ and, for large $x_{12}$, it reads:
\be
\label{2loclead}
P_2^{eq}(\phi_1,t;\phi_2,t;x_{12})=\Phi_0(\phi_1) \Phi_0(\phi_2) +\Phi_1(\phi_1) \Phi_1(\phi_2)(a H x_{12})^{-2\lambda_1/H}+\ldots\;.
\ee
Note that we removed the $\epsilon$ dependence from the second term since it contributes at the relative order $O(\sqrt{\lambda}\log\eps)$ and consequently we should not keep it at this order. In fact the next order calculation should cancel the $\eps$ dependent contribution. The first term corresponds to the disconnected contributions to the two-location $n$-point functions, while the second term produces the leading connected contribution. For example, { for symmetric potentials,} the large distance behavior of the two-point function is controlled by this term. At large $x_{12}$ we get
\bea
\label{2ptflead}
&&\langle\hat\phi(\vx_1,t)\hat\phi(\vx_2,t)\rangle=\langle\phi_1\phi_2\rangle\l(1+O(\sqrt{\lambda})\r)=\nn\\
&&\qquad=C_1^2\l(a H x_{12}\r)^{-2\lambda_1/H}\l(1+O\l(\sqrt{\lambda},\l(a H x_{12}\r)^{2(\lambda_1-\lambda_2)/H}\r)\r)\;,
\eea
where $C_1$ is the same as in \eqref{2ptfeqtime}. Again, for an even potential one can replace $\lambda_2$ with $\lambda_3$ in the error estimate. Notice that, since we are focussing on the perturbation theory in~$\sqrt{\lambda}$, we neglected to stress that there are order $\epsilon^2$ and $\delta$ corrections. We leave this implicit below unless it is unclear from the context. 

The expression in (\ref{2ptflead}) allows us to compute the leading contribution at small~$k$'s to the momentum-space two-point function. Fourier transforming \eqref{2ptflead} we get\footnote{Since the integral in the Fourier transform is divergent at large $x_{12}$ it should be understood in the sense of distributions, see Appendix~\ref{app:2-point-Fourier} for more details.}
\be
\label{P2k}
\langle\hat\phi(\vk,t)\hat\phi(-\vk,t)\rangle'= \frac{ \lambda_1}{H}C_1^24 \pi^2 \,k^{-3+2\lambda_1/H}(a H)^{-2\lambda_1/H}\l(1+O\l(\sqrt{\lambda},\l(k/a H\r)^{2(\lambda_2-\lambda_1)/H}\r)\r)\;.
\ee
Note an extra factor of $\lambda_1$, and hence $\sqrt{\lambda}$, as opposed to the coordinate-space expression. 

For our purposes we have to do slightly better than \eqref{2loclead}. First, we would like to make sure we have a controlled approximation and estimate the corrections to the leading result. Second, we will be interested in the Fourier transform of the distribution at $k\sim \eps a H$ in order to estimate various terms in equation \eqref{1ptfull} related to the background inhomogeneities. The absolute magnitude of the distribution at these $k$'s is suppressed by $\sqrt{\lambda}$ which implies that it can get a relative order one contribution from the subleading sudden approximation. Indeed, the ratio of timescales that controls the sudden approximation is of order $\sqrt{\lambda}$ and, consequently, one expects corrections to the leading sudden approximation to contribute at this order. Moreover, these corrections are localized mostly around $x \eps a H \sim 1$ so they are important for $k\sim \eps a H$. {Additionally, for $k$ very close to $\eps a H$, the corrections from late times also give contributions of order $\sqrt{\lambda}$ as we discuss them below.}

Let us now carefully setup the systematic sudden approximation expansion. Substituting \eqref{latetime} into \eqref{Shroed} we get the following equation for $A_{nm}(t)$:
\be
\dot A_{nm}(t)=-(\lambda_n+\lambda_m)A_{nm}(t)-\sum_{p,q}j_0(\eps a H x_{12}) H'_{nm,pq}A_{pq}(t)\;,
\label{ShroedA}
\ee
where $H'_{nm,pq}$ is the matrix element of $H'$ between the Eigenfunctions of $H_1$. 
We separate the time evolution of $ P_2$ in three regions: $t<t_i$, $t_i<t<t_f$ and $t_f<t$, where
\be
\label{tif}
t_i=t_0-\Delta\; H^{-1},\qquad t_f=t_0+\Delta\; H^{-1}\,,
\ee
and $t_0$ is given in \eqref{t0}.
See figure~\ref{fig:Aoft} for a visualization. We introduce three time intervals in order to take into account that the interaction is being switched off not completely suddenly but during the period of time from $t_i$ to $t_f$. We would like to take $\Delta$ as large as possible, so that $j_0$ is very close to 1 and 0 in regions I and III respectively; however, we must take $\Delta\ll\lambda^{-1/2}$ so that the intermediate time interval is short enough, which is necessary for the sudden perturbation theory to work. In this sense $\Delta$ is a parameter similar to $\epsilon$: no physical quantity can depend on it and we are free to chose its value to render perturbation theory maximally efficient. Since the residual size of the $H'$ interaction in regions I and III is of order $e^{-\Delta}$, it is possible to choose $\Delta$ in such a way that corrections coming from those regions are smaller than any power of $\lambda$. Consequently, we can assume that $A_{nm}$'s evolve in these regions with the Hamiltonians $H_0=H'+H_1$ and $H_1$. Interesting power-law corrections in $\lambda$ come from region II. 
\begin{figure}[htbp] 
	\centering
		\includegraphics[width=.7\linewidth,angle=0]{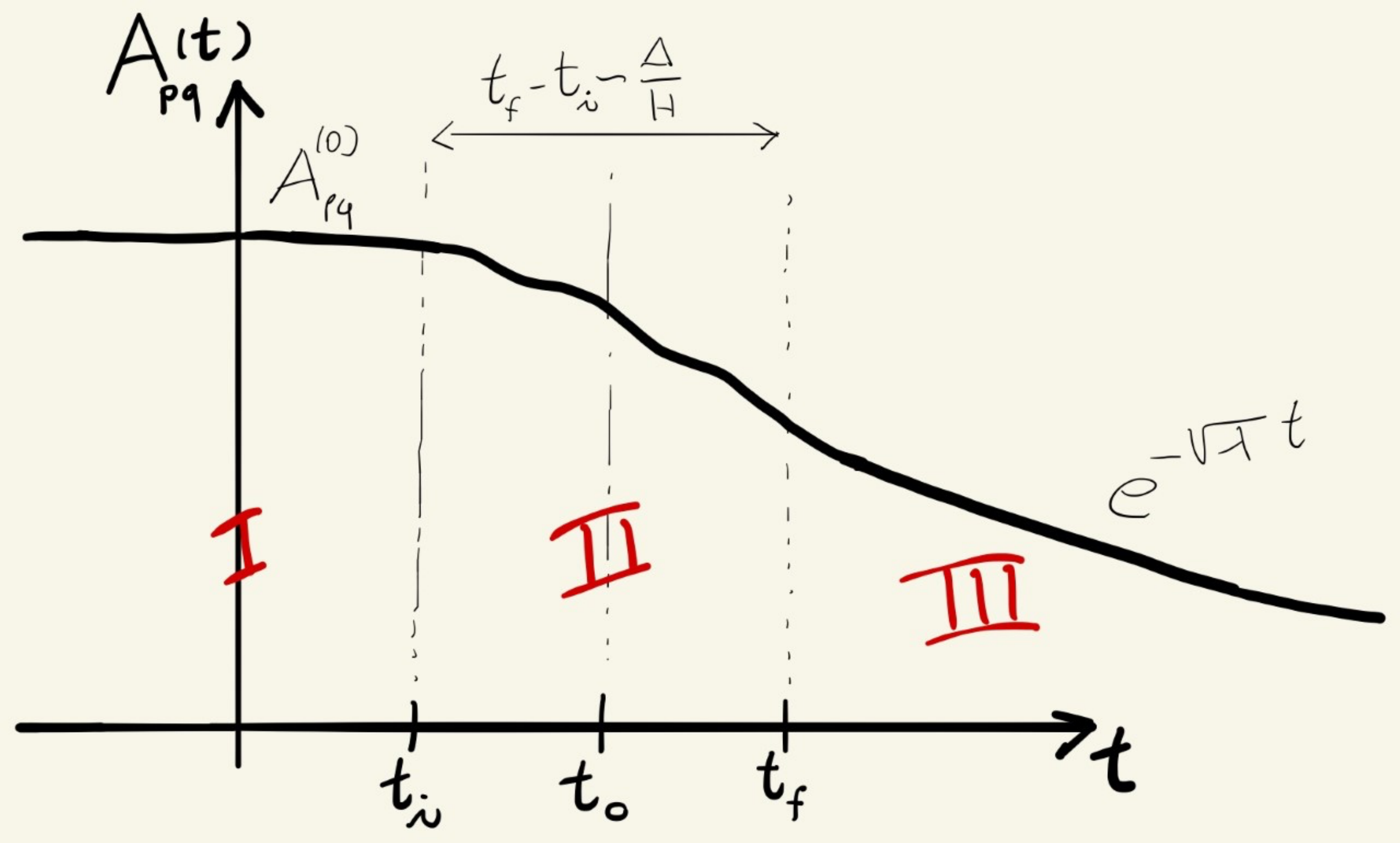}
	\caption{$A_{nm}(t)$, regions I, II, III.}
	\label{fig:Aoft}
\end{figure}

In region II we solve \eqref{ShroedA} perturbatively. That is we write $A_{nm}(t)=A_{nm}^{(0)}+A_{nm}^{(1)}(t)$, where $A^{(0)}_{nm}$ is given by~(\ref{Anm0}), and  $A_{nm}^{(1)}(t)/A_{nm}^{(0)}\sim\sqrt{\lambda}$. For $A_{nm}^{(1)}(t)$ we get: 
\be
A_{nm}^{(1)}(t)=-(\lambda_n+\lambda_m){ \int_{t_i}^t dt'\;}A_{nm}^{(0)} -\sum_{p,q}\int_{t_i}^t d\, t' j_0\l(\eps a(t') H x_{12}\r) H'_{nm,pq}A_{pq}^{(0)}\;.
\label{suddenA1}
\ee
Now let us use the fact that the initial state $\sum_{p,q} A_{pq}^{(0)} \Phi_{p}(\phi_1)\Phi_{q}(\phi_2)$ was annihilated by $H_0+H'$:
\be
(H_0+H')\sum_{p,q}A_{pq}^{(0)}  \Phi_{p}(\phi_1)\Phi_{q}(\phi_2)=0\;.
\ee 
We can multiply this expression by $\mu(\phi_1)\Phi_{n}(\phi_1)\mu(\phi_2)\Phi_{m}(\phi_2)$ on the left and integrate over $\phi$'s to get
\be
(\lambda_n+\lambda_m) A_{nm}^{(0)}+\sum_{p,q} H'_{nm,pq}A_{pq}^{(0)}=0\;.
\ee
Substituting this into \eqref{suddenA1} we see that the evolutions of different coefficients decouple:
\be
A_{nm}^{(1)}(t)= (\lambda_n+\lambda_m) \int_{t_i}^t d\, t' \l[j_0\l(\eps a(t') H x_{12}\r)-1\r] A_{nm}^{(0)}\;.
\label{AII}
\ee
In particular, at this order they stay diagonal in $m,n$ due to \eqref{Anm0}.
The last expression also makes it manifest that $A_{nm}^{(1)}(t)$ is of relative order $\sqrt{\lambda}$.

Finally, in the region III we get at subleading order
\be
A_{nm}(t)=\l(A_{nm}^{(0)}+A_{nm}^{(1)}(t_f)\r)e^{-(\lambda_n+\lambda_m)(t-t_f)}(1+O(\lambda))\;.
\label{AIII}
\ee
The $O(\lambda)$ in this expression has to be understood in a sense that we have included the $O(\sqrt{\lambda})$ corrections coming from sudden perturbation theory. There are also other perturbative corrections, still of order $\sqrt{\lambda}$, to the Eigenvalues $\lambda_n$ that are coming from the corrections to the operator $\Gamma_\phi$, as we will compute in section~\ref{sec:Eigenvalues}. Expressions \eqref{AII} and \eqref{AIII} are valid independently of the order to which the Eigenvalues are computed.

To summarize, we get the following expression for the two-point distribution in the equilibrium state which is also plotted on figure~\ref{fig:Aoft}:  
\bea
\label{2ptfinal}
&&  P_2^{eq}(\phi_1,t;\phi_2,t;\vx_{12})=\\
&&\qquad\delta(\phi_1-\phi_2) P^{eq}_1\l(\frac{\phi_1+\phi_2}{2}\r),\quad t<t_i \quad (I) \nn \\ \nn
&&\qquad\sum_n \Phi_n(\phi_1)\Phi_n(\phi_2)\l(1+2\lambda_n  \int_{t_i}^t d\, t' \l[j_0\l(\eps a(t') H x_{12}\r)-1\r] \r), \quad t_i<t<t_f \quad (II)\\ \nn
&& \qquad\sum_n\Phi_n(\phi_1)\Phi_n(\phi_2)\l(1+2\lambda_n  \int_{t_i}^{t_f} d\, t' \l[j_0\l(\eps a(t') H x_{12}\r)-1\r] \r) e^{-2\lambda_n(t-t_f)} , \quad t_f<t \quad (III)
\eea

Until now we have been treating the distribution as a function of $t$ at fixed $x_{12}$, however,~\eqref{2ptfinal} is valid for any $t$ and $x_{12}$ so we can equivalently treat it as a function of $x_{12}$ at fixed $t$, which is what we need to compute the Fourier transform. Note that $t_i$ and $t_f$ depend on $x_{12}$ through $t_0$, see equations \eqref{tif} and \eqref{t0}. As a function of $x_{12}$, $ P_2^{eq}$ is also given piecewise on three intervals: $x_{12}<x_i$, $x_i<x_{12}<x_f$ and $x_f<x_{12}$, with $x_i=e^{-\Delta}(a(t)\eps H)^{-1}$ and $x_f=e^{\Delta}(a(t)\eps H)^{-1}$. When $x_{12}$ belongs to each of those intervals, $t$ correspondingly belongs to region I, II or III defined above. It is again convenient to study the contribution of each Eigenfunction independently, however, now we will keep the dependence of the coefficients $A_{nn}(t,x_{12})$ on $x_{12}$ explicit.\footnote{As we showed above, at the order we are computing only $A_{nm}$ with $n=m$ contribute.} We are interested in computing the Fourier transform $\tilde A_{nn}(k)=\int d^3 x\, e^{-i \vk \cdot \vx_{12}}A_{nn}(x_{12})$ with $A_{nn}(x_{12})$ given by
\bea
\label{Annx}
&& A_{nn}(x_{12})=\\
&&\qquad1,\quad x_{12}<x_i \nn \\ \nn
&&\qquad 1+2\lambda_n  \int_{t_i(x_{12})}^t d\, t' \l[j_0\l(\eps a(t') H x_{12}\r)-1\r] , \quad x_i<x_{12}<x_f\\ \nn
&& \qquad \l(1+2\lambda_n  \int_{t_i(x_{12})}^{t_f(x_{12})} d\, t' \l[j_0\l(\eps a(t') H x_{12}\r)-1\r] \r) e^{-2\lambda_n(t-t_f(x_{12}))} , \quad x_f<x_{12}
\eea
Consequently, the Fourier transform of $A_{nn}(x_{12})$ will split into three integrals:
\bea
\label{threeints}
\tilde A_{nn}(k)=\l(\int_0^{x_i} d^3 x_{12} \,+\int_{x_i}^{x_f} d^3 x_{12} +\int_{x_f}^{\infty} d^3 x_{12} \r)e^{- i\vk\cdot\vx_{12}}\ A_{nn}(x_{12})
={\cal{I}}_{I}+{\cal{I}}_{II}+{\cal I}_{III}\;.\nn\\
\eea
This expression allows us to compute several $\sqrt{\lambda}$ corrections to the spectrum at small $k\ll \epsilon a(t) H$, more precisely, to the prefactor in Eq.~\eqref{P2k}. However, the most interesting corrections in this regime come from the corrections to the Eigenvalues, which we compute in section \ref{sec:Eigenvalues}. We, therefore, will not compute the corrections to the prefactor in this regime. Importantly, we have shown that the corrections coming from the sudden perturbation theory are small and consequently the result in \eqref{P2k} is a reliable approximation. 

Instead, the corrections from sudden perturbation theory play the most significant role at $k\sim\eps a H$ for which ${\cal I}_{II}$ gives the main physical contribution. We thus focus on this case. The computation of the integrals is straightforward but rather tedious and hence is relegated to the Appendix~\ref{app:2-point-Fourier}. 
{Notice that, for modes that are very close to the cutoff, $|k/\Lambda-1|\sim e^{-\Delta}$, the corrections to time evolution at late times in region III are also important. They can be computed by simply integrating perturbatively  equation \eqref{Shroed} at late times. Details are also presented in Appendix~\ref{app:2-point-Fourier}.} Keeping the leading contribution when $\lambda\to0$, $\Delta\to \infty$ and $\sqrt{\lambda}\Delta\to0$, we get
\bea
\label{Annk}
\tilde A_{nn}(k)=\frac{4\pi^2\lambda_n}{H}k^{-3}\theta_\Delta(\eps a(t) H-k)\l(1+O\l(\sqrt{\lambda}\Delta,e^{-\Delta}\r)\r)\;.
\eea
Even though we indicated a potential $O\l(\sqrt{\lambda}\Delta\r)$ correction coming from a naive estimate of the next term in the sudden perturbation theory, similarly to what happens for the dependence on $\epsilon$ and $\delta$, all the $\Delta$ dependence will get cancelled at the next order and the actual correction to the result is of the relative order $O\l(\sqrt{\lambda}\r)$. { $\theta_\Delta$ in~(\ref{Annk}) represents  { $\theta$-function smeared on a scale $k/\Lambda-1\sim e^{-\Delta}$, which is given explicitly in App.~\ref{app:2-point-Fourier}}.

From \eqref{Annk}, \eqref{latetime} and \eqref{2loclead}} the Fourier transform of the two-locations equilibrium distribution, $\tilde{P}^{eq}(\phi_1,t;\phi_2,t;\vk)$ { at all wavenumbers} can be easily derived:
\be
\label{2ptfourier}
\tilde{ P}^{eq}(\phi_1,t;\phi_2,t;\vk)=\theta_\Delta(\Lambda(t)-k)\frac{4\pi^2}{H k^{{3}} }\sum_{n=1}^{\infty} \left(\frac{k}{a H}\right)^{2\lambda_n/H}\Phi_n(\phi_1)\Phi_n(\phi_2){\lambda_n}\l(1+O\l(\sqrt{\lambda}\r)\r)\, .
\ee
Note that we have dropped the contribution coming from $n=0$ since it corresponds to the disconnected piece and is proportional to $\delta^{(3)}(\vk)$. This expression allows us to compute correlation functions of the form $\langle\hat \phi_\ell^n(\vk,t)\hat \phi_\ell^m(-\vk,t)\rangle$. We see that at the leading non-trivial order all of these correlation functions are proportional to {the smeared theta-function $\theta_\Delta(\Lambda(t)-k)$}. Of course this will not be true to higher orders, for example, it is evident that the correlator $\langle\hat \phi_\ell^2(\vk,t)\hat \phi_\ell^2(-\vk,t)\rangle$ has some support at $\Lambda\lesssim k\lesssim2\Lambda$. What we observe here is that, as we had anticipated earlier, in order to see this effect one needs to pay additional powers of $\sqrt{\lambda}$. One could arrive at this conclusion by analyzing the perturbation theory of short modes, however, it is an important, and somewhat non-trivial, consistency check that we were able to derive the $\theta$-function by studying the differential equation for the distribution of the long modes.

\subsubsection{$n$-location distribution}
\label{nlocation}
In this section, we have focussed on the perturbative computation of two-locations $n$-point functions. It is quite clear how to generalize it to an higher number of locations. The main idea is that, at very early times, all points start completely correlated, and then, generically one at the time, and in the same sudden way as we used in this section, each point becomes uncorrelated from the rest and starts evolving independently. It is quite clear at this point how in this way one can generalize the calculation to an arbitrary number of locations. See~\cite{Creminelli:2008es} for an early example of this computation in the context of slow-roll eternal inflation. 
{ We are now going to explain this in some more detail. 

Imagine that we wish to compute the equal-time $n$-location distribution {$P_n(\phi_1,\,\vec x_1,\, t\,;\ldots;\;\phi_n,\,\vec x_n ,\,t)$, which in this section, for shortness, we will denote as $P_n(\phi_1,\ldots\phi_n;t,\{\vx_{ij}\})$}.
 For simplicity, let us focus on the regime where the distances between the two nearest points are hierarchically organized, {\it i.e.}, up to relabeling of the points, we have {$x_{ij}\gg x_{jk}$, for any $i,j,k$ such that $i<j$ and $i<k$. As before, we fix all $\vx_i$'s and study evolution of $P_n$ with time. }
In  this situation, let $t_{0,i}$ be the point when the long field at $\vx_i$, $\phi_i$, begins to behave independently from the other ones (out of which, the field values $\phi_{1},\ldots, \phi_{i-1}$ behave all independently, while the field values $\phi_{i+1},\ldots,\phi_{n}$ are still completely correlated). {We will work at leading order in sudden perturbation theory and, as before, take 
 $t_{0,i}=-H^{-1}\log(\epsilon H x_{i i+1})$. }
 
 {Other hierarchical kinematical configurations, where, for example, points first separate into groups, can be studied in an analogous way. In case when distances between some pairs of points are comparable, in order to be sensitive to the ratio of these distances, one needs to include subleading sudden perturbation theory corrections, similarly to what was done in the previous section for $P_2$.}

As anticipated in section~\ref{sec:perturbatie-structure}, up to quantum corrections, the $n$-location probability distribution satisfies a straightforward generalization of~(\ref{Shroed}): 
\bea\label{Shroedn}
&&\frac{\d}{\d t}P_n(\phi_1,\ldots\phi_n;t,\{\vx_{ij}\})=\\ \nn
&& \qquad  =-\left(H_1^{(3)}+{\frac{1}{2}}{\sum_{i\neq j}} j_0(\eps a H  x_{ij}) H'_{ab}\right)P_n(\phi_1,\ldots\phi_n;t,\{\vx_{ij}\})\ ,
\eea
where
\be
H_1^{(n)}=-\Gamma_{\phi_1}-\ldots-\Gamma_{\phi_n}\ ,\qquad H'_{ij}=-\frac{H^3}{4\pi^2}\frac{\d^2}{\d{\phi_i}\d{\phi_j}}\ . 
\ee

At leading order in  sudden perturbation theory, at each time, eq.~(\ref{Shroedn})  can be simply recast as the {\it independent} evolution of some field values: the ones that are already fully decoupled from the others, and the ones that are still fully correlated among them, and that are treated as one single field value.  
Let us start by considering the evolution at times earlier than $t_{0,1}$. Clearly, in this regime all field values are fully correlated, and the distribution is the equilibrium one
\be
P_{n}(\phi_1,\ldots,\phi_n;t<t_{0,1},\{\vx_{ij}\})=\left( \prod_{q=2}^{n} \delta^{(1)}(\phi_{1}-\phi_q)\right) P_1^{eq}(\phi_1)\ .
\ee  
Next, let us consider the evolution in the interval $t\in[t_{0,1},t_{0,2}]$, where the evolution of the point $\phi_1$ is uncorrelated to all the other one.  The boundary condition at $t=t_{0,1}$ is such that also the field $\phi_1$ is fully correlated with the others. We therefore have the same solution as in the two-location case we just studied:
\bea\label{eq:oneout}\nn
&&P_{n}(\phi_1,\ldots,\phi_n;t_{0,1}<t<t_{0,2},\{\vx_{ij}\})=\sum_p e^{-2\lambda_p (t-t_{0,1})} \Phi_p(\phi_1)\Phi_p(\phi_{2})   \left( \prod_{q=3}^{n} \delta^{(1)}(\phi_{2}-\phi_q)\right)\\ 
&&\qquad\qquad\qquad=\sum_p \left(\epsilon a(t) H x_{12}\right)^{-2\lambda_p/H} \Phi_p(\phi_1)\Phi_p(\phi_{2})   \left( \prod_{q=3}^{n} \delta^{(1)}(\phi_{2}-\phi_q)\right) \ .
\eea
In the subsequent interval $,t_{0,2}<t<t_{0,3}$, the field $\phi_{2}$ begins to evolve in an independent way from the other field values. The general solution in the interval takes the form
\bea
&&P_{n}(\phi_1,\ldots,\phi_n;t_{0,2}<t<t_{0,3},\{\vx_{ij}\})=\\ \nn
&&=\sum_{p_1,p_{2},p_3} A_{p_1,p_2,p_3}(t_{0,1},t_{0,2}) e^{-(\lambda_{p_1}+\lambda_{p_2}+\lambda_{p_3}) (t-t_{0,2})}\; \Phi_{p_1}(\phi_1)\Phi_{p_2}(\phi_2)\Phi_{p_3}(\phi_3) \left( \prod_{q=4}^{n} \delta^{(1)}(\phi_{3}-\phi_q)\right) ,
\eea
with $A_{p_1,p_2,p_3}(t_{0,1},t_{0,2})$ is fixed by imposing this expression to be equal to (\ref{eq:oneout}) evaluated at $t=t_{0,2}$. By using the orthonormality of the Eigenfunctions, we can solve for $A_{p_1,p_2,p_3}(t_{0,1},t_{0,2})$ and write 
\bea
\label{P3time}
&&P_{n}(\phi_1,\ldots,\phi_n;t_{0,2}<t<t_{0,3},\{\vx_{ij}\})=\left( \prod_{q=4}^{n} \delta^{(1)}(\phi_{3}-\phi_q)\right) \ \times \\ \nn
&&\qquad \times\ \sum_{p_1,p_{2},p_3} \hat C_{p_1,p_2,p_3}\, e^{-2\lambda_{p_1} (t_{0,2}-t_{0,1})} e^{-(\lambda_{p_1}+\lambda_{p_2}+\lambda_{p_3}) (t-t_{0,2})}\; \Phi_{p_1}(\phi_1)\Phi_{p_2}(\phi_2)\Phi_{p_3}(\phi_3) \  ,
\eea
where 
\be
\label{OPE}
\hat C_{p_1,p_2,p_3}\equiv\int d\phi\, \mu^2(\phi)\; \Phi_{p_1}(\phi)\Phi_{p_2}(\phi)\Phi_{p_3}(\phi)\,
\ee
 is typically an order one number. If $n=3$, we would be done. Intuitively, we see that the decay of the correlation function between 
time $t_{0,2}$ and $t_{0,1}$, associated to the decay of the correlation between $\phi_1$ and the other field values, is now multiplied by another exponential decay in the time interval $t-t_{0,2}$, now associated to the fields $\phi_1$ and $\phi_2$ being independent from the others. Also, as for the two-location distribution, all the $x$ dependence comes from the crossing times $t_{0,i}$'s. Notice, importantly, that this results shows that the three-point function will be unsuppressed with respect to the two-point function, again confirming that  our non-perturbative solution cannot be interpreted as the dynamical generation of a mass term.

{Before we proceed with the general case, let us study the result for the three-point distribution in slightly more detail. It is of particular interest the late-time limit when the three points are already separated. In this case, the expression \eqref{P3time} with $n=3$ is the relevant one. Substituting for $t_{0,1}$ and $t_{0,2}$ we can recover the explicit $x$-dependence to get
\bea
&& P_{3}(\phi_1,\phi_2,\phi_3;t,\vx_{12},\vx_{23})
 =\sum_{p_1,p_{2},p_3} \frac{\hat C_{p_1,p_2,p_3}\Phi_{p_1}(\phi_1)\Phi_{p_2}(\phi_2)\Phi_{p_3}(\phi_3)}{(a H x_{12})^{\frac{2 \lambda_{p_1}}{H}}(a H x_{23})^{\frac{ \lambda_{p_2}+\lambda_{p_3}-\lambda_{p_1}}{H}}} \,.
\eea
Note that in our approximation we cannot distinguish between $x_{12}$ and $x_{13}$, as subleading $O(\sqrt{\lambda})$ corrections are necessary to do this. We thus can conclude that our three point function is consistent with the one that we expect from dS invariance, namely
\bea
&& P_{3}(\phi_1,\phi_2,\phi_3;t,\vx_{12},\vx_{23})
 =\sum_{p_1,p_{2},p_3} \frac{\hat C_{p_1,p_2,p_3}\Phi_{p_1}(\phi_1)\Phi_{p_2}(\phi_2)\Phi_{p_3}(\phi_3)}{(a H x_{13})^{\frac{ \lambda_{p_1}+\lambda_{p_3}-\lambda_{p_2}}{H}}(a H x_{12})^{\frac{ \lambda_{p_1}+\lambda_{p_2}-\lambda_{p_3}}{H}}(a H x_{23})^{\frac{ \lambda_{p_2}+\lambda_{p_3}-\lambda_{p_1}}{H}}} \,.\nn\\
\eea
In principle, it is possible to compute the subleading corrections and check  the above expression, but we leave this for the future. Instead, in section \ref{sec:generic-2-point} we will check explicitly dS-invariance of the two-point function.
}

Let us proceed with the higher-point functions. Iterating, in the generic interval $[t_{0,j-1},t_{0,j}]$, with $1\leq  j\leq n$, with {$t_{0,n}=\infty$}, we have
\bea
&&P_{n}(\phi_1,\ldots,\phi_n;t_{0,j-1}<t<t_{0,j},\{\vx_{ij}\})=\\ \nn
&&\quad=\sum_{p_1,\ldots ,p_j} A_{p_1,\ldots,p_j}(t_{0,1},\ldots, t_{0,j-1}) e^{-\sum_{i=1}^j\lambda_{p_i}( t-t_{0,j-1})}\; \left( \prod_{i=1}^{j}\Phi_{p_i}(\phi_i)\right)  \left( \prod_{q=j+1}^{n} \delta^{(1)}(\phi_{j}-\phi_q)\right)\ ,
\eea
where $A_{p_1,\ldots,p_j}(t_{0,1},\ldots, t_{0,j-1})$ {can be expressed recursively through integrals of Eigenfunctions similarly to \eqref{OPE}, multiplied by some exponentials of time intervals.  Each time $t_{0,i}$, with $i<j$, appears in the exponents as }$e^{-\left(\sum_{i'\leq i} \lambda_{i'} \right)(t_{0,i}-t_{0,i-1})}$, where $\lambda_i$'s are some of the Eigenvalues.
This time dependence corresponds to the decay of the correlation function as more and  more points begin to evolve independently. We do not expect any further suppression than this, showing that the theory is completely non-Gaussian.

\section{Stability of rigid dS space and attractiveness of the Bunch Davies vacuum}
\label{sec:stability}

In this section we discuss the implications of our results for states that are different than the BD vacuum and also for the stability of de Sitter space in the rigid limit.
As already mentioned at the beginning of section~\ref{sec:nonequil1pt}, the Fokker-Planck-like equation that we have derived requires some knowledge of the wave function in order to be able to express the action of  $\hat \Pi $ on $\Psi[\phi]$ as the multiplication of $\Psi$ by a function of $\phi$. Our Fokker-Planck-like equation applies therefore only  to those states  that have the same  expression for $\hat \Pi\,\Psi[\phi]$. Within this class of states, we have shown in section~\ref{sec:nonequil1pt} how the BD solution is an attractor. Since correlation functions in the  BD state decay, this tells us that rigid de Sitter space, with the BD vacuum, is stable.  It is now interesting to consider more  general states.

There  are several ways to consider different states than the  BD vacuum. Perhaps a rather general one is to realize that there is a large family of states in the vicinity of the BD ones whose wave function can be obtained by performing a path integral with the BD boundary conditions in the infinite past, and with appropriate insertions of operators along the path. Schematically, we can define such a state as
\be\label{eq:generic_state2}
\psi_{{\cal{O}}_{1,\ldots, n}}={ N_{{\Psi}}^{-1}}\int {\cal{D}}\phi \; {\cal O}(x_{s,1}^\mu,x_{s,2}^\mu,\ldots)\, e^{i S}\ ,
\ee
where $ {\cal O}(x^\mu_{s,1},x_{s,2}^\mu,\ldots)$ is made of products of $\phi$ fields at spacetime points {$x^\mu_{s,1}, x^\mu_{s,2},\ldots$,} { and $N_\Psi$ is a normalization constant}. Therefore, correlation functions in such a state can be constructed out of correlation functions in the BD vacuum.  {This teaches us two important points. First is that the decay of connected correlation functions in the BD vacuum {with time}  that we derived earlier on implies that the connected correlation functions  will decay in this additional state as well~\footnote{We remind the reader that connected correlation functions  decay because the space and time dependence of connected correlation functions is dictated by the strictly positive Eigenvalues of the operator $-\Gamma_{\phi}$ ({for correlators of operators inserted at both different time and space see section~\ref{sec:generic-2-point}}).}. We take the fact that correlation functions decay in time as the statement that de Sitter space is stable in the rigid limit, which is the limit we are working in in this paper. 

The second point is that if wee keep the physical ({\it i.e.} the de Sitter-invariant) distance between the points of the correlation function constant, while we send the distance between the insertion points of the operator that define the states, $x^\mu_{s,i}$ and the points of evaluation of the correlation function to infinity, then correlation functions converge to the same ones as the BD vacuum. This shows that the BD vacuum is a late-time attractor, at least for the states that can be obtained by~(\ref{eq:generic_state2}). Additionally, since, as we will show in section~\ref{sec:generic-2-point},  correlation functions in the BD vacuum linearly realize the de Sitter symmetry, we also conclude that this symmetry is not spontaneously broken in the same limit.}

It is instructive to show that the BD vacuum is an attractor for some state in its vicinity  by considering the following alternative construction. As we have seen, though correlation functions cannot be computed perturbatively, the result of  acting with $\hat \Pi $  on $\Psi[\phi]$ and write it as a multiplication of $\Psi$ by a function of $\phi$ can be computed from a perturbatively-computed wave function. Since this relation is what specified our Fokker-Planck equation to the BD vacuum, one could consider to construct a similar equation for a different state. However, for states `sufficiently close' to the BD vacuum, we can see that the answer quickly converges to the one obtained with the BD vacuum. Here, by `sufficiently close' we mean states where correlation functions preserve the same scaling as the BD vacuum; in particular, the typical size of $\phi_\ell$ is of order $\lambda^{-1/4}$. To show this, let us consider a general modification to the wave function of the BD vacuum valid for long wavelength modes, as these are the ones that become non-perturbative:
\be
\Psi[\phi]=\Psi_{BD}[\phi]\; {\rm Exp}\left(i\int d^3x \frac{\tilde\varepsilon(\lambda,\phi_\ell(\vec x),\eta)}{\eta^3} \right)\ ,
\ee
where we neglected gradients of the field, { as they are suppressed}. At late times, the leading component of the BD wave function is the potential term $\sim i \lambda\phi_\ell^4/\eta^3$, which, in $\lambda$ counting, scales as $i/\eta^3$. So, in order to preserve the same scaling for~$\phi_\ell$, we assume that  $\tilde\varepsilon(\lambda,\phi_\ell(\vec x),\eta)$ is at most of order one.

Let us consider the Schrodinger equation for this wave function, eq.~(\ref{eq:shrodinger}). Given that the BD vacuum is a solution, each single term of this equation will vanish if $\tilde\varepsilon$ vanishes. Therefore, we obtain, schematically,
\bea\nn
&&i \frac{\d}{\d \eta}\Psi[\phi]=\left\{\int d^3k \left[-\frac{\eta^2}{2H^2}\frac{\delta^2}{\delta\phi_{\vec k}\delta\phi_{-\vec k}}+\frac{H^2}{2\eta^2}k^2 \phi_{\vec k}\phi_{-\vec k}\right]+\int d^3 x \frac{H^4}{\eta^4} \frac{\lambda \phi(\vec x)^4}{4}\right\} \Psi[\phi] \\ \nn
&&\Rightarrow\ -\frac{\d_\eta\tilde\varepsilon(\lambda,\phi_\ell(\vec x),\eta)}{\eta^3}+3\frac{\tilde\varepsilon(\lambda,\phi_\ell(\vec x),\eta)}{\eta^4}=O(\sqrt{\lambda})\;\times\; \tilde\varepsilon(\lambda,\phi_\ell(\vec x),\eta)+O(\sqrt{\lambda})\;\times\; \tilde\varepsilon(\lambda,\phi_\ell(\vec x),\eta)^2\ .
\eea
The right-hand size contains the results when at least one of the functional derivatives $\delta/\delta\phi$ acts on $\tilde\varepsilon$. Because of the scaling of $\phi_\ell$, these functional derivatives are suppressed by a factor of $\sqrt{\lambda}$. Notice that this scaling applies only to long wavelength fields. At leading order, the right-hand side can therefore be dropped. The solution for  $\tilde\varepsilon(\lambda,\phi_\ell(\vec x),\eta)$ is therefore
\be
\tilde\varepsilon(\lambda,\phi_\ell(\vec x),\eta)=\tilde\varepsilon(\lambda,\phi_\ell(\vec x),\eta_0)\,\frac{\eta^3}{\eta_0^3}\ \to\ 0\ ,
\ee
{where $\eta_0$ is time at which we perturbed the state.}
This is nothing but the fact the slow rolling solution is an attractor for modes outside the horizon. This result shows that for a significant basin of attraction, the wave function of the long modes, and therefore correlation functions, tend to the one of the BD vacuum. The BD state is therefore an attractor,{ for a class of states we just described. In Appendix \ref{app:Wigner} we show a slightly different formalism that allows us to consider an even wider class of states and again find that the BD state is an attractor.}

\section{Subleading-order calculations for $\lambda \phi^4$ }
\label{sec:sublead}
\subsection{Subleading $P_1$}
\label{sec:subleadP1}
Our next step is to derive the next-to-leading order corrections to the equation governing the single-field distribution $P_1$. Remember that Eq.~\eqref{1pt} still contained all the terms of relative order $O(\sqrt{\lambda})$. We, however, neglected those terms while going to \eqref{1pteqtime}.
Our goal in this section is to repeat the same steps, keeping track of the relevant subleading terms. Naturally, there are corrections to the diffusion term, which come from the fact that the correlator of short modes of $\phi$, $\langle\hat\phi(\vk_s,t)\hat\phi(-\vk_s,t)\rangle_{\phi_\ell}$ with $k_s=\Lambda(t)$ has perturbative corrections, and corrections to the drift term, which come from the expectation value of the momentum operator and are due to the subleading terms in the wave function.

Let us study these corrections separately before combining them into the full equation. To make the expressions slightly shorter in this section we work with the $\lambda \phi^4$ potential.

\subsubsection{$\phi\;$-$\phi$ correlator}
Let us now turn to the correlator $\l\la\hat\phi(\vk_s,t)\hat\phi(-\vk_s,t)\r\ra'_{\phi_\ell}$ for $k_s=\Lambda(t)$. At order $\sqrt{\lambda}$, the only effect is the correction to the mass of short modes produced by the background of the long modes. Because of the smoothness of the window function, we can take the short modes to be local operators that depend on the long wavelength fields at the same location, up to corrections of higher order in $\sqrt{\lambda}$ and $\delta$. We prove this intuitive fact in App.~\ref{app:locality-space}. Therefore the effective Hamiltonian of short modes contains the term $\int d x^3 \frac{1}{2}\delta m_s^2\phi_s^2$ with 
\be
\label{deltam2}
\delta m_s^2= 3\lambda \phi_1^2\;.
\ee
The result for the massive correlator is well-known~(see~(\ref{eq:massive2pt})). Expanding to subleading order in $\delta m_s^2$ and leading order in $\eps$, we get 
\be
\label{diffsl}
\l\la\hat\phi(\vk_s,t)\hat\phi(-\vk_s,t)\r\ra'_{\phi_1}=\frac{H^2}{2 \Lambda(t)^3}\l(1+\l( \log\l(\eps/2\r)-\psi\l(3/2\r)\r)\frac{2 \delta m_s^2}{3 H^2}\r)+O(\lambda)
\;.
\ee
Our notation stresses that at this order the correlator depends only on $\phi_1$.

\subsubsection{Subleading momentum}

From (\ref{Pidef}), in order to compute the subleading contribution from the momentum, we need to compute the subleading part of the imaginary part of exponent of the  wave function. There are  {four} main contributions at the order we are concerned. 

There is a tree-level contribution from the diagram on the left of Fig.~\ref{fig:subleadingdiagrams}. The resulting expression reads
\bea
&&\left({\rm Fig.~\ref{fig:subleadingdiagrams},\; left}\right)=i^2\frac{\lambda^2}{4^2} \left(\frac{4\cdot4}{2}\right)  (-i)^2\int^0_{-\infty}d\Delta\eta_1 \int^0_{-\infty}d\Delta\eta_2\\ \nn
&&\qquad \left(\left(-\frac{1}{H\eta_1}\right)^4\left(-\frac{1}{H\eta_2}\right)^4  \left[G\left(\eta_1,\eta_2,\left|\sum_{i=1,2,3}\vec k_i\right|\right) \times\right.\right. \\ \nn  
&&\qquad\ \left.\left. \times\ K(\eta_1,k_1)K(\eta_1,k_2)K(\eta_1,k_3)K(\eta_2,k_4)K(\eta_2,k_5)K(\eta_2,k_6)\right]_{\eta_2=\eta_1-i\Delta\eta_2,\;\eta_1=\eta-i\Delta\eta_1}\right.+\\ \nn
&&\qquad \  \left.+\left\{\eta_1\leftrightarrow\eta_2\right\}\right)\ ,
\eea
where $G$ and $K$ are given in \eqref{Bulk2Bulk} and \eqref{eq:Bulk2Boundary} correspondingly.
The evaluation of this diagram is highly facilitated by the fact that with our power counting rules, we can focus on all wavelengths being long. The integral is therefore convergent in the past even without accounting for the exponential  suppression from the $i\varepsilon$-prescriptions, so one can take the limit $k\to 0$ in all the above expressions before taking the time integral. We obtain
\bea\nn
&&\log\Psi\ \supset\ - i a^3 \frac{\lambda^2}{54 H^3}\int \left(\prod_{i=1,\ldots,6}\frac{d^3k_i}{(2\pi)^3}\right) (2\pi)^3\delta^{(3)}\left(\sum_{i=1,\ldots,6}\vk_i\right)\ \times \phi_{\vec k_1}\ldots  \phi_{\vec k_6}\left(1+O({\eps})\right) , \\ 
\eea
which leads  to the following  expression for the momentum
\be
\Pi_1\supset -\frac{\lambda^2  \phi_1^5}{9  H^3} \left(1+O(\epsilon,\sqrt{\lambda})\right)\ ,
\ee
{where by $\phi_1$ and $\Pi_1$ we, as before, denote long fields and momenta at $\vx_1$. }
\begin{figure}[htbp] 
	\centering
		\includegraphics[width=.9\linewidth,angle=0]{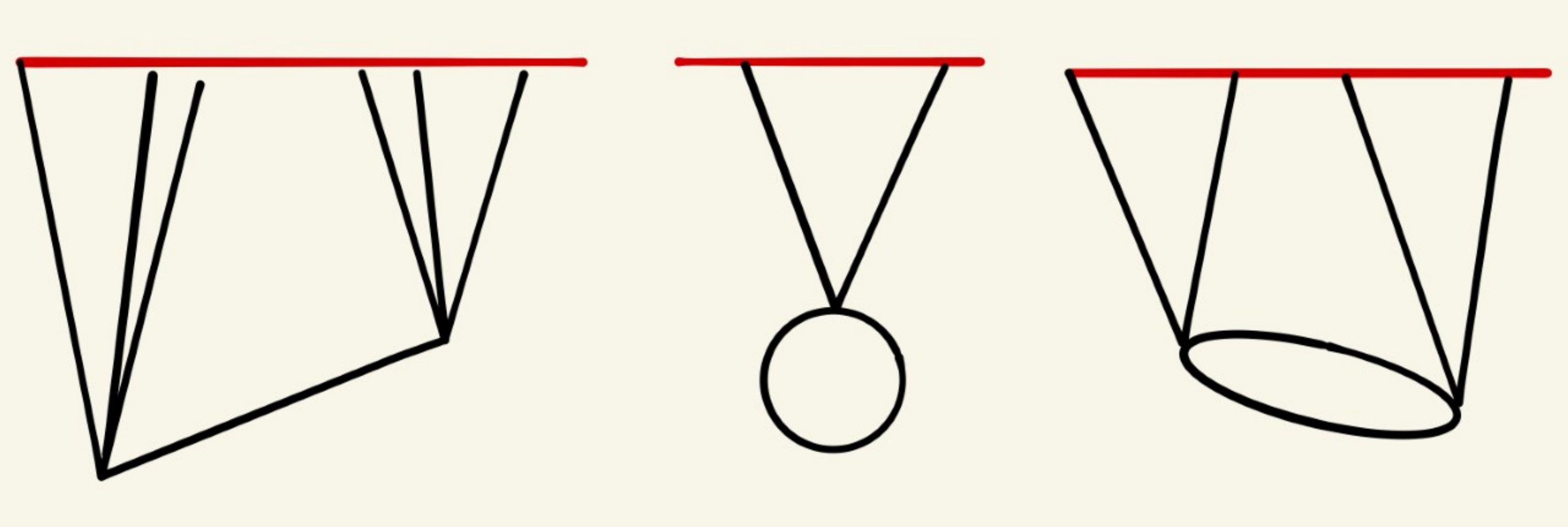}
	\caption{\small Witten diagrams contributing to the exponent of the wave function that are  needed to evaluate the momentum operator at subleading order. {\it Left:} a tree level diagram contributing to a term in $\lambda^2\phi^6$; {\it center:} a one-loop diagram contributing to a term in $\lambda\phi^2$; {\it right:} a one loop diagram contributing to a term in $\lambda^2\phi^4$, and which turns out to be negligible.}
	\label{fig:subleadingdiagrams}
\end{figure}

Next, we evaluate the one-loop diagram shown in the center  of Fig.~\ref{fig:subleadingdiagrams} that contributes to the $\phi^2$-term  in the exponential of the wave function. It reads
\bea
&&\left({\rm Fig.~\ref{fig:subleadingdiagrams},\; center}\right)=i\frac{\lambda}{4} \left(4\cdot3\right)  (-i)\int^0_{-\infty}d\Delta\eta_1 \int \frac{d^3q}{(2\pi)^3}\\ \nn
&&\qquad\qquad \left[ \left(-\frac{1}{H\eta_1}\right)^4  G\left(\eta_1,\eta_1,q\right)K(\eta_1,k_1)K(\eta_1,k_2)\right]_{\eta_1=\eta-i\Delta \eta_1}=-i a(\eta)^3 c_1\,\lambda H (1+O(\epsilon))  \ ,
\eea
where $c_1$ is  some real numerical constant.\footnote{{This diagram is UV divergent and, consequently, $c_1$ depends on the UV regulator. Below we will see that it will combine with other UV-divergent contributions to contribute to the finite physical mass term. The prescription for the UV regularization was discussed in section \ref{sec:wavefunctionperturbativity}}. } The last step follows from the fact that we can, again, take the limit $k_1\to 0$ and $k_2\to 0$ before taking the time integral. This contributes to the wave function as 
\bea\nn
&&\log\Psi\ \supset\ - i c_1 a^3 \lambda H \int \left(\prod_{i=1,2}\frac{d^3k_i}{(2\pi)^3}\right) (2\pi)^3\delta^{(3)}\left(\sum_{i=1,2}\vk_i\right)\ \times \phi_{\vec k_1}  \phi_{\vec k_2}\left(1+O(\eps)\right)\ , \\
\eea
which leads  to the following  expression for the momentum
\be\label{eq:pi1}
\Pi_1\supset -2c_1  \lambda H\phi_1 \left(1+O(\sqrt{\lambda},\eps)\right)\ .
\ee

Another relevant diagram is associated to the same tree-level diagram that contributes to the wave function that we considered in~(\ref{eq:wavefuncquartic}), in Fig.~\ref{fig:quartic_tree_level}, and that, again, was computed in~\cite{Anninos:2014lwa}. In this case, however, we can take two modes to be long (and taken in the limit of vanishing wavenumber, {so that they are enhanced by $\lambda^{-1/4}$}), and two modes short. The diagram reads 
\bea
&&\left({\rm Fig.~\ref{fig:quartic_tree_level}}\right)\equiv\lambda\,a(\eta)^3\,H\, {\cal{I}}(k_3,k_4,\eta)=-\lambda\frac{i e^{-i \eta (k_3+k_4)}}{24
   \eta^3 H^4 (\eta k_3+i) (\eta k_4+i)}\;\times \\ \nn
  &&\qquad\times\ \left(2   e^{i \eta (k_3+k_4)} \left(\eta^2 \left(k_3^2-k_3
   k_4+k_4^2\right)-i \eta
   (k_3+k_4)+1\right)\right.\\ \nn
   &&\quad\qquad\left. -\eta^3  
   \left(k_3^3+k_4^3\right) (2 i \text{Ci}((k_3+k_4)
   \eta)-2 \text{Si}((k_3+k_4) \eta)+\pi )\right)\l(1+O(\eps,{\sqrt{\lambda}})\r)\ ,
\eea
where Ci and Si are respectively the CosineIntegral and the SineIntegral. This contributes to the wave function as 
{\bea
&&\log\Psi\ \supset - i \lambda a^3\int_0^{\Lambda} \left(\prod_{i=1,2}\frac{d^3k_i}{(2\pi)^3}\right)\int_\Lambda^\infty \left(\prod_{i=3,4}\frac{d^3k_i}{(2\pi)^3}\right) \times\\ \nn
&& \qquad\qquad \times(2\pi)^3\delta^{(3)}\left(\sum_{i=3,4}\vk_i\right)\, H\, {\cal{I}}(k_3,k_4,\eta) \times \phi_{\vec k_1}\ldots  \phi_{\vec k_4}\left(1+O(\epsilon,\sqrt{\lambda})\right), 
\eea
and to the momentum as
\be\label{eq:pi2}
\Pi_1\supset-\lambda H \phi_1  \int_\Lambda^\infty \frac{d^3q}{(2\pi)^3}\;{\rm Re}\left[{\cal{I}}(q,q,\eta)\right]\left\langle\phi(\vec q) \phi(\vec q)\right\rangle' \left(1+O(\epsilon,\sqrt{\lambda})\right)\ .
\ee}
Notice that this particular  contribution  shows that in order to compute correlation functions of long fields, one needs to keep in the wave function also short modes, {\it i.e.} {if we kept only the long, $k\eta\ll 1$, limit of the wave function, we would get an order-one-wrong calculation of this subleading term in the momentum operator and hence an $O(\sqrt{\lambda})$ mistake in the calculation of correlators of the long modes, even on very long distances~\footnote{ 
Of course, this is just  a manifestation of the fact that if one computes of a physical quantity by regularizing the theory with a given physical cutoff $\Lambda^{\rm UV}_{\rm ph}$ (not to be confused with $\Lambda$), one should consistently keep all modes up to  {$\Lambda^{\rm UV}_{\rm ph}$} until the final physical observables is computed, because sensitivity to short modes can occur at several different stages of the calculation.
}.}

Both terms in (\ref{eq:pi1}) and (\ref{eq:pi2}) are degenerate in form with yet a last  contribution, the  one from a mass  term, which  indeed acts as a counterterm for the two diagrams above. It is therefore useless to compute the explicit result of the loop  integrals, and we  simply define the sum of these diagrams and of the mass counterterm to give the following result:
\be\label{eq:pi3}
\Pi_1\supset-\left[\frac{\bar m^2}{3H} -\frac{\lambda H}{4\pi^2}\l( \log\l(\eps/2\r)-\psi\l(3/2\r)\r)\right]  \phi_1  \left(1+O(\epsilon,\sqrt{\lambda})\right)\ .
\ee
Notice that this offers a definition of the {``physical''} mass, $\bar m$, where we pulled off the factor of $\frac{\lambda H}{4\pi^2}\l( \log\l(\eps/2\r)-\psi\l(3/2\r)\r)$ because it will simplify some forthcoming results. We stress that we take $\bar m^2$ to be much smaller than $\sqrt{\lambda} H^2$, but potentially larger or equal than $\lambda H^2$~(\footnote{We see explicitly the mass generation, as anticipated in section \ref{sec:strategy}.}), which justifies keeping this contribution.

{ Finally, another diagram that naively could contribute at the order we are concerned is represented on the right of Fig.~\ref{fig:subleadingdiagrams}. Given the discussion in section \ref{sec:wavefunctionperturbativity} we expect this diagram to be further subleading. We show it explicitly in section \ref{sec:subleading-tadopoles}.}

\subsubsection{Subleading equation for the single-point distribution\label{sec:subleading1point}}
We are now in a position to combine together the pieces computed above and write down the full equation governing the subleading time evolution of the one-point distribution:
\be
\label{1pteqsublead}
\frac{\d}{\d t}P_1(\phi,t)=\l[\frac{H^3}{8\pi^2}\frac{\d^2}{\d\phi^2}\l(1+a \phi^2\r)+\frac{\d}{\d \phi}\l(\frac{\lambda\phi^3}{3 H}+b\phi+c\phi^5\r)\r]P_1(\phi,t)\l(1+O(\eps^2,\delta,\lambda)\r)\,,
\ee
where we defined for shortness
\bea
\label{abc}
&& a=\frac{2 \lambda}{H^2}\l( \log\l(\eps/2\r)-\psi\l(3/2\r)\r)\,,\qquad c=\frac{\lambda^2}{9 H^3}\,,\qquad b=\frac{\bar m^2}{3H} -\frac{\lambda H}{4\pi^2}\l( \log\l(\eps/2\r)-\psi\l(3/2\r)\r)\;.\nn\\
\eea

Naturally, the leading order pieces in the equation agree with \eqref{1pteqphi}. Let us find the equilibrium solution of the subleading equation.
Using the leading solution given in \eqref{1pteqsol}, it is straightforward to get it:
 \bea\label{eq:subleading_sol_1loc}
 &&P_1^{eq}(\phi_1)=N e^{-\frac{ 2\pi^2 \lambda \phi_1^4}{3 H^4}}\l(1-\lambda\frac{\phi_1^2}{H^2}\l(\log\l(\eps/2\r)-\psi\l(3/2\r)+\frac{4 \pi^2}{3}\frac{\bar m^2}{\lambda H^2}\r)+\r.\nn\\
 &&\l.\qquad+\,\frac{8 \pi^2}{9}\frac{\lambda^2 \phi_1^6}{H^6}\l(\log\l(\eps/2\r)-\psi\l(3/2\r)-\frac{1}{6}\r)+O(\lambda)\r)\,,
\eea
where $N$ is, again, a  normalization constant, given by
\be
N^{-1}=\lambda^{-\frac{1}{4}}H\frac{6\sqrt{2}\pi\Gamma\l(\frac{1}{4}\r)+\sqrt{3}\sqrt{\lambda}\l(1-8\pi^2\frac{\bar m^2}{\lambda H^2}\r)\Gamma\l(\frac{3}{4}\r)}{4 \l(6 \pi^2\r)^{\frac{3}{4}}}
\;.
\ee
Note that the subleading terms depend on $\log\eps$. This is natural to expect because the correlation functions of the long modes only are supposed to have this dependence, however, it should cancel out once we compute the correlators of the full operator $\phi=\phi_\ell+\phi_s$. Indeed, let us compute the correlator of $\phi(x^{\mu}_1)^{2n}$ in the equilibrium state. At the subleading order a term with two short modes contributes. Their correlator can be easily computed given the known wave function and, to leading order, that is in the free massless approximation reads: 
\be
\label{phis2ofx}
\l\la\phi_s(x^{\mu})^{2}\r\ra=-\frac{H^2}{4 \pi^2}\l(\log\l(\eps/2\r)-\psi\l(3/2\r)\r)+H^2 V_2\, ,
\ee
where $V_2$ is a UV divergent piece that depends on the UV cutoff, but not on $\eps$. We also separated the  digamma function, $\psi$, from it in order to shorten some future formulas since the separation between finite and infinite pieces is arbitrary anyway. What is important is that the coefficient of  $\log\epsilon$ in \eqref{phis2ofx} is controlled by the IR part of the integral and hence is calculable. Using this result we get
\bea
\label{corrsublead}
&&\l\la\phi(x^{\mu}_1)^{2n}\r\ra=\l\la\phi_1^{2n}\r\ra+\frac{2n(2n-1)}{2}\l\la\phi_1^{2n-2}\r\ra\l\la\phi_s(x^{\mu})^{2}\r\ra+O\l(\lambda^{-n/2+1}\r)=\nn\\
&&\qquad=\lambda^{-\frac{n}{2}}H^{2n}\left(\frac{3}{2}\right)^{n/2}\Bigg[ \frac{ \pi ^{-n} \Gamma
   \left(\frac{n}{2}+\frac{1}{4}\right)}{\Gamma \left(\frac{1}{4}\right)}\,+
\nn\\
&&\quad\qquad+\sqrt{\lambda }\,\frac{6^{-3/2} \pi ^{-n-1}}{\Gamma
   \left(\frac{1}{4}\right)^2}  \Bigg(\l(3-24\,\pi^2\frac{\bar m^2}{\lambda H^2}- n\l(2-48 \pi^2 V_2\r)\r)
   \Gamma \left(\frac{1}{4}\right) \Gamma \left(\frac{n}{2}+\frac{3}{4}\right)-\nn\\
 &&\quad\qquad+   \l(3-24\,\pi^2\frac{\bar m^2}{\lambda H^2}\r) \Gamma
   \left(\frac{3}{4}\right) \Gamma \left(\frac{n}{2}+\frac{1}{4}\right)\Bigg)+O(\lambda)\Bigg]\,.
\eea
The leading piece is of course equal to \eqref{1ptphinlead}. The subleading piece is, to the best of our knowledge, computed here for the first time. As we have expected, the $\eps$ dependence cancelled between the leading short mode and the subleading long mode contributions. This is an important consistency check of our procedure. Note also that the cutoff-dependent part of the operator $\phi_s^2$, defined in \eqref{phis2ofx} and denoted by $V_2$, appears in the answer. This is related to the renormalisation of the composite operator, or in other words, we should add the contribution of the operator $\phi^{2n-2}$ with an infinite prefactor, whose finite part is arbitrary and, in principle, $n$ dependent. This is the only type of mixing that can be important at this order in $\sqrt{\lambda}$. {The UV-sensitivity} makes this result somewhat formal, however, we still feel it is an important exercise which demonstrates the technology we developed to deal with the IR issues. In section \ref{sec:Eigenvalues} we will compute the two-location correlation functions at large-distances at subleading order in $\sqrt{\lambda}$ using basically the same equation. The connected part of those correlators is free from the UV ambiguities and depends only on $\lambda$ and $\bar m^2$. 

\subsection{Corrections to the Eigenvalues and large-distance correlators\label{sec:Eigenvalues}}
In this section we will study subleading in $\sqrt{\lambda}$ corrections to the large-distance and late-time behavior of the correlation functions. As discussed in section~\ref{sec:two-point}, this is controlled by the Eigenvalues of the evolution operator that governs the time dependence of the single-field distribution. This time evolution at subleading order is given in equation \eqref{1pteqsublead}. Consequently, we need to solve the following Eigenvalue equation:
\be
\label{1pteqsubleadleo}
\l[\frac{H^2}{8\pi^2}\frac{\d^2}{\d\phi^2}\l(1+a \phi^2\r)+\frac{\d}{\d \phi}\l(\frac{\lambda\phi^3}{3 H}+b\phi+c\phi^5\r)\r]\Phi_n(\phi)=-\l(\lambda_n+\delta\lambda_n\r)\Phi_n(\phi)\,,
\ee
The terms proportional to $a$, $b$ and $c$, as well as $\delta\lambda_n$, are of relative order $O(\sqrt{\lambda})$, and we work to linear order in these parameters. It is convenient to bring the differential operator to a more canonical form. We first make the change of variable, $\phi=\phi'+a\phi'^3/6$, after which the equation becomes 
\bea
&&\l[\frac{H^3}{8 \pi ^2}\frac{\d^2}{\d\phi'^2}+\l( \frac{\lambda\phi'^3}{3 H} +\frac{3 a H^3 \phi' }{8 \pi ^2}+b \phi' +c \phi'^5 
  \r)\frac{\d}{\d\phi'}+\r.\nn\\
&& \qquad \l. +\l(\frac{\lambda  \phi'^2}{H}+\frac{a H^3}{4 \pi ^2}+\frac{a \lambda  \phi'^4}{3 H}+b+5 c \phi'^4\r)\r]\Phi_n(\phi')=-\l(\lambda_n+\delta\lambda_n\r)\Phi_n(\phi')\,.\nn\\
\eea
The coefficient of the two-derivative term is now a constant.
In order to get rid of the first-derivative term we define 
\bea
\tilde\Phi_n(\phi')=e^{\frac{\pi ^2 \lambda  \phi'^4}{3 H^4}+\frac{3 a \phi'^2}{4}+\frac{2 \pi ^2 b \phi'^2}{H^3}+\frac{2 \pi
   ^2 c \phi'^6}{3 H^3}} \Phi_n(\phi')\,,
\eea
For which the equation takes its final form:
\bea
&&\l[\frac{H^3}{8 \pi ^2}\frac{\d^2}{\d\phi'^2}+\l(\frac{a H^3}{16 \pi ^2}-\frac{a \lambda  \phi'^4}{6 H}-\frac{4
   \pi ^2 b \lambda  \phi'^4}{3 H^4}+\frac{b}{2}-\frac{4 \pi ^2
   c \lambda  \phi'^8}{3 H^4}+\frac{5 c \phi'^4}{2}-\frac{2 \pi
   ^2 \lambda ^2 \phi'^6}{9 H^5}+\frac{\lambda  \phi'^2}{2 H}\r)\r]\times\nn\\
 && \qquad\qquad\qquad \qquad\qquad\qquad \qquad \qquad\qquad\qquad \times \tilde\Phi_n(\phi')=-\l(\lambda_n+\delta\lambda_n\r)\tilde\Phi_n(\phi')\,.
\eea
Substituting \eqref{abc} we get
\bea
&&\l[-\frac{H^3}{8 \pi ^2}\frac{\d^2}{\d\phi'^2}+W_0(\phi')+W_1(\phi')\r] \tilde\Phi_n(\phi')=\l(\lambda_n+\delta\lambda_n\r)\tilde\Phi_n(\phi')\;,\nn\\
&&W_0(\phi')\equiv\frac{2 \pi ^2 \lambda ^2 \phi'^6}{9 H^5}-\frac{\lambda  \phi'^2}{2 H}\,,\\
 &&W_1(\phi')\equiv\frac{4 \pi ^2 \lambda ^3 \phi'^8}{27 H^7}+\frac{4 \pi ^2
   \lambda  \bar m^2 \phi'^4}{9 H^5}-\frac{5 \lambda ^2 \phi'^4}{18 H^3}-\frac{\bar m^2}{6
   H}  \,.\nn
\eea
Here $W_0\sim\sqrt{\lambda}H$ is the leading potential term. The bound states in this potential have energies corresponding to the leading Eigenvalues $\lambda_n$. $W_1\sim\lambda H$ (assuming $\bar m^2\sim \lambda H^2$) is instead a small perturbation. Before we proceed, note that all dependence on $\epsilon$ cancelled out after our manipulations. This means that the Eigenvalues are independent of it, as they should be, since at large distances the contribution of short modes can be neglected. 

Corrections to the Eigenvalues can be computed using the usual formulas of perturbation theory for the Schrodinger equation:
\be
\delta\lambda_n=\int d \phi\, \tilde \Phi^2_n(\phi) W_1(\phi)=\int d \phi\, \mu(\phi) \Phi^2_n(\phi) W_1(\phi)\;,
\ee
where in the last step we went back to the original Eigenfunctions of the operator $\Gamma_\phi$ defined in \eqref{Phin}. Consequently, $\delta \lambda_n\sim H \lambda$.

The correction to the $n=0$ Eigenvalue vanishes, as can be easily checked analytically: therefore there is still an  equilibrium state at this order. Since we do not know the analytic expressions for any higher Eigenfunctions, corrections to other Eigenvalues can be only computed numerically, which is, nevertheless, straightforward. As we anticipated, the Eigenvalues only depend on $\lambda$ and on the value of the physical mass, $\bar m$, but not on any UV-cutoff or $\eps$-dependent quantities. As an example, the equal time two-point function at large distances behaves as
\be\label{eq:subleadingtimedependenceleo}
\l\la\hat \phi(\vx_1,t)\hat\phi(\vx_2,t)\r\ra\sim(a H x_{12})^{-\frac{2}{H}\l(\lambda_1+\delta\lambda_1\r)}\;.
\ee 

 We end this subsection by emphasizing one reason for why in our formalism we encounter no more secular divergences from the subleading terms we have added to our equation, for example in (\ref{1pteqsubleadleo}). Indeed, if we were to naively solve~(\ref{1pteqsubleadleo}) perturbatively in these corrections, we would encounter new secular divergences. This would be the same phenomenon that we observe when we solve perturbatively in the mass of a free field. This is also evident from (\ref{eq:subleadingtimedependenceleo}), where the Taylor expansion at late times in $\delta\lambda_1$ would break after $O(H/\delta \lambda_1)$ Hubble times. Instead, what we do in our formalism is to {find the new Eigenvalues and Eigenfunctions and keep the latter in the exponents, similar to what one does in time-independent perturbation theory in Quantum Mechanics. In particular, the corrected Eigenvalues stay positive due to the smallness of $\sqrt{\lambda}$. It is easy to see that the equations for higher-$n$ point functions are amenable to a similar perturbative analysis safe from secular divergences. In particular, at late times all $n$-point distributions are controlled by the same Eigenvalues, guaranteeing control of all the correlation functions at large spacetime separations.   }

\subsection{Estimates of neglected corrections}
\label{sec:subleading-tadopoles}
In the former sections, several contributions were assumed to be small. We now have the formalism  to estimate, and, if need be, to compute them. This is what we are going to do in this section, {thus additionally confirming that further subleading corrections are organized in a hierarchical perturbative manner}.\\

{\bf {$k^3\log k$ terms}:} We start by estimating the size of the contribution to the momentum $\Pi$ that appear in the wave function and that are proportional to $k^3$, see \eqref{eq:wavefunction_schematic}. In fact, due to the presence of additional logarithms, one might be worried that these terms are enhanced  by additional powers of~$1/\sqrt{\lambda}$ with respect to the naive scaling, and so ultimately {could not be computed perturbatively (although the overall size of these corrections is of order $\eps^3$). In section \ref{sec:wavefunctionperturbativity} we already anticipated that this is not the case and now we are going to confirm this anticipation.}

Let us focus on the {quartic-in-fields} term for definiteness { and start with analyzing its contribution to $\Pi$. Since details of the momenta dependence are not important for our power-counting argument, let us assume that there is a contribution of the form $\lambda  k_1^3\log \l(k_1\eta\r)\phi_{\vk_1}\phi_{\vk_2}\phi_{\vk_3}\phi_{-(\vk_1+\vk_2+\vk_3)}$. After Fourier transform,} and neglecting numerical coefficients, it contributes to $\Pi$ schemetically as
\bea
&&\Pi[\phi,\vec x]\supset \frac{\lambda}{H^4 a(t)^3} \phi(\vx)^2 \int d^3{\Delta x}\; \phi(\vec x+\vec{\Delta x}) \int^{\Lambda(t)} d^3k_1 \; e^{i \vec k\cdot \vec{\Delta x}}\, k_1^3 \;\log\l(k_1\eta\r)\\ \nn
&&\qquad\sim  \frac{\lambda}{H^4a(t)^3} \phi(\vx)^2 \int_{1/\Lambda(t)} d^3{\Delta x}\; \phi(\vec x+\vec{\Delta x}) \frac{\log\l(\Delta x\, a(t)\, H \r)}{\Delta x^6} \ ,
\eea
where the second passage holds only for $\Delta x\gtrsim 1/\Lambda(t)$,  as this contribution originates from the wave function expanded in $k\eta\lesssim 1$. 

To see how large is this contribution, let us compute the correlation with $\phi_\ell(\vec x)$ (as the expectation value of $\Pi$ vanishes by symmetry). We obtain, restricting to the long modes inside the momentum,
{ \bea\nn
\label{UVdomination}
&&\langle\phi_\ell(\vx_1)\Pi[\phi_\ell,\vx_1]\rangle\supset \frac{\lambda}{H^4 a(t)^3} \int d\phi_1\,d\phi_2\; \phi_1^3 \phi_2 \int_{1/\Lambda(t)} d^3 \Delta x\;\frac{\log\l(\Delta x\, a(t)\, H \r)}{\Delta x^6}  P_2(\phi_1,t;\phi_2,t;\Delta \vx)\\
&&\qquad\sim \frac{\lambda}{H^4 a(t)^3} \int d\phi_1\,d\phi_2\; \phi_1^3 \phi_2 \int_{1/\Lambda(t)} d^3 \Delta x\;\frac{\log\l(\Delta x\, a(t)\, H \r)}{\Delta x^6}\ \times \\ \nn
&&\qquad\qquad\qquad\qquad\times \ \left(\Phi_0(\phi_1)\Phi_0(\phi_2)+\Phi_1(\phi_1)\Phi_1(\phi_2) \left(\Delta x\,\Lambda(t)\right)^{-\lambda_1/H}+\ldots \right)\\ \nn
&&\qquad\sim \frac{\lambda}{H} \,\epsilon^3 \log(\epsilon)\, \langle\phi_1^4\rangle ,
\eea 
where in the last passage we used that fact that the integral in $\Delta x$ is dominated at short distances, where {$P_2$} is $\Delta x$-independent to leading order and equal to $\delta(\phi_1-\phi_2) P_1^{eq}(\phi_1)$}~(\footnote{{Note that this is one   example of a contribution of the two-location distribution $P_2$ to the evolution equation for $P_1$.}}). With respect to the leading term, {$\langle\lambda \phi_\ell(\vx)^3\phi_\ell(\vx)\rangle\sim1$}, this contribution is suppressed by a factor of $\epsilon^3\log(\epsilon)$. Notice in particular that the fact that the integral in $\Delta x$ is peaked  at the  shortest distances makes the logarithm not enhanced by a factor of $1/\sqrt{\lambda}$: we obtain, instead, $\log\epsilon\ll 1/\sqrt{\lambda}$. The same conclusion trivially holds if we consider other contributions proportional to $k^3$, but with an higher  number of fields. Each successive term adds a factor of {$\lambda\phi_{k_i}\phi_{k_j} \log(k\eta)$ where the log is always protected by the factor of $k^3$ present in the corresponding vertex}, which results in a relative suppression of order $\sqrt{\lambda}\log\epsilon\ll1$ with respect to the former one. 

{ Let us make some general comments on the perturbative structure of our formalism. The convolutions of products of fields with two- and higher-point distributions that appear in our expressions are similar to computing Feynman diagrams with some sort of dressed propagators that at low momenta behave 
 as $\sim k^{-3+2\lambda_n/H}\sim k^{-3+\sqrt{\lambda}}$. Note that if, as it happens for computation of correlation functions from the wave function in usual perturbation theory, some logarithms were not protected by powers of $k$, those logarithms would become IR-dominated and invalidate the perturbative expansion.  However, the diagrammatic structure of our expansion is different and the logarithms {\it do} remain protected by factors of $k$'s, so that the resulting integrals are dominated by the UV where the logarithms are not enhanced. For our purposes, we use perturbation theory{, originating from Taylor expanding $\Psi^*\Psi$ in the powers of short modes around the Gaussian,} only to compute short modes expectation values. Consequently, we might encounter terms of the form $k_\ell^3 \log (k_\ell\eta) /k_s^3\leq k_\ell^3 \log (k_\ell\eta) /\eps^3$,  which still maintain the required property when integrals over $k_\ell$ are taken, similarly to what happens in \eqref{UVdomination}. Of course some integrals over the long momenta are indeed IR dominated but then they do not have log-enhancements and simply reproduce the negative powers of $\lambda$ responsible for the $\phi_\ell(\vx)\sim\lambda^{-1/4}$ behavior of the long field, which is already taken into account in our power-counting.}
 
In case the reader finds the above argument for suppression of the $k^3 \log k$ corrections too much dependent on the subtle properties of the wave function, (s)he might want to study the formalism of Appendix \ref{app:Wigner}, in which the wave function is used only to estimate approximate expectation values of fields and momenta in the equilibrium state, thus perturbativity may be more manifest by construction.
\\

{\bf Effect of $\left[\ldots\right]_{\Lambda}$:} In eq.~(\ref{1ptfull}), the drift term appears in the form $\left[\Pi\l[\phi,\vec x,t\r]\right]_{\Lambda(t)}$, where the parenthesis represent the effect of going to Fourier space and back to real space including only the modes up  to  $\Lambda(t)$. In the main text, we have neglected this effect and simply used directly the field at the same  location, because it was assumed the difference to be subleading. Let  us indeed now check that the correct implementation of this Fourier transform gives a correction of order $\lambda$, and therefore it can be neglected at the order we are computing. For example,  let us focus on the  leading drift term of~(\ref{1ptfull}). For the neglected correction to the leading drift, we have
\bea
&&\Delta{\rm Drift.}\quad\supset\quad \int {\cal D}\phi_\ell \;\delta(\phi_\ell(\vec x_1)-\phi_1)  \left(\left[ \phi_\ell^3(\vec x_1)\right]_{\Lambda}-\phi_\ell^3(\vec x_1)\right) P_\ell[\phi_\ell,t]=\\ \nn
&&\quad=\int {\cal D}\phi_\ell \;\delta(\phi_\ell(\vec x_1)-\phi_1)  \left(\left(\int^\Lambda \frac{d^3k}{(2\pi)^3}\int d^3x_2\; e^{i\vec k\cdot \left(\vec x_1-\vec x_2\right)} \phi_\ell^3(\vec x_2)\right)\right.\\ \nn 
&&\qquad\qquad\qquad\qquad\quad\qquad\left.-\left(\int \frac{d^3k}{(2\pi)^3}\int d^3x_2\; e^{i\vec k\cdot \left(\vec x_1-\vec x_2\right)} \phi_\ell^3(\vec x_2)\right)\right) P_\ell[\phi_\ell,t]=\\ \nn
&&\quad=\int d\phi_2\;   \left(\left(\int^\Lambda \frac{d^3k}{(2\pi)^3}\int d^3x_2\; e^{i\vec k\cdot \left(\vec x_1-\vec x_2\right)} \phi^3_2\right)+\right.\\ \nn 
&&\qquad\qquad\qquad\left.-\left(\int \frac{d^3k}{(2\pi)^3}\int d^3x_2\; e^{i\vec k\cdot \left(\vec x_1-\vec x_2\right)} \phi^3_2\right)\right) P_2(\phi_1,t;\phi_2,t;\vx_{12})=\\ \nn
&&\quad=\int d\phi_2\;   \left(\left(\int^\Lambda \frac{d^3k}{(2\pi)^3}\phi^3_2\, P_2(\phi_1,t;\phi_2,t;\vk)\right)-\left(\int \frac{d^3k}{(2\pi)^3} \phi^3_2\,P_2(\phi_1,t;\phi_2,t;\vk)\right) \right) =\\ \nn
&&\quad=\int d\phi_2\;   \int^\infty_\Lambda \frac{d^3k}{(2\pi)^3}\phi^3_2\, P_2(\phi_1,t;\phi_2,t;\vk)  = \phi_1^3 P_1(\phi_1,t)\,\times \,O(\lambda,e^{-\Delta})\ ,
\eea
where in the second step we used the Fourier-space representation of $\delta^{(3)}(\vec x_1-\vec x_2)$ in writing $\phi_\ell^3(\vec x_1)=\int d^3 x_2\;\delta^{(3)}(\vec x_1-\vec x_2)\;\phi_\ell^3(\vec x_2)$, and where in the last step we used the fact that {$P_2[\phi_1,t;\phi_2,t;\vk]\propto \theta_\Delta(k-\Lambda(t))$ up to terms of order $O(\lambda)$, as derived in Appendix \ref{app:2-point-Fourier}}. We therefore see that the corrections due  to the Fourier transform enter at the $O(\lambda)$ subleading order. Therefore, though we have provided a formalism to compute them, they can be neglected  at the order at which we work in this paper.\\

{\bf Tadpole Diffusion:}  So  far, we  have neglected to include the tadpole term from the diffusion term. The  reason is that, in the case of a smooth  widow function, it gives a correction that is subleading by  a  full factor of $\lambda$, and so negligible for the order at which we work in this paper. In  App.~\ref{app:sharp-diffusion}, we show the tadpole term that enters in the diffusion term is bounded by 
\bea\nn\label{eq;tadpoleborad1main}
&&\langle\Delta\phi\rangle_{\phi_1} P_1^{eq}(\phi_1)\lesssim \lambda^{\frac{5}{4}}P_1^{eq}(\phi_1)+O\l(e^{-\Delta}\r)\ ,
\eea
consequently its contribution to the time evolution operator is at most of order 
\be
\frac{\d}{\d \phi_1} \l(\langle\Delta\phi\rangle_{\phi_1} P_1(\phi_1,t)\r)\sim \lambda^{\frac{3}{2}} P_1(\phi_1,t)\,,
\ee
which means that indeed we could consistently neglect it. In App.~\ref{app:sharp-diffusion} we also show that if a sharp window function is used, {the strong breaking of locality that is induced implies that the tadpole term is enhanced by a factor of $1/\sqrt{\lambda}$ with respect to what we have in (\ref{eq;tadpoleborad1main}), and so it contributes at the first subleading order}.
\\

{{\bf Loop corrections to the momentum:}}
Finally, yet another term that could contribute is the one-loop diagram that corrects the $\lambda\phi^4/\eta^3$ term of the exponent of the wave function, represented on the right of Fig.~\ref{fig:subleadingdiagrams}. This terms could in principle go as $\lambda \phi_\ell^4 (\lambda\log(k\eta))/\eta^3$, and so be suppressed just  by $\sqrt{\lambda}$ with respect to the leading term. As we discussed in section~\ref{sec:wavefunctionperturbativity}, the logarithms must actually be protected by a prefactor of $k^3$, and so this contribution it has either suppressed by $\epsilon^3$ or it has no logarithmic enhancement. Below we verify this explicitly for this specific case. The  diagram reads
\bea\nn\label{eq:diagramquarticloop}
&&\left({\rm Fig.~\ref{fig:subleadingdiagrams},\; right}\right)\equiv9\lambda^2 \int_{-\infty}^{\eta} \frac{d\eta_1}{(-H\eta_1)^4} \int_{-\infty}^{\eta_1} \frac{d\eta_2}{(-H\eta_2)^4} K(\eta_1,k_1)K(\eta_1,k_2)K(\eta_2,k_3)K(\eta_2,k_4)\ \times\\ 
&&\qquad\times\  \int \frac{d^3q}{(2\pi)^3}\; G\left(\eta_1,\eta_2,q\right) G\left(\eta_1,\eta_2,|\vec k_1+\vec k_2-\vec q|\right)\ .
\eea
We are only interested in the limit $k_i\eta\ll1$ when all the fields are long, and the only potentially relevant contribution has a $\log(k\eta)$. Since the momentum integral is IR convergent, the lower limit of integration of the momentum integral can be set to zero, while the upper limit is given by some Lorentz-invariant regulator, which is time dependent in comoving momenta. It is simpler to change variables to `physical' momenta: from $q\to q_{\rm ph}=q \, a(\eta_2)$. At this point, the boundaries of the momentum integral are time independent and so one can perform the time integrals first. In the limit of $q_{\rm ph}/H\ll1$, the  integrand takes a simple form
\be
\propto 9\lambda^2 \int d^3 q_{\rm ph}\int^{\eta}d\eta_1 \int^{\eta_1}d\eta_2 \; \frac{\left(\eta_1^3-\eta^3\right)^2}{\eta_1^6\eta_2^7}\sim \lambda^2\int d^3 q_{\rm ph} \frac{1}{\eta^3}\ ,
\ee 
and the result of the integration has no logarithmic enhancement. For generic $q_{\rm ph} /H$, one can change variables to $\tilde\eta_2=\eta_2/\eta_1$ and $\tilde\eta_1=\eta_1/\eta$. Now the limits of integration for $\eta_1$ and $\eta_2$ are equal to one at late times. In the limit $k_i\eta_j\to0$, the integrand is just a factor of $1/\eta^3$ times a function that is $k_i$ and $\eta$  independent. The integrals in $\tilde\eta_1$ and $\tilde\eta_2$ are convergent, and so we conclude that there is no logarithmic factor in this limit. If we expand in $k_i\eta_j\ll1$, and extract three factors of $k_i$ from the propagators, the integrand in  $\tilde\eta_1$ gets additional factors of $\tilde \eta_1$, and becomes logarithmically divergent at early times, up to the time $1/(k\eta)$ where the propagator cuts it off, giving rise to a $k_i^3\log({ k_i}\eta)$ contribution. We therefore confirm that this diagram contributes only  at subleading order {$O(\eps^3)$ or $O(\lambda)$} and so is negligible.

\section{Generic 2-point function and dS-invariance  \label{sec:generic-2-point}}
We now proceed to the calculation of general two-point functions for which the two points are neither at the same spatial location nor at equal time. The latter two cases were treated in section \ref{sec:two-point}. The goal of this section is to demonstrate a method suitable for the computation { in this kinematical regime,} and also to check that the two-point function of the field is de Sitter invariant, that is it depends only on the interval
\be
z^2=\frac{2}{H^2}\l(\cosh H(t_1-t_2)-1\r)-e^{H(t_1+t_2)}(\vx_1-\vx_2)^2\,.
\ee
This will be a confirmation that de Sitter invariance is preserved by the BD vacuum.

We will thus focus on the kinematical regime where this check is the most non-trivial for as low order in perturbation theory as possible. Note that, since in our approach space and time are treated very asymmetrically, the fact that the leading order two-locations and two-times distributions at large separations are controlled by the same exponent is already an indication of dS invariance \cite{Starobinsky:1994bd}, although { it is a very weak one, as this condition is not even, a priori,  a necessary one}. Let us first identify this { simplest but non-trivial} kinematical regime. For convenience let us fix $x_{12}=|\vx_1-\vx_2|$. Then $t_1$ and $t_2$ can belong to one of the three regions defined by equation \eqref{tif}, see figure~\ref{fig:Regions}. Without loss of generality we pick $t_2>t_1$. We also want to focus on the regime where long modes dominate. In this case it is the simplest to go to large separations at time $t_2$, so that the leading Eigenvalue, $\lambda_1$, dominates the connected correlator. This requires $t_2$ to belong to the region III. We thus expect the correlators to behave as $(z^2)^{-\lambda_1/H}\approx e^{-(t_2-t_1) \lambda_1}\l(1-e^{2 H t_1}x_{12}^2\r)^{-\lambda_1/H}$. In order for the nontrivial dependence on both $x_{12}$ and $t_2-t_1$ to be visible at subleading order in $\sqrt{\lambda}$, we should pick $e^{2 H t_1}x_{12}^2\sim1$. This requirement fixes $t_1$ to belong to the region I. Then we get
\be
\label{xdependence}
(z^2)^{-\lambda_1/H}\approx  e^{-(t_2-t_1) \lambda_1}\l(1-\lambda_1H^{-1} \log\l(1-e^{2 H t_1}x_{12}^2\r)+O(\lambda)\r).
\ee
Our goal in the rest of this section will be to reproduce the second, $O\l(\sqrt{\lambda}\r)$, term in this expression.
\begin{figure}[htbp] 
	\centering
		\includegraphics[width=.4\linewidth,angle=270]{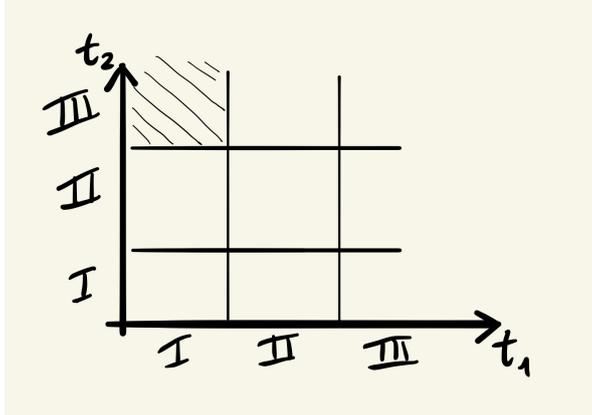}
	\caption{Kinematic regimes for a generic two-point function}
	\label{fig:Regions}
\end{figure}

Let us now outline the method we are going to use for this computation. We focus on the two-point function $\la\hat\phi(\vx_1,t_1) \hat\phi(\vx_2,t_2)\ra$ in the BD state. There are contributions from fields that are long at both locations, but also contributions from fields that are short at time $t_1$ but become long at time $t_2$, { and finally, at least in principle, also from fields that are short at both locations}. 

As far as the contribution of long modes is concerned, up to corrections of order $\epsilon^3$ we can treat them classically and consequently their probability distribution satisfies the equations presented in section \ref{sec:perturbatie-structure}. Quantum corrections will be discussed in section~\ref{sec:quantum}.
To compute the long-long component we will need the generic two-point equilibrium distribution. We are going to first use an auxiliary three-point distribution, $P^{eq}_3(\phi_1, t_1, \vx_1; \phi_2', t_1, \vx_2;\phi_2, t_2, \vx_2)$, and integrate it over  $\phi_2'$ to get the distribution\\$P^{eq}_2(\phi_1, t_1, \vx_1;\phi_2, t_2, \vx_2)$ that we are interested in, see figure~\ref{fig:4pt} for illustration. This allows us to use exactly the same logic as in section~\ref{sec:two-point}. Namely, $P^{eq}_3$ satisfies the following equation at leading order:
\bea
\label{eq:3pt}
&&\frac{\partial}{\d t_2}P^{eq}_3(\phi_1, t_1, \vx_1; \phi_2', t_1, \vx_2;\phi_2, t_2, \vx_2)=\Gamma_{\phi_2} P^{eq}_3(\phi_1, t_1, \vx_1; \phi_2', t_1, \vx_2;\phi_2, t_2, \vx_2)\ , 
\eea
and the following initial conditions
\be
\label{ic3pt}
P^{eq}_3(\phi_1, t_1, \vx_1; \phi_2', t_1, \vx_2;\phi_2, t_1, \vx_2)=\delta(\phi_2-\phi_2') P_2^{eq}(\phi_1,t_1;\phi_2',t_1;x_{12})\,.
\ee

We will also need to compute the long-short component. This is present due to the modes that are short at time $t_1$ and that become long at some time earlier than $t_2$. In the kinematic regime we specified the short-short contribution can be neglected, as they contribute as $(\epsilon a(t_2) x)^{-1}$, whose relative contribution is smaller than $e^{-\Delta}$ and therefore, as discussed in section~\ref{sec:twolocations}, smaller than any power of $\sqrt{\lambda}$.
\begin{figure}[htbp] 
	\centering
		\includegraphics[width=.6\linewidth,angle=0]{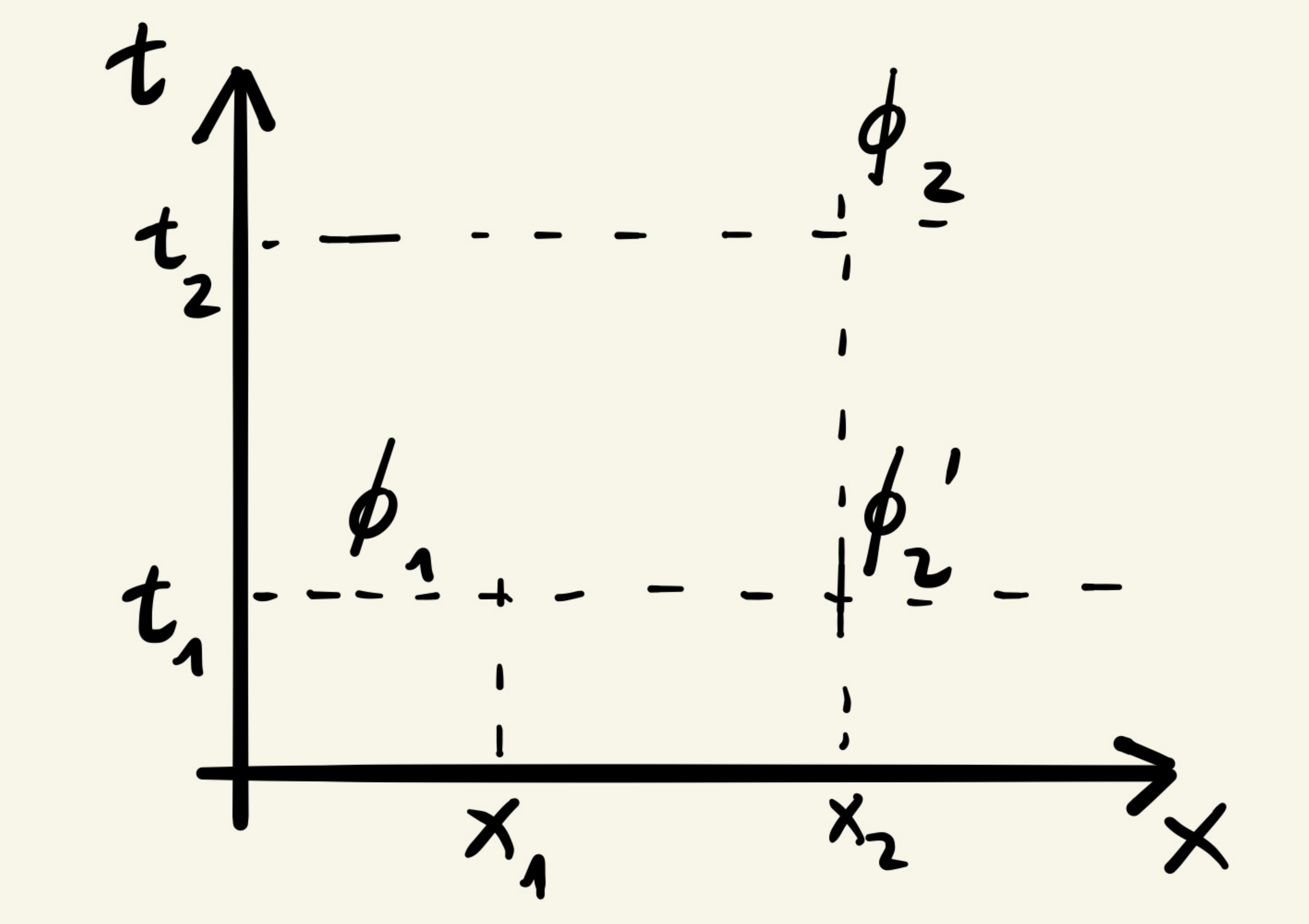}
	\caption{Locations of the auxiliary three-point distribution.}
	\label{fig:4pt}
\end{figure}

We start with the long-long contribution. Since $t_1$ is in the region I, $P_2^{eq}(\phi_1,t_1;\phi_2',t_1;x_{12})=\delta(\phi_1-\phi_2') P_1^{eq}(\phi_1)$. We can now decompose the dependence on $\phi_2$ of the initial conditions in equation \eqref{ic3pt} into Eigenstates of $\Gamma_{\phi_2}$:
\bea\nn
\delta(\phi_2-\phi_2')\delta(\phi_1-\phi_2') P_1^{eq}(\phi_1)=\delta(\phi_2-\phi_1)\delta(\phi_1-\phi_2')P_1^{eq}(\phi_1)=\sum_n\Phi_n(\phi_2)\Phi_n(\phi_1)\delta(\phi_1-\phi_2')\,, \\ 
\eea
where we used the expression for the $\delta$-function given in \eqref{completeness} as well as the fact that $P_1^{eq}(\phi)=\mu^{-1}(\phi)$. In this form it is manifest that the time evolution is the same as for  
the one location two-times distribution, and we get
\bea
\label{long-long-pt}
&&P^{eq}_2(\phi_1, t_1, \vx_1;\phi_2, t_2, \vx_2)=\int d \phi_2' \;P^{eq}_3(\phi_1, t_1, \vx_1; \phi_2', t_1, \vx_2;\phi_2, t_2, \vx_2)  =\\ \nn
&&\int d \phi_2' \sum_n\Phi_n(\phi_2)\Phi_n(\phi_1)\delta(\phi_1-\phi_2')e^{-\lambda_n(t_2-t_1)}=\sum_n\Phi_n(\phi_2)\Phi_n(\phi_1) e^{-\lambda_n(t_2-t_1)}\,.
\eea
As we see, this result is independent of $x_{12}$. 
We thus switch to the short-long contribution that is supposed to introduce the correct $x_{12}$ dependence anticipated in \eqref{xdependence} from de Sitter invariance.

The logic of this part of the calculation is the following: we decompose the short field operator at time $t_1$ into Fourier modes. For each mode, $ \phi({\vk})$, there is a moment of time, that we call $t_k$, when the mode passes from being `short' to `long', that is 
\be\label{eq:tkdef}
k=\Lambda(t_k)= \eps  a(t_k) H\,.
\ee
If $t_k<t_2$, this long mode is correlated with the long field at time $t_2$, as it is part of the Fourier modes that constitute it. We need therefore to include the contribution to the correlation from all these modes. This can be computed using the Fourier transform of the long-modes two-point distribution. Let us implement these steps. As we said, the short-long contribution will receive contribution only from the short modes that become long by the time $t_2$:
\be
\label{short-long}
\langle  \phi_s(\vx_1,t_1)\phi_\ell(\vx_2,t_2)\rangle=\int_{\Lambda(t_1)}^{\Lambda(t_2)} \frac{d^3 k}{(2\pi)^3}\,e^{i \vk\cdot\vx_1}\langle  \phi(\vk,t_1)\phi_\ell(\vx_2,t_2)\rangle\,,
\ee
where, as usual, the limits of integration are imposed on the absolute value of the momentum. We now relate the short mode at $t_1$ to the corresponding mode at $t_k$ using the free field evolution, neglecting perturbative-in-$\lambda$ corrections, since we are interested in the leading-order result stemming from this term. Using the bulk-to-boundary propagator $K$, we get:
\be
\label{modeevol}
 \phi(\vk,t_1)=\frac{e^{\frac{i k}{a(t_1) H}}\l(i+\frac{k}{a(t_1) H}\r)}{e^{\frac{i k}{a(t_k) H}}\l(i+\frac{k}{a(t_k) H}\r)}\, \phi(\vk,t_k)={e^{\frac{i k}{a(t_1) H}}\l(1-i\frac{k}{a(t_1) H}\r)} \phi(\vk,t_k)+O(\eps)\,.
\ee
On the other hand, treating $\phi(\vk,t_k)$ as a Fourier transform of the long field at time $t_k$, we get
\bea
\label{phik}
\langle \phi(\vk,t_k) \phi_\ell(\vx_2,t_2)\rangle=\int d\phi_kd\phi_2\int d^3 x_k \,e^{-i \vk \cdot \vx_k}\,\phi_k\phi_2\, P_2(\phi_k,\vx_k,t_k;\phi_2,\vx_2,t_2)\;.
\eea
Naturally, this Fourier transform will be dominated by $|\vx_k-\vx_2|\sim 1/\Lambda(t_k)$. We thus need to compute the two-point distribution in the kinematic region II-III of figure~\ref{fig:Regions}, where we replace $t_1$ with $t_k$ on the $x$-axis. The computation is similar to the one we just did for the long-long contribution. Again we introduce an auxiliary three-point distribution, $P^{eq}_3(\phi_k, t_k, \vx_k; \phi_2', t_k, \vx_2;\phi_2,\vx_2, t_2)$, however now at the initial time $t_2=t_k$ we match it to the two-locations distribution in region $(II)$ given in \eqref{2ptfinal}. We get
\bea
 &&P_2^{eq}(\phi_k,\vx_k,t_k;\phi_2,\vx_2,t_k)=\int d\phi_2'\; P^{eq}_3(\phi_k, t_k, \vx_k; \phi_2', t_k, \vx_2;\phi_2,\vx_2, t_k)=\\
 &&=\int  d\phi_2' \sum_n\Phi_n(\phi_k)\Phi_n(\phi_2')A_{nn}(t_k,x_{k2})\delta(\phi_2-\phi_2')=\sum_n\Phi_n(\phi_k)\Phi_n(\phi_2)A_{nn}(t_k,x_{k2})\,, \nn
\eea
where $A_{nn}$ is given in the second line \eqref{Annx}. 
We now need to time evolve it up to the time $t_2$ with $\Gamma_{\phi_2}$, which leads to 
\be\
\label{2ptphik}
P_2^{eq}(\phi_k,\vx_k,t_k;\phi_2,\vx_2,t_2)=\sum_n A_{nn}(t_k,x_{k2}) e^{-\lambda_n(t_2-t_k)}\Phi_n(\phi_k)\Phi_n(\phi_2)\;.
\ee
Consequently, the $x$-dependence is exactly the same as in the two-location distribution. 
Substituting \eqref{modeevol}, \eqref{phik} and \eqref{2ptphik} into \eqref{short-long} we finally get 
\bea
\langle\phi_s(\vx_1,t_1)\phi_\ell(\vx_2,t_2)\rangle=C_1^2\int_{\Lambda(t_1)}^{\Lambda(t_2)} \frac{d^3 k}{(2\pi)^3}\,e^{i \vk\cdot\vx_{12}}{e^{\frac{i k}{a(t_1) H}}\l(1-i\frac{k}{a(t_1) H}\r)}\tilde A_{11}(t_k,k) e^{-\lambda_1(t_2-t_k)}\nn\,,\\
\eea
where we kept only the contribution from the leading Eigenvalue and $C_1$ is given below~\eqref{2ptfeqtime}. This expression involves the Fourier transform of $A_{nn}$, which is given in \eqref{Annk}.
 If we look back at \eqref{Annk}, we notice that it contains the {smeared} $\theta$-function with the step exactly at the position where we are evaluating it. This is not surprising because we are studying the mode that just became long. By physical reasoning we set $\theta_\Delta(0)=1$ in this case~\footnote{More in detail, we can evolve the short modes to a time {$(1+\varepsilon) t_k$, taking $1\gg\varepsilon\gg e^{-\Delta}$, which is slightly after it became long. This is ok because perturbation theory does not break down suddenly. This justifies replacing $\theta_\Delta(0)$ with 1 to a very high precision. Alternatively, one can directly use the formulas in App.~\ref{app:2-point-Fourier} and do a more careful matching between long and short modes, which leads to the same answer for the correlator.}}. The angular integration is straightforward and gives 
\bea\label{eq:longshortfinal}
\langle\phi_s(\vx_1,t_1)\phi_\ell(\vx_2,t_2)\rangle=\lambda_1C_1^2 e^{-t_2 \lambda_1}\int_{\Lambda(t_1)}^{\Lambda(t_2)} \frac{d k}{H k}\, \frac{2 \sin k x_{12}}{k x_{12}} e^{\frac{i k}{a(t_1) H}}\l(1-i\frac{k}{a(t_1) H}\r)\l(\frac{k}{H}\r)^{\lambda_1/H}\nn\,,\\
\eea
where we substituted $t_k$ in terms on $k$ using~(\ref{eq:tkdef}) and dropped terms of order $ \sqrt{\lambda}\log\eps$ as they are higher order. This integral is bounded in the UV and from the upper limit of integration we get a contribution $\sim(\eps a(t_2)x_{12})^{-1}$, which is much smaller than any power of $\sqrt{\lambda}$ according to our choice of kinematics, as discussed earlier for the short-short contribution. From the IR limit of integration to leading order in $\lambda$ we get what is the leading contribution
\be
\langle\phi_s(\vx_1,t_1)\phi_\ell(\vx_2,t_2)\rangle=-\lambda_1H^{-1}C_1^2 e^{-(t_2-t_1) \lambda_1 }\l(\log\l(-1+a^2(t_1)x_{12}^2\r)+2(-1+\gamma_E+\log\eps)\r)\,,
\ee
where $\gamma_E$ is the Euler constant and we assumed that $a^2(t_1)x_{12}^2\sim 1$ when doing the $\lambda$ expansion. The $x$-dependence is exactly the one that we anticipated, while the constant piece is unimportant for us since it can be absorbed into the many-other $O(\sqrt{\lambda})$ corrections to the $x$-independent part of the two-point function that we do not compute~\footnote{For example, they include the ones from the corrections to the Eigenfunctions.}.  Combining this result with the leading long-long contribution obtained from \eqref{long-long-pt} we get
\bea\label{eq:generictwospacetimes}
&&\langle\phi(\vx_1,t_1)\phi(\vx_2,t_2)\rangle=C_1^2 e^{-(t_2-t_1) \lambda_1}\l(1-\lambda_1H^{-1}\log\l(-1+a^2(t_1)x_{12}^2\r)\r)\times\nn\\
&&\qquad\qquad\times \l(1+O\l(\sqrt{\lambda},e^{-(t_2-t_1)(\lambda_2-\lambda_1)}\r)\r)\,,
\eea
where the $O(\sqrt{\lambda})$ corrections do not involve coordinate dependence. This result is consistent with \eqref{xdependence} and hence de Sitter invariance.

\section{Implications of thermality in the static patch\label{sec:thermality}}
\begin{figure}[htbp] 
	\centering
		\includegraphics[width=.4\linewidth,angle=270]{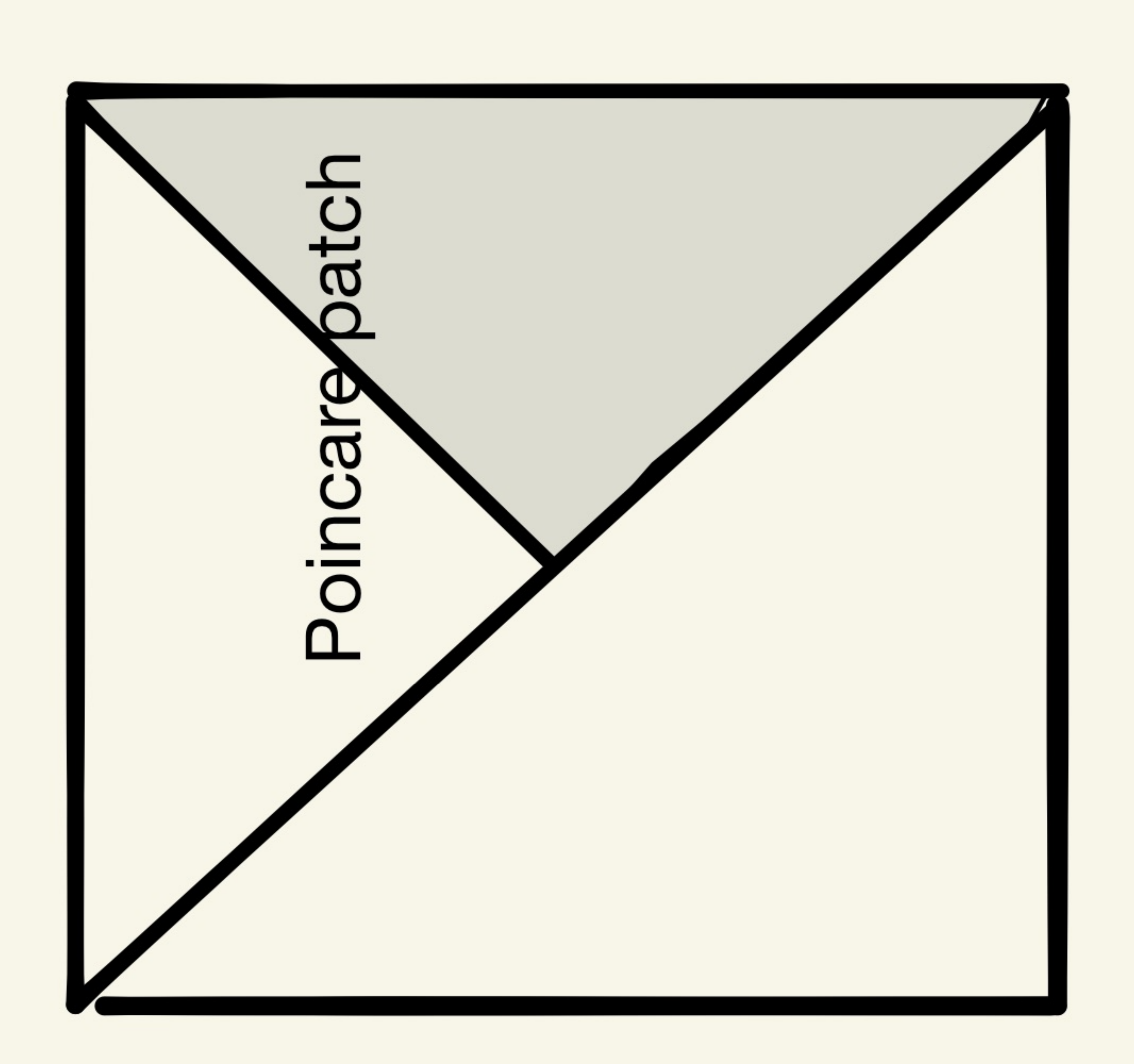}
	\caption{{Penrose diagram exhibiting global { {\it (full square)}}, Poincare { {\it (upper-right triangle)}}, and static {\it (darkened triangle) }patches of dS space.}}
	\label{fig:Static}
\end{figure}
It is often said that the equilibrium state of a quantum field theory in de Sitter space should be thermal. As applied to the full spacial slice in the FRW coordinates we consider here this statement cannot be literally correct. Indeed, the state that we are studying is a pure state and is described by a wave function which we succeeded to find order by order in perturbation theory. Instead, a more precise statement is that physical observables localized within a single static patch can be described with the help of a thermal density matrix \cite{Gibbons:1977mu}~(see Fig.~\ref{fig:Static}). A simple way to understand this is to note that the static patch of de Sitter space, upon the Wick rotation, becomes a sphere with a periodic Euclidean time coordinate which signals the thermal behavior.\footnote{The global de Sitter space also becomes a sphere upon the Wick rotation, however, Euclidean time in this case is not periodic as it goes from one pole of the sphere to the other.} In this regard, global and static patches of de Sitter space are similar to the full Minkowski space and its Rindler patch, as nicely explained in \cite{Goheer:2002vf}. Thermality of de Sitter space is the basis for its most fundamental properties. In particular, a gravitational theory in de Sitter space is supposed to have a finite entropy. One of the more basic consequences of the existence of the thermal description, visible even in the $M_{\rm pl}\to\infty$ limit that we study, is that the correlation functions restricted to the static patch should obey the KMS condition~\cite{Kubo:1957mj, Martin:1959jp}. This condition is a certain form of periodicity in the imaginary time, which for the two-point function of fields in flat space takes the following form:
\be
\l \la\hat \phi(\vx_1,t_1)\hat\phi(\vx_2,t_2+i\beta)\r\ra=\l \la\hat \phi(\vx_1,t_1)\hat\phi(\vx_2,t_2)\r\ra^{\dagger}\,,
\ee
where $\beta$ is the inverse temperature. De Sitter space is ``thermal'' with temperature given by 
\be
\beta=2\pi H^{-1}\;.
\ee
Our approach allows us to compute correlation functions inside the static patch in particular, and, consequently, we expect the KMS condition to hold. Let us check that it is indeed the case. For general two-point functions in de Sitter, periodicity is in the direction of the time-like Killing vector field, which exists everywhere { within a single} static patch. {As usual, we work in FRW slicing. }For simplicity, we set $\vx_1~=~\vx_2~=~0$ . Then arbitrary $t_1$ and $t_2$ belong to the static patch of an observer located at $\vx=0$, and the KMS condition takes the same form as in flat space, since the killing vector is just translation in time. This static patch is also fully localized within the expanding patch of de Sitter that we consider. Next, we take the limit of large $t_2-t_1$, so that the correlators of long modes are dominated by the leading Eigenvalue. This simplifies the calculation and also makes the interesting non-perturbative dynamics most manifest. The calculation is similar to the one of section \ref{sec:generic-2-point}, but there are several additional subtleties. The long-long contribution, as before, reads 
\be
\langle\hat\phi_\ell(0,t_1)\hat\phi_\ell(0,t_2)\rangle=C_1^2 e^{-(t_2-t_1) \lambda_1}\,,
\ee
{ where, as usual, we focused on the leading Eigenvalue.}
This expression on its own violates the KMS conditions by terms of order $O(\lambda_1)$. { But indeed we still} need to add the short-long contribution. This is, analogously to \eqref{eq:longshortfinal},
\bea
\label{lsKMS}
\langle\hat\phi_s(0,t_1)\hat\phi_\ell(0,t_2)\rangle=\sum_n\lambda_nC_{1,n}^2 e^{-t_2 { \lambda_n}}\int_{\Lambda(t_1)}^{\Lambda(t_2)} \frac{d k}{H k}\, 2 e^{\frac{i k}{a(t_1) H}}\l(1-i\frac{k}{a(t_1) H}\r)\l(\frac{k}{H}\r)^{{ \lambda_n}/H}\nn\,,\\
\eea
where $C_{1,n}=\int d \phi\, \phi\, \Phi_n(\phi)$.
The lower limit of integration gives
\be
\label{lsKMS2}
H^{-1}\sum_n \lambda_n C_{1,n}^2 e^{-(t_2-t_1) \lambda_n}\l(i\pi+2(1-\gamma_E-\log\eps)\r)\,,
\ee
where only the imaginary part is important for us. What is different to the situation in section~\ref{sec:generic-2-point}, is that now the upper limit of integration gives an unsuppressed contribution:
\be
I(\Lambda_2)=-2 H^{-1}\sum_n \lambda_n C_{1,n}^2 e^{i \frac{\eps a(t_2)}{a(t_1)}}\;.
\ee
This contribution is due to the modes that become long right before the moment of time~$t_2$ and, consequently, we expect it to cancel against the short-short contribution. Let us check that this is indeed the case. First we massage $I(\Lambda_2)$:
\bea
&&I(\Lambda_2)=- 2 H^{-1} e^{i\frac{\eps a(t_2)}{a(t_1)}} \sum_n \lambda_n \int d\phi\, d\phi'\, \phi\phi' \Phi_n(\phi)\Phi_n(\phi')=\nn\\
&&\quad= 2H^{-1}e^{i\frac{\eps a(t_2)}{a(t_1)}} \sum_n \int d\phi\, d\phi' \phi\phi' \Gamma_\phi \Phi_n(\phi)\Phi_n(\phi')=\nn\\
&&\quad=2H^{-1}e^{i\frac{\eps a(t_2)}{a(t_1)}}  \int d\phi\, d\phi' \phi\phi' \Gamma_\phi \delta(\phi-\phi') P_1^{eq}(\phi')=2H^{-1}e^{i\frac{\eps a(t_2)}{a(t_1)}}  \int d\phi\, \phi \Gamma_\phi \l(\phi  P_1^{eq}(\phi)\r)=\nn\\
&&\quad= 2 H^{-1}e^{i\frac{\eps a(t_2)}{a(t_1)}}  \int d\phi\, \phi \frac{H^3}{8\pi^2} \frac{\d}{\d\phi} N e^{-{\frac{8 \pi^2 V(\phi_1)}{3 H^4}}}=- e^{i\frac{\eps a(t_2)}{a(t_1)}}  \frac{H^2}{4\pi^2}\;,
\eea
where we used \eqref{1pteqsol}, \eqref{Gammaphi} and some properties of the Eigenfunctions.

Now we switch to the contribution of the short modes, which reads~\footnote{Short modes are deep inside horizon at time $t_1$: since we assumed $e^{-\sqrt{\lambda}{H(t_2-t_1)}}\ll1$, it follows that $H(t_2-t_1) \gg 1/\sqrt{\lambda}\gg |\log \eps|$, consequently, $k>\eps a(t_2) H$ implies $k\gg a(t_1) H$. We neglect subleading corrections in $a(t_1)/(\epsilon a(t_2))\ll1$ as in the rest of this section.}
\bea
&&\int_{\Lambda(t_2)} \frac{d^3k}{(2\pi)^3}\frac{H^2}{2 k^3} e^{\frac{i k}{a(t_1)H}}\l(- i  \frac{k}{a(t_1) H}\r)\l(1+ i  \frac{k}{a(t_2) H}\r)e^{-\frac{i k}{a(t_2)H}}=\\ \nn
&&\qquad\qquad\qquad= e^{i\frac{\eps a(t_2)}{a(t_1)}}  \frac{H^2}{4\pi^2}\left(1+O\l(\frac{a(t_1)}{\epsilon\, a(t_2)},\epsilon\r)\right)\;.
\eea
The cancellation occurs as expected. {Now, at large time separations, we can keep just the leading Eigenvalue contribution in \eqref{lsKMS} { (or rather directly in \eqref{lsKMS2})} and check explicitly that our two-point function indeed} satisfies the KMS condition, at least at the order that we computed:
\bea
&&\langle\hat\phi (0,t_1)\hat\phi(0,t_2+2\pi i H^{-1})\rangle=C_1^2 e^{-(t_2 +2\pi i H^{-1}-t_1) \lambda_1}(1+\pi i \lambda_1 H^{-1})+\ldots=\nn\\
&&=C_1^2 e^{-(t_2 -t_1) \lambda_1}(1-\pi i \lambda_1 H^{-1})+\ldots=\langle\hat\phi (0,t_1)\hat\phi(0,t_2)\rangle^\dagger\,,
\eea
where dots stand for the subleading terms and irrelevant real pieces. 

The KMS condition is obviously not the only consequence of thermality. In particular, the Eigenvalues of the operator $\Gamma_\phi$ should be related to the equilibration time of the thermal system corresponding to the static patch. We leave further exploration of these relations for future work.

\section{large-$N$\label{sec:largeN}}

All the techniques developed in this paper can be trivially generalized to an arbitrary number of fields. A particularly useful generalization is the $O(N)$ symmetric model with a large number of fields $N$. In this section we consider the corresponding model. One advantage of the large-$N$ approximation is that the Eigenvalue problem which determines the behavior of correlation functions can be solved analytically by semi-classical methods. Namely, we consider the potential
\be
V(\phi_i)=\frac{\lambda}{4 N}(\phi_i\phi_i)^2+\frac{m^2}{2}\phi_i\phi_i\, \qquad i=1\ldots N\;,
\ee
where we chose the usual large-$N$ scaling of the coupling constant to ensure a proper $1/N$ expansion. The prime purpose of this section is to enable a comparison with other approaches to perturbation theory in dS space which might become available in the future, see e.g.~\cite{Nacir:2016fzi,LopezNacir:2016gfj,LopezNacir:2018xto}  for existing results. Since large-$N$ calculations are often done in space-time dimensions other than 4, to facilitate the comparison with the other approaches we also work in an arbitrary number of dimensions. We will denote the number of spacial dimensions by $d$. In the remainder of this section, we work at leading order in $1/N$ and at subleading order in $\sqrt{\lambda}$; in this section we also set $H=1$.

Let us focus on the $N$-field analog of the subleading equation for the single-point distribution, \eqref{1pteqsublead}. Needless to say, the derivation proceeds in exactly the same way as in the single-field case so we will not repeat the individual steps here.
The whole equation for the time-evolution of the one-point distribution is given by
\bea
\label{1pteqsubleadN}
&& \frac{\d}{\d t} P_1(\vec\phi_1,t)=\Gamma^{(N)} P_1(\vec\phi_1,t)=\frac{\d}{\d\phi_{1i}}\l(\frac{1}{d}\frac{\d}{\d \phi_{1i}}V_{eff}\l(\vec\phi_1\r) P_1(\vec\phi_1,t)\r)+\frac{\d^2}{\d\phi_{1i}^2}\l(D P_1(\vec\phi_{1},t)\r)+\nn\\
&&\qquad\qquad\qquad\qquad\qquad\qquad\qquad\qquad\qquad\qquad\qquad\qquad+\,O(\lambda^\frac{3}{2})\cdot P_1(\vec\phi_1,t)=0\;,
\eea
here $\vec\phi_1$ is a collection of $N$ long fields at some space-time point and
\bea
&& D=\frac{\Gamma\l(\frac{d+2}{2}\r)}{2\, d \pi^{\frac{d+2}{2}}}, \qquad V_{\rm{eff}}\l(\vec \phi\r)=\frac{\lambda}{4 N}(\phi_i\phi_i)^2+\frac{\bar m^2}{2}\phi_i\phi_i+\frac{\lambda^2}{ 2 d^2 N^2}(\phi_i\phi_i)^3\,,
\eea
and $\bar m^2\ll\sqrt{\lambda}$ is the effective physical mass squared of the long modes.\footnote{As before we chose $\bar m^2\ll\sqrt{\lambda}$ to simplify our power-counting. It is straightforward to repeat all the calculations for an arbitrary value of the mass, and moreover for an arbitrary weakly coupled large-$N$ potential.}
We dropped the $\log\eps$ terms in this equation since they will cancel in all physical observables analogously to the single-field case for which we checked it explicitly.

The operator $\Gamma^{(N)}$ controls the equilibrium one-point functions, and its Eigenvalues determine the late-time and large-distance behavior of the two-point functions. The next step is, by means of variable redefinitions analogous to those done in section \ref{sec:Eigenvalues},  to bring this operator  to the canonical form. The corresponding canonical operator (with the same Eigenvalues) reads:
\bea
&&\Gamma^{(N)}_c=-2 D \l(-\frac{1}{2 }\frac{\d^2}{\d\phi_{1 i}\d\phi_{1 i}}+N W\l(\frac{\phi_{1i}\phi_{1i}}{N}\r)\r)\,,\\ \nn
&&W(\rho)=\frac{\lambda^2 \rho^3}{8 d^2 D^2}-\frac{\lambda \rho}{4 d D}-\frac{\bar m^2}{4 d D}+\frac{\lambda \bar m^2}{4 d^2 D^2}\rho^2+\frac{3 \lambda^3 \rho^4}{4 d^4 D^2}\;.
\eea
Now the Eigenvalues can be read off using the same semiclassical techniques as used in large-$N$ quantum mechanics \cite{Yaffe:1981vf}, section IV~(\footnote{The same expressions can be obtained by treating the problem as a 1-dimensional large-$N$ QFT and using more traditional methods, described, for example, in \cite{Moshe:2003xn}.}). 
First, one determines the vev of $\rho$ by solving the equation
\be
8\rho_0^2 W'(\rho_0)=1\,,
\ee
which gives
\be
\rho_0=\l(\frac{{d D}}{{\lambda}}\r)^{1/2}-\frac{3D}{ d}-\frac{\bar m^2}{2 \lambda}{+O(\sqrt{\lambda})}\;.
\ee
Naturally, the characteristic value of $\rho$ is of order $\lambda^{-1/2}$: the same power of $\lambda$ as we get for $\phi_\ell^2$ in the single-field case. The leading part has been computed in various different ways, see e.g. \cite{Beneke:2012kn,Nacir:2016fzi}.
For the ground state and the smallest Eigenvalues in the vector and singlet channels we get, correspondingly,
\bea
&&\lambda_0=2D N\l(\rho_0 W'(\rho_0)+W(\rho_0)\r)=0\,,\\
&&\lambda_v=\frac{D}{\rho_0}=\l(\frac{D \lambda}{d}\r)^{1/2}+\frac{6D \lambda+d \bar m^2}{2 d^2}\,,\\
&&\lambda_s=2 D\l(8 W'(\rho_0)+4\rho_0W''(\rho_0)\r)^{1/2}=4 \l(\frac{D \lambda}{d}\r)^{1/2}+\frac{12 D\lambda}{d^2}\,.
\eea
The "ground state" energy calculation is merely a cross-check of our equations. It is supposed to be zero to all orders since it corresponds to the equilibrium distribution. The smallest vector-channel Eigenvalue controls the large-distance behavior of the two point function {$\langle\phi_i(x)\phi_j(y)\rangle\sim \delta_{ij}z^{-2 \lambda_v}$}, where $z$ is the dS-invariant distance between $x$ and $y$. Its leading piece matches the result of \cite{Nacir:2016fzi,LopezNacir:2016gfj,LopezNacir:2018xto} who computed it to leading order in both $1/N$ and $\sqrt{\lambda}$. The subleading in $\sqrt{\lambda}$ piece is scheme-dependent, in a sense that it depends on the renormalized value of the mass. {Consequently, when comparing with other methods, one should exercise caution and preferably use the same scheme. The same applies to 
the expectation value $\langle\phi_i(x)^2\rangle$. It is important to remember that, at subleading order, the short modes contribute and hence this one-point function is given by $N \rho_0+\langle\phi_{s,i}(x)^2\rangle$, see section~\ref{sec:subleading1point} for a discussion in the single-field case. The short modes contribution to the one-point function is also divergent.}

Interestingly, the smallest singlet Eigenvalue, $\lambda_s$ is UV-insensitive even at subleading order. It contributes to the large-distance behavior of correlators like $\langle\phi_i(x)\phi_i(x)\phi_j(y)\phi_j(y)\rangle$. Thus it provides a very concrete and robust consequence of the subleading terms in our equation. It would be very interesting to compute the corresponding correlator with some other techniques and compare the results, see \cite{ONdS}.

Often, large-$N$ calculations can be performed for a finite value of the coupling $\lambda$ and using only the $1/N$ expansion. It seems that this can also be done in our approach, since all what is needed is to have some perturbative control over the short modes for which the large-$N$ expansion should suffice.

\section{Outline of $\hbar$ corrections\label{sec:quantum}}
As far as the single-time probability distributions and correlation functions are concerned, our treatment has been fully quantum mechanical. However, as stressed in sections~\ref{sec:perturbatie-structure} and~\ref{sec:twotimes}, we used an intuitive classical approximation for computing distributions of long fields involving multiple times. In this section we will show how to implement a complete quantum-mechanical treatment of these objects and justify more rigorously the classical approximation. Let us focus on the time-ordered two-point function of some operators ${\cal O}_1(t_1)$ and ${\cal O}_2(t)$ for $t_1<t$~(\footnote{We will assume that $\op$'s are made only of long modes, as it was already shown how to treat short modes and short modes becoming long at some time between $t_1$ and $t$ in section \ref{sec:generic-2-point}.}). It is useful to start with the path-integral representation for the wave functions and the correlator given by
\bea
\label{Pquant}
&&\l \langle { \cal O}_1(t_1) { \cal{O}}_2(t)\r\rangle=\int{\cal D} \phi\; \Psi^*_{\op_1}[ \phi,t;t_1] \Psi[\phi,t]\; {\cal O}_2(t)\,,\\\nn
&& \Psi^*_{\op_1}[\phi,t;t_1]= \int^{\phi,t} {\cal D} \{\phi'\}\;e^{-i S}\; {\cal O}_1(t_1)\ ,\qquad  \Psi[\phi,t]=\int^{\phi,t}{\cal D} \{\phi \}\; e^{i S}\ .
\eea
Here the notation ${ \cal D} \{\phi\}$ is introduced to distinguish path integrals over four dimensional `histories' of the field versus three-dimensional path integrals over fields at some given time.\footnote{We assume the same $i\varepsilon$ prescription as in the rest of the paper, which projects the initial state at $t=-\infty$ on the BD vacuum.} Other than that, \eqref{Pquant} is the usual representation, in which the path integral over ${ \cal D} \{\phi\}$ computes the `ket' state at time $t$, integral over ${ \cal D} \{\phi'\}$ computes the `bra' state with the insertion of an operator ${\cal O}_1(t_1)$, after which we insert $ {\cal O}_2(t)$ and `glue' the path integrals at time $t$ by doing the ${\cal D} \phi$ integral~\footnote{Notice that, as conventional, if the ${\cal O}$'s involve the momentum, one should use the path integral formulation that involves both conjugate variables.}. Let us now introduce the following distribution
\bea\label{PO1}
P_{\op_1}\l[{\phi,t;t_1}\r]=\Psi^*_{\op_1}[{\phi,t;t_1}] \Psi[\phi,t]\,.
\eea
For $\op_1=1$, this simply reduces to $\eqref{eq:Pdefinition}$, while for a non-trivial $\op_1$, $P_{\op_1}$ is the quantum mechanical analog of the two-times distribution $P^{[2]}$ defined above {in}~\eqref{Pnfunct}, { multiplied by $\op_1(\phi')$ and integrated over $\phi'$}:
\bea
P_{\op_1}\l[{ \phi,t;t_1}\r]=\int \MD \phi'\; P^{[2]}\l[{ \phi,t;\phi',t_1}\r]\;\op_1[\phi']\;\l(1+O(\hbar)\r)\;.
\eea
Importantly, in the same way as the wave functional, {and as it will become more clear shortly}, the distribution $P_{\op_1}$, can be computed in perturbation theory without encountering dangerous IR divergences. These are, of course, still present in~\eqref{Pquant}, but only when we take the final integral over ${\cal D} \phi$. For the two-times correlators, we thus can employ the same strategy  as we did for the single-time ones. Instead of trying to evaluate the strongly-coupled path integral, we derive the equation that $P_{\op_1}$ satisfies, and solve it using perturbation theory in all our parameters. 

\begin{figure}[htbp] 
	\centering
		\includegraphics[width=.3\linewidth,angle=0]{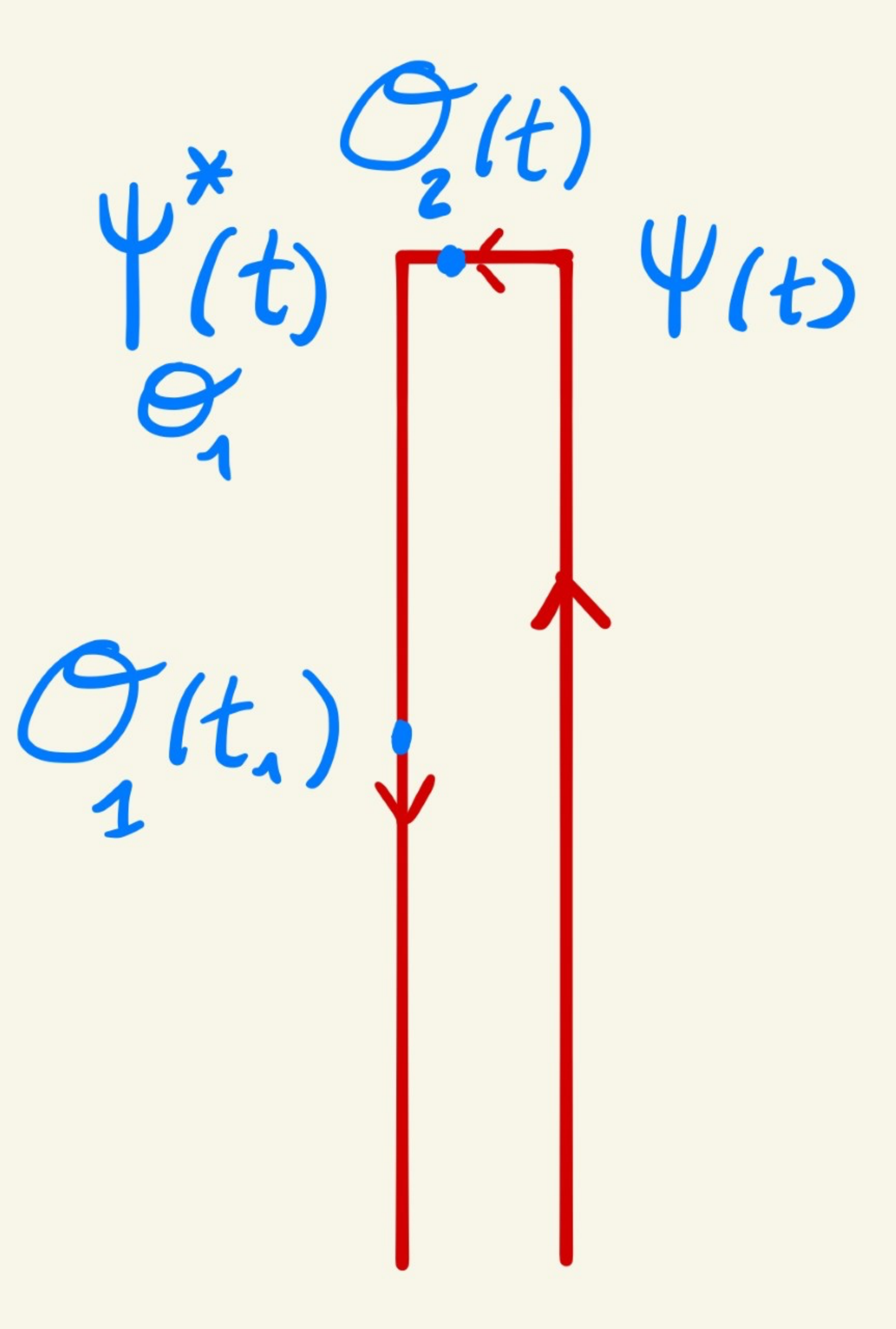}
	\caption{{Path integral representation of $\l\langle { \cal O}_1(t_1) { \cal{O}}_2(t)\r\rangle$.{}}}
	\label{fig:Quantum}
\end{figure}

For $t<t_1$ the operator insertion does not matter so $P_{\op_1}=P$. For $t>t_1$ instead, in direct analogy with \eqref{continuityeq}, we get
\bea
\label{quantumcontinuity}
&&\frac{\d }{\d t}P_{\op_1}\l[{ \phi,t;t_1}\r]= -\frac{i}{2 a^3} \int  d^3 x \frac{\delta }{\delta\phi(\vec x)}\left(\Psi^*_{\op_1}[{ \phi,t;t_1}]  \frac{\delta}{\delta\phi(\vec x)}\Psi[\phi,t]-\Psi[\phi,t]\frac{\delta}{\delta\phi(\vx)} \Psi^*_{{\cal{O}}_1}[{ \phi,t;t_1}] \right)\,.\nn\\
\eea
Let us now envisage how the calculation of $\Psi^*_{{\cal{O}}_1}$ in perturbation theory looks like. { It helps the intuition to realize that the evaluation of the path integral defining $\Psi^*_{{\cal{O}}_1}$ in (\ref{PO1}) is identical to the one of the wave function $\Psi^*$ of a different action than the one we study here, where we add to our action a `fake' translation-breaking vertex of the form $\delta(t-t_1){{\cal{O}}_1}$. The Feynman rules to compute  $\Psi^*_{{\cal{O}}_1}$ are therefore the same as the ones for the calculation of the wave function, with the additional condition that the vertex in ${{\cal{O}}_1}$ must be used once and only once in every diagram. 
This makes it evident that that the same reasons that allows us to compute the wave function perturbatively in an expansion in $\sqrt{\lambda}$ and $\epsilon$, allows us to compute  $\Psi^*_{{\cal{O}}_1}$.} 

To be specific, let us take, for example, $\op_1(t_1)=\phi_\ell(\vx_1)\Big |_{t_1}$, so that it contains only fields with momentum ~$k<\Lambda(t_1)$. 
To leading order in $\sqrt{\lambda}$, the evolution of $\op_1$ is just that of a free massless field which evolves according to \eqref{eq:Bulk2Boundary}, so, up to corrections of order $\eps$, it is constant in time. Thus we can write 
\be
\Psi^*_{{\cal{O}}_1}[{\phi,t;t_1}]=\op_1[\phi]\Psi^*[t,\phi]\l(1+O\l(\sqrt{\lambda},\eps\r)\r)\,,
\ee
where the $O\l(\sqrt{\lambda},\eps\r)$ terms can be computed in perturbation theory. Consequently, the leading correction to the evolution of the probability distribution, as compared to the classical treatment of the long modes distribution, comes from the variational derivative in \eqref{quantumcontinuity} acting on~$\op_1$:
\bea
\label{quantumevolution}
&&\frac{\d }{\d t}P_{\op_1}\l[{ \phi,t;t_1}\r]= - \int  d^3 x \frac{\delta }{\delta\phi(\vec x)}\l(\Pi[\phi,\vx,t]P_{\op_1}\l[{ \phi,t;t_1}\r]{-}\frac{i}{2 a^3}\frac{\delta \log \l(\op_1[\phi]\r)}{\delta\phi(\vx)} P_{\op_1}\l[ \phi,t;t_1\r]\r)\,.\nn\\
\eea
It is now clear that the new term is suppressed by powers of $\eps$ because of the factor of $a^{-3}$. Indeed, the leading term in the equation does not have this suppression because the phase of $\Psi$ is large and of order $a^3$, which is also the reason for the semiclassical approximation to apply.
We can now define the long distribution 
\be
P_{\op_1,\ell}\l[\phi_\ell, t; t_1\r]=\int {\cal{D}}\phi \;\delta\left[\phi_{\ell }(\vec x)-\int^{\Lambda(t)}\frac{d^3 k}{(2\pi)^3}\; e^{i\vec k\cdot\vec x}\; \phi(\vec k)\right]\;  P_{\op_1}\l[{ \phi,t;t_1}\r]\,,
\ee
take the derivative of $P_{\op_1,\ell}\l[\phi_\ell, t; t_1\r]$ with respect to $t$, and, using equation \eqref{quantumevolution}, repeat all the manipulations of sections \ref{sec:EFT} and \ref{sec:leadeq}. Let us focus on the simplest case of the one-location distribution, which will be enough to compute, for example, the correlation function $\l\langle\phi_\ell(\vx_1,t_1)\phi_\ell(\vx_1,t)\r\rangle$. The new term will go through all our manipulations straightforwardly and will produce the following correction in the evolution equation
\bea
&&\frac{\d P_{\op_1,1}\l(\phi_1,t;t_1\r)}{\d t}\supset \frac{i}{2 a(t)^3}\frac{\d}{\d\phi_1}\l(\frac{\delta \phi_\ell(\vx_1)\Big |_{t_1}}{\delta\phi_\ell(\vx_1)} P_{1}(\phi_1,t)\r)=\frac{\d}{\d\phi_1}\l(\l(\int^{\Lambda(t_1)} d^3k\r)\frac{i}{2 a(t)^3} P_{1}(\phi_1,t)\r)\sim\nn\\
&&\qquad\qquad\qquad\qquad\sim\;\eps^3 e^{-3H(t-t_1)}\frac{\d}{\d\phi_1}P_{1}(\phi_1,t)\,,
\eea
where we used momentum space to compute the functional derivative: $\frac{\delta\phi_\ell(\vec x_2)}{\delta\phi_\ell(\vec x_1)}=\delta^{(3)}(\vec x_2-\vec x_1)=\int^{\Lambda(t)}\frac{d^3k}{(2\pi)^3}\; e^{i \vk\cdot(\vx_2-\vx_1)}$. As expected, the correction is suppressed at least by $\eps^3$, and is even further suppressed at large time separations. 

Finally, let us discuss the boundary conditions for $P_{{\cal{O}}_1}[{ \phi,t;t_1}]$ at $t=t_1$. Since the evolution of $P_{\op_1}$ for $t<t_1$ is the same as for $P$, they simply read 
\be
P_{{\cal{O}}_1}[{ \phi,t;t_1}]\Big |_{t\to t_1}=\op_1[\phi]P[\phi,t],
\ee
which is { what would be implied on $P_{\op_1}$ from} \eqref{Pnfunct}, without any quantum mechanical corrections. The generalization of $P_{{\cal{O}}_1}$ to the case of several times is straightforward, at least when all operators are time-ordered.

\section{Outline of gradient corrections\label{sec:gradients}}

Similarly to what we did in the former section for the quantum corrections, here we give the essential steps that allow  us to include the gradient corrections. There are two places where we neglected them.  The simplest is when we computed the expectation value of the short modes in the background of the long, and we neglected to gradients of order $\epsilon$. Since these expectation values are computed using ordinary perturbation theory, it does not contain any conceptual challenge to include them.

The conceptually most interesting gradient correction occurs in the drift term, which is sensitive to the laplacian of $\phi$:
\bea\label{eq:drift_grad}
&&\frac{\d}{\d t} P_\ell[\phi_\ell,t]\supset\quad\l.-\int d^3 x \frac{\delta}{\delta \phi_\ell(\vx)}\l( \left\langle\Big[ \Pi[\phi,\vx,t]\Big]_{\Lambda(t)}\right\rangle_{\phi_\ell}P_\ell[\phi_\ell,t]\r)\nn\r.\\ 
&&\ \ \qquad\qquad \supset\quad -\int d^3 x \frac{\delta}{\delta \phi_\ell(\vx)}\l(\left( \frac{1}{3H}\frac{\d_i^2}{a(t)^2} \phi_\ell(\vec x)\right)P_\ell[\phi_\ell,t]\r)\ .
\eea
{Indeed, since in our procedure we eventually keep track of long fields only at finitely many spacial points, it may not be obvious how to systematically compute their gradients.}

It is now useful to write the laplacian of $\phi$ as
\be
\d_i^2\phi_\ell(\vec x)= \lim_{\Delta x\to 0}\frac{1}{\Delta x^2} \sum_{i=1}^3\left(\phi_\ell(\vec x+\Delta x\, \hat x^i)+\phi_\ell(\vec x-\Delta x\, \hat x^i)-2\phi_\ell(\vec x)\right)\ .
\ee
We then obtain
\bea\label{eq:drift_grad2}
&&\frac{\d}{\d t} P_\ell[\phi_\ell,t]\supset\quad \frac{1}{3H}\ \times \\ \nn
&&\times - \sum_{i=1}^3 \lim_{\Delta x\to 0}\frac{1}{a(t)^2 \Delta x^2}  \int d^3x\frac{\delta}{\delta \phi_\ell(\vx)}  \left(\left(\phi_\ell(\vec x+\Delta x\, \hat x^i)+\phi_\ell(\vec x-\Delta x\, \hat x^i)-2\phi_\ell(\vec x)\right)\ P_\ell[\phi_\ell,t]\right)  .
\eea

Let us now derive the equation for the one-location probability distribution. Applying the definition~(\ref{eq:n-location-definition}), we obtain
\bea\label{eq:drift_grad3}
&&\frac{\d}{\d t} P_1(\phi_1,t)\supset \quad- \frac{1}{3H}\sum_{i=1}^3 \lim_{\Delta x\to 0}\frac{1}{a(t)^2 \Delta x^2}   \int\MD\phi_\ell\; \delta^{(1)}\l(\phi_{\ell }(\vx_1)-\phi_{1}\r)\ \times\\ \nn
&& \qquad\qquad\qquad\quad\times \int d^3x\frac{\delta}{\delta \phi_\ell(\vx)}  \left(\left(\phi_\ell(\vec x+\Delta x\, \hat x^i)+\phi_\ell(\vec x-\Delta x\, \hat x^i)-2\phi_\ell(\vec x)\right)\ P_\ell (\phi_\ell,t)\right) \ .
\eea
We can now perform the integral over the values of $\phi_\ell$ at all locations except at the locations~$\vec x_1\pm \Delta x\, \hat x^i$. For this term, we obtain the following expression in terms of the three-locations probability distribution:
\bea\label{eq:drift_grad4}
&&\frac{\d}{\d t} P_1(\phi_1,t)\supset \quad \frac{1}{3H}\frac{\delta}{\delta \phi_1} \sum_{i=1}^3 \lim_{\Delta x\to 0}\frac{1}{a^2 \Delta x^2}   \int d\phi_{+}d\phi_{-} \\ \nn
&& \quad    \left(\left(\phi_{+}+\phi_{-}-2\phi_1\right)\ P_3(\phi_1,\,\vec x_1,\, t\,;\;\phi_+,\,\vec x_1+ \Delta x\, \hat x^i,\,t\,;\;\phi_-,\,\vec x_1 -\Delta x \,\hat x^i,\,t)\right) \ .
\eea
We have now obtained an expression that is prone to a perturbative expansion. In fact, once we compute, at zeroth-order in the gradients, the three-location distribution $P_3(\phi_1,\,\vec x_1,\, t\,;\;\phi_+,\,\vec x_1+ \Delta x\, \hat x^i,\,t\,;\;\phi_-,\,\vec x_1 -\Delta x \,\hat x^i,\,t)$, then we can plug it in~(\ref{eq:drift_grad4}) and compute the $O(\epsilon^2)$ corrections from the gradients for the one-location distribution $P_1$. 

Since only the coincidence limit of  $P_3$ matters, in practice the solution can be found more easily than solving for $P_3$ in all kinematical regimes. { In fact, $P_3$ solves the  equation given in~(\ref{Shroedn}) for $n=3$.}
Similarly to the case of $P_2$, at coincidence $j_0$ goes to one and, {assuming for simplicity the BD state }, we have that it exists a solution for $P_3$ of the form 
\bea
&&P_3(\phi_1,\,\vec x_1,\, t\,;\;\phi_+,\,\vec x_2,\,t\,;\;\phi_+,\,\vec x_3,\,t)\Big|_{\Delta x\to 0}=\\ \nn
&&\qquad\qquad\qquad = \delta(\phi_1-\phi_2)\,\delta(\phi_2-\phi_3)\; P^{eq}_1\l(\frac{\phi_1+\phi_2+\phi_3}{3}\r)+O(\Delta x^2)\ .
\eea
This part of the solution gives a vanishing contribution when plugged in~(\ref{eq:drift_grad4}). The part of the solution of order $\Delta x^2$ is obtained by the correction in $H'$ from Taylor expanding: $ j_0(\eps a H \Delta x_{ab})\simeq 1 -(\eps a H \Delta x_{ab})^2/4$. One can solve for this part by using ordinary perturbation theory for the differential equation~{(\ref{Shroedn})}, as the interaction is small for all the kinematical regime of interest. By plugging this perturbed solution in~(\ref{eq:drift_grad4}), one obtains a term that goes as $\sim \epsilon^2 H^3 \frac{\d^{3} ({\phi_1} P_1^{eq})}{\d\phi_1^{3}}$, to be compared with  $H^3 \frac{\d^2  P_1^{eq}}{\d\phi_1^2}$. Therefore the correction from gradients can be computed and is of order~$\epsilon^2$, as expected. This procedure can be iterated up to desired precision. We further see that, in order to compute the gradient corrections, we have a perturbative structure that, order by order, involves higher-locations probability distributions, evaluated at a lower perturbative order. This is reminiscent of the perturbative structure we have in $\sqrt{\lambda}$ when we {account for finite momenta of the long modes.}

\section{Summary and Outlook}

Perturbatively-computed correlation functions {of light fields with a potential}
 in de Sitter space are IR-divergent, and therefore ordinary perturbation theory cannot be trusted. In this paper, we have developed a formalism that allows us to compute correlation functions for {this class of theories} in the limit of rigid de Sitter space in a controlled way. {For simplicity we focus on the $\lambda \phi^4+m^2\phi^2$ potential with a small mass $m^2\ll\sqrt{\lambda}H^2$, hence the title. Other stable potentials, as well as inclusion of heavier fields, is a straightforward generalization. }

{\bf Main ideas:} Several conceptual ingredients needed to be introduced in order to develop our approach.  Maybe the main one has been to introduce an artificial cutoff at {physical} scales longer than the Hubble scale, $\epsilon H$, and divide the modes in shorter and longer than this scale. For `short' modes, perturbation theory holds, while for `long' modes a non-perturbative treatment is required. This non-perturbative treatment is obtained by deriving an evolution equation for the probability distribution of these modes. Since modes evolve from `short' to `long', this functional equation takes the form of a Fokker-Planck-like equation, with a drift term and diffusion term that in turn contains a tadpole contribution as well.  The coefficients of this equation are determined by computing expectation values of operators involving long and short modes, for fixed configuration of the long modes. Since short modes are perturbative, these expectation values can be computed using perturbation theory.  This {initially-obtained}  Fokker-Planck-like equation is not closed, because it involves not just the probability distribution of the field, but also {the momentum operator}. In order for it to become closed, we need to be able to express the action of the momentum operator on the wave function in terms of a multiplication of the wave function by a known function of $\phi$. This can be easily done if the wave function is in turn known. We then show that the wave function is not affected by neither IR nor dangerous secular divergences, and so can be computed using  perturbation theory, which we do by specializing to the Bunch-Davies~(BD) vacuum or to a state close to that. This resulting functional equation for the probability distribution {is still impossible to solve directly.}
 A crucial simplification then comes from the fact that the de Sitter horizon makes the gradients irrelevant over long distance scales. Out of the probability distribution for the field, we therefore define  probability distributions for the field value at a fixed {number of spacial} points. Their {time evolution follows a system of linear partial differential equations with finitely many variables. This system has a form roughly similar to the one of the Boltzmann hierarchy, as, {order by order in perturbation theory}, the equations for the lower-point distributions decouple from those for the higher-point ones. At this stage, the problem is conceptually solved, since a system of finitely-many partial differential equations can in principle be solved numerically. Nevertheless, we proceed to study the system analytically. First, we formally solve the leading equation for the single-point distribution in terms of Eigenvalues and Eigenfunctions of a certain differential operator.} We then identify several expansion parameters that allow us to {set up a perturbation theory around the leading solution.} The parameters we expand in are the square root of the coupling constant, $\sqrt{\lambda}$, as well some {exponentially} small parameters that we introduce to control our approximations, the most important of which being  $\epsilon$, the scale of the cutoff between `long' and `short' modes. This parameter $\epsilon\ll1$ controls the corrections from gradient terms and from quantum mechanical effects. We also introduce two other `control' parameters.  One is $e^{-\frac{1}{\sqrt{\lambda}}}\ll\delta\ll 1$,  that imposes that the cutoff between short and long modes is not too sharp, so that the long-wavelength dynamics is kept quasi local in space.  The other is $e^{-\frac{1}{\sqrt{\lambda}}}\ll e^{-\Delta}\ll 1$, that controls how we can solve the Fokker-Planck equation using sudden perturbation theory. We have explained, in a rather explicit way, how to compute corrections in all of {the expansion} parameters. At leading order in all of these parameters, our formalism reduces to the one of Starobinsky and collaborators~\cite{Starobinsky:1986fx,Starobinsky:1994bd}. Our construction can be thought as a rigorous derivation of those {findings}, plus the derivation of a more general formalism that allows us to find arbitrary accurate {results}, at least in principle. 

{\bf Main results:} Using this formalism, we find {that there exists} a non-perturbative equilibrium distribution for the density of fields at $n$-points, whose properties can be computed from the Fokker-Planck-like equation in a perturbative expansion in powers of $\sqrt{\lambda}$, $\epsilon$, $\delta$ and $e^{-\Delta}$. {We find that  the correlation functions are absolutely non-Gaussian.} 
The nature of the formalism is such that all correlation functions involving a general number of fields, but involving only a given number of spacetime locations, can be essentially evaluated with the same formula.   A consistency check of our calculation is that the dependence on the control parameters $\epsilon$, $\delta$ and $e^{-\Delta}$ should cancel order by order in perturbation theory. We have quite extensively checked that this is the case for $\epsilon$, and, to a less exhaustive degree, for $\delta$, while the same check for $e^{-\Delta}$ was left to future work. For these cancellations to happen, it is important to include the contribution of the short modes in physical observables.\\

We find the following main results:
\begin{enumerate}
\item {\it Equilibrium:} Correlation functions at a single spacetime point, $\langle\phi(\vx,t)^n\rangle$, tend to a steady state, associated to an equilibrium distribution of the form $P_1^{eq}(\phi)\sim e^{-{V(\phi)}/H^4}$, where the typical magnitude of the field value is $H/\lambda^{1/4}$, {for $V=\lambda \phi^4$}. Let us call this as $\langle\phi(\vx)^n\rangle_{eq}$.

\item {\it Stability:} Correlation functions at different spacetime points  decay towards the one obtained from equilibrium at all points: $\langle\phi(\vx_1,t_1)^{n_1}\ldots \phi(\vec x_n,t_n)^{n_n}\rangle\ \to\ \langle\phi(\vx_1)^{n_1}\rangle_{eq}\cdot\ldots\cdot\langle\phi(\vx_n)^{n_n}\rangle_{eq}$, {when all $t_i\to\infty$ but $x_i$ kept fixed}.  We argue that this shows that the BD vacuum is a stable vacuum for de Sitter space, in the rigid limit.

{\item{\it State-independence:} A related but stronger statement is true. Consider a state which is a small perturbation of the BD state at some time $t_0$, than for all $t_i-t_0\to\infty$, and physical distances $z_{jk}$ fixed, even though arbitrarily large, correlation functions in this state approach those in BD: $\langle\phi(\vx_1,t_1)^{n_1}\ldots \phi(\vec x_n,t_n)^{n_n}\rangle\ \to\ \langle\phi(\vx_1,t_1)^{n_1}\ldots \phi(\vec x_n,t_n)^{n_n}\rangle_{eq}$. This shows that the BD vacuum is also an attractor.
}

\item {\it de Sitter invariance:} For the two-point function, the decay of correlation functions {at long distances} is approximately given by $z^{-c\sqrt{\lambda}}$, where $z$ is the de Sitter invariant distance between the two points {and $c$ is some constant dependent on the details of the potential}. Therefore the de Sitter symmetry is linearly realized and not spontaneously broken in the BD vacuum.

\item {\it Thermality:} If we restrict to the static patch, we have checked that our correlation functions satisfy the KMS condition~\cite{Kubo:1957mj, Martin:1959jp} that must be obeyed by thermal ensembles, confirming a behavior that is expected to hold on general grounds.

\item {\it Large-N:} We extend our formalism to include the case of a $O(N)$ symmetric $\lambda/N(\vec\phi\cdot\vec\phi)^2 $ theory, finding that in the large-$N$ limit, some results lead to simpler analytical expressions.

{\item {\it Subleading corrections:} we explicitly compute the $O(\sqrt{\lambda})$ corrections to the single-point distribution and to the large-distance behavior of the two-point functions for the $\lambda \phi^4$ potential.}

\end{enumerate}

{\bf Outlook:} We now  briefly mention some future avenues that are opened by our findings.

\begin{itemize}
\item A clear limitation of our analysis is that we work in the rigid limit where $\mpl\to \infty$ and $H$ is kept constant. We believe, however, that  our formalism is equally applicable to dynamical gravity, coupled to matter fields. Of course, in the presence of gravity, in order to make correlation functions in dS, or approximately dS, space well defined, one needs a preferred choice of time slicing. This can be provided, for example, by an inflaton. Then, our formalism is suitable for studying the dynamics of another light scalar field $\phi$ during inflation. Since we already understand how to deal with the self-interactions of a field, as far as coupling to gravity is concerned, there does not seem to be much difference between  $\lambda \phi^4$ theory and a free particle with $m^2\sim\sqrt{\lambda}H^2$ coupled to gravity. In both cases it is clear that for 
$\sqrt{\lambda}\gg H^2/\mpl^2$, the effects of gravity on correlators are perturbative. On the other hand, for $\sqrt{\lambda}\lesssim H^2/\mpl^2$, the situation is more subtle and the  non-perturbative nature of our formalism may become relevant. We thus believe that this corresponding development of our formalism is a natural next step. In addition, it would be interesting to compare the results with other approaches to gravity in dS, {\it e.g.}~\cite{Dvali:2017eba,Dvali:2018jhn}.

Another natural application of the formalism is in the regime where the backreaction of the inflaton on the expansion of the universe is significant as, for example, in the case of `slow roll eternal inflation' studied in~\cite{Creminelli:2008es,Dubovsky:2008rf,Dubovsky:2011uy,Lewandowski:2013aka}.

\item At a more technical level, it would be instructive to do explicit higher-order calculations using our formalism. At the sub-sub-leading order in $\sqrt{\lambda}$, we expect interesting effects, related to interactions between long fields at different spacial points, to contribute to physical observables. The same is true about corrections proportional to our auxiliary parameters, which we only outlined but did not compute since they can be made exponentially small in the coupling.

\item It would be interesting to study some more phenomenological applications of the formalism, for example in relation with spontaneous symmetry breaking in dS (which formally does not exist) and with (pseudo-) Goldstone bosons. In particular, in some models of axionic dark matter, the details of the distribution of the axion field during inflation can have important consequences for the abundance and substructure of dark matter, as studied in the recent works \cite{Guth:2018hsa,Graham:2018jyp} and \cite{Arvanitaki:2019rax}.

\item We also hope that our conceptually simple results may eventually assist in understanding deeper questions akin to holography for cosmological spacetimes. In fact, in our non-gravitational case, the long-distance limit of correlators that we compute does define some non-local and non-unitary Euclidean conformal theory in three dimensions. We also emphasized and made explicit a very non-perturbative relation between these correlators and the wave function of the BD state, which in turn, in the late time limit, can be associated with another, different, CFT.

\end{itemize}

We leave this, and more, to future work.

\section*{Acknowledgements}{We thank Nima Arkani-Hamed, 
Masha Baryakhtar, Matt Baumgart, Paolo Creminelli, Lorenzo Di Pietro, Sergei Dubovsky, Shota Komatsu, Matthew Lewandowski, Juan Maldacena, Mehrdad Mirbabayi, Sasha Polyakov, Fedor Popov, Slava Rychkov, Eva Silverstein, Raman Sundrum, Ken Van Tilburg, Mark Wise, and Edward Witten for discussions. VG is a Marvin L. Golberger member at IAS. VG and LS are supported by the Simons Foundation Origins of the Universe program (Modern Inflationary Cosmology collaboration). LS is supported by the NSF award 1720397.}
\appendix

\section{Wigner distribution and the phase space description of long modes }
\label{app:Wigner}

\subsection{Phase space formalism}
The formalism for calculation of the correlation functions of long modes presented in the main text consisted roughly of two steps: the first is the perturbative calculation of the wave function and the second is the derivation of the equations for the probability distributions of the long modes. Consequently, strictly speaking, it resulted in calculation of correlation functions in the particular state in which wave function was computed. In practice we focused on the BD state and since, as we showed, correlators in this state decay at large space-like and time-like separations, our results remain intact for all states that differ from BD by insertion of some local operators in the distant past. Moreover, in section \ref{sec:stability} we showed that the deformation of the leading imaginary long-modes part of the BD wave function relaxes back to the BD state in times of order $H^{-1}$. In this Appendix we would like to present a different formalism which appears to be more state-independent. Another nice feature of this formalism is that it is manifestly IR finite since one never needs to take the path integral over the long modes. The drawback of this formalism, however, is that it is significantly more complicated. In fact, in order to simplify the logic somewhat, we will still refer to some results obtained above and with the use of the BD wave function.\footnote{Of course we always assume the Minkowski vacuum in the deep UV. }

The formalism discussed here crucially relies on the semiclassical nature of the state of long modes, the feature not-so-obviously necessary in the previous calculations.
To benefit from semiclassicality we need to choose a description of the system in which the classical limit is most transparent. Importantly, the classical background in our case is not close to any particular classical solution of the equations of motion, rather it corresponds to a distribution over the classical phase space of the system. In general, the information about this distribution is contained in the density matrix of the system. In the ordinary quantum treatment, this is for example given in the $\phi$-representation by a functional of two field variables: $\langle\phi(\vx)|\hat\rho(t)|\phi'(\vx)\rangle=\rho[\phi,\phi',t]$. The classical phase space, however, is rather parametrized by the field and its conjugate momentum, $(\phi(\vx),\pi(\vx))$, which we will for convenience denote by $\gamma_a(\vx)$, $a=1,2$, or sometimes simply $\gamma(\vx)$ when we refer to both. Consequently, it is tempting to perform some sort of Fourier transform on the density matrix in order to introduce the momentum variable. A convenient (and clearly clever) choice of the Fourier transform is known as the Wigner transform, and has the following form:
\be\label{eq:wigner-probability-definition}
P^{(w)}[\gamma,t]=\int \MD\chi e^{2 i \int d^3 x\frac{  \pi(\vx) \chi(\vx)}{\hbar}}\rho[\phi-\chi,\phi+\chi,t]\ .
\ee
$P^{(w)}$ is usually called the Wigner distribution. Let us briefly review some of its important properties. The von Neumann equation for the density matrix (which is of course equivalent to the Shroedinger equation in the case of a pure state) under the Wigner transform maps into the following evolution equation for $P^{(w)}$:
\bea
\label{WignerMoyal}
&&\frac{\d}{\d t} P^{(w)}[\gamma,t]=-\left\{\left\{P^{(w)}[\gamma,t], \Ham^{(w)}[\gamma,t] \right\}\right\}\ ,
\eea
where the double curly bracket denotes the Moyal bracket, given by
\bea\label{eq:moyal}
\left\{\left\{A[\gamma,t], B[\gamma,t] \right\}\right\}&
=A[\gamma,t]\cdot \frac{2}{ \hbar} \sin\left(\frac{\hbar}{2}\int{ d^3x\left(\overset{\leftarrow}{\d}_{\phi(\vx) }\overset{\rightarrow}{\d}_{\pi(\vx) }-\overset{\leftarrow}{\d}_{\pi(\vx) }\overset{\rightarrow}{\d}_{\phi(\vx)} \right)}\right)\cdot B[\gamma,t]=&\nn\\
&=\left\{A[\gamma,t], B[\gamma,t] \right\}+{\cal{O}}(\hbar^2)
\label{Moyal}
\eea
and, as usual, the Poisson brackets are given by 
\be
\left\{A[\gamma,t], B[\gamma,t] \right\}=\int d^3x\,\l(\frac{\d A[\gamma,t]}{\d \phi(\vx)}\frac{\d B[\gamma,t]}{\d \pi(\vx)}-\frac{\d A[\gamma,t]}{\d \pi(\vx)}\frac{\d B[\gamma,t]}{\d \phi(\vx)}\r)\;.
\ee
We stress that the equation \eqref{WignerMoyal} contains the full information about evolution of any quantum system. In the classical limit, where we can formally take $\hbar$ to zero, the Moyal bracket reduces to the Poisson bracket, and we get simply the Liouville equation which describes the evolution of the classical distribution on phase space. This is one way to see that in the classical limit the Wigner distribution coincides with the classical probability distribution. This is the main advantage of working with the Wigner transform.

The Hamiltonian in \eqref{WignerMoyal} is in fact the Wigner transform of the Hamiltonian operator, which is defined as
\be
\Ham^{(w)}[\gamma,t]=\int \MD\chi \; e^{2 i\int d^3x\; \frac{ \pi(\vec x) \chi(\vx)}{\hbar}}\; \la\phi-\chi|\hat \Ham(t) |\phi+\chi\ra.
\ee
Analogously the Wigner transform can be defined for any operator, and then expectation values can be computed as~\footnote{The Wigner transform of $\hat\phi$ and $\hat\pi$ is simply $\phi$ and $\pi$. The Wigner transform of an operator which consists of a product of $\hat\phi$ and $\hat\pi$ is a polynom in $\phi$ and $\pi$, which actually corresponds to the original operator defined with some specific ordering of  $\hat\phi$ and $\hat\pi$.  Since ordering ambiguity in quantum field theory corresponds to a specific choice of counterterms, and since in our formalism we will add all counterterms, we can simply assume that the operators that we Wigner-transform are already ordered in the ordering chosen by the Wigner transformation.}
\be
\la\hat\op(t)\ra=\int\; \MD \gamma\; P^{(w)}[\gamma,t]\;\op^{(w)}[\gamma,t]\ .
\label{Wignerop}
\ee

For simplicity of notation, we will drop the superscript ${}^{(w)}$ in $\op^{(w)}$ in what follows, and we will simply write $\langle\op\rangle$  for  expectation values.
Note also that the integral of $P^{(w)}$ over momenta gives just the probability distribution for the fields used throughout the main text:
\be
\int\MD\pi\, P^{(w)}\l[\gamma,t\r]=P[\phi,t]\,.
\ee
\subsection{Wigner distribution for long modes}
The following discussion largely parallels that in section \ref{sec:functionequation}, \ref{sec:perturbatie-structure}, and \ref{sec:Singlepoint}, so we will drop some of the details. We proceed by defining the long-fields Wigner distribution:
\be
P^{(w)}_\ell[\gamma_\ell,t]=\int {\cal{D}}\gamma \;\delta^{(2)}\left[\gamma_{\ell }(\vec x)-\int_0^{\Lambda(t)}\frac{d^3 k}{(2\pi)^3}\; e^{i\vec k\cdot\vec x}\; \gamma(\vec k)\right]\; \; P^{(w)}[\gamma,t]\, ,
\ee
and derive the time evolution equation for it. As before, it consists of two term:
\be
\frac{\d}{\d t} P^{(w)}_\ell [\gamma_\ell,t]= {\rm Diffusion} + {\rm Drift}\; .
\ee
The diffusion term now has additional contributions from the expectation values of the momenta as well as the mixed contributions:
\bea
&&{\rm Diffus.}= \l( \int d^3 x\;\frac{\d}{\d\gamma_{\ell ,a}(\vx)} \,\l( \left\langle-\dot\Delta \gamma_{a}(\vx)\right\rangle_{\gamma_\ell}\; P^{(w)}_\ell[\gamma_\ell,t]\r)\;+\r.\\ \nn
&&\l.\qquad +\;\int d^3 x \;\int d^3 x'\;\frac{\d^2}{\d\gamma_{\ell ,a}(\vx)\d\gamma_{\ell ,b}(\vx')} \;\l( \left\langle\dot\Delta \gamma_{a}(\vx) \Delta \gamma_{b}(\vx')\right\rangle_{\gamma_\ell}\; P^{(w)}_\ell[\gamma_\ell,t]\r)\r)\ \times\\ \nn
&&\qquad\times\ \left(1+O(\delta)\right)\;.
\eea

The drift term is now derived from the Moyal evolution \eqref{WignerMoyal}. Brackets containing short modes all turn out to be total derivatives and consequently do not contribute. Since long modes are classical, up to power law in $\eps$ corrections, we can replace the Moyal bracket with the Poisson one. The resulting drift term reads:
\bea\label{eq:drift-Wignerl}
&&{\rm Drift}=\l[-\int d^3 x\frac{\d}{\d \phi_\ell(\vx)} \;\l( \left\langle\left[\frac{\d \Ham[\gamma,t]}{\d \pi(\vx)}\r]_{\Lambda(t)}\right\rangle_{\gamma_\ell}P^{(w)}_\ell [\gamma_\ell,t]\r)+\nn\r.\\
&&\l.+\int d^3x\frac{\d}{\d \pi_\ell(\vx)} \;\l( \left\langle\l[\frac{\d \Ham[\gamma,t]}{\d \phi(\vx)}\r]_{\Lambda(t)}\right\rangle_{\gamma_\ell}P^{(w)}_\ell [\gamma_\ell,t]\r)\r]\l(1+O(\hbar\eps^3,\delta)\r)\;.
\eea

Next we proceed by defining the single-point Wigner distribution as\footnote{$n$-point Wigner distributions can be analogously defined.}
\be
P^{(w)}_1(\gamma_1,t)=\int \MD\gamma_\ell\,\delta^{(2)}\l(\gamma_1-\gamma_\ell(\vx_1)\r)P^{(w)}\l[\gamma_\ell,t\r]\,.
\ee
We are now in a position to derive the equation which controls its time evolution. To do so we need to compute correlation functions of $\phi(\vk)$ and $\pi(\vk)$ for $k\sim\Lambda(t)$ in the background of fixed long modes. For a change we can think of doing this directly in the Heisenberg picture bypassing the wave function calculation. The reason we can do it is because all momenta and time integrals are finite since the long modes are kept fixed and treated simply as background fields. The leading order calculation is straightforward since short fields are either independent of the long ones, or at worst depend on the long fields at the same moment of time. At higher orders in $\sqrt{\lambda}$, however, correlators of short modes depend on long modes at different times. At the technical level, this requires introduction of multi-times Wigner distributions which appear in the equation at subleading orders. This feature particularly complicates the formalism we are developing in this Appendix. Since for us it serves mostly complimentary purposes, we will not dwell into these details and present simply the leading-order equations.

In what follows we assume that the state is not to far from the state described by the wave function \eqref{eq:wavefunction_schematic}. What we mean by this is that the characteristic value of the long field is $\phi_\ell \sim\lambda^{-1/4} H$, and of the velocity is $v_\ell\sim-\lambda\phi_\ell^3/(3H)$. As usual,  we will assume that very short modes, $k\gg H$ are in the Minkowski vacuum, and we don't need to make any assumptions for the modes of intermediate lengths, $k\gtrsim\Lambda(t)$, other than that their amplitude is not too large. Then we can treat long modes as background fields and compute the expectation values of the short modes using just the usual in-in perturbation theory. Since the calculation is done with an explicit IR cutoff, these calculation is under perturbative control. In particular, for the momentum operator at some short wavelength~$\vk_s$ we have 
\be
\label{piphirelation}
\hat\pi(\vk_s,t)=a^3\frac{d}{dt}\hat{\phi}(\vk_s,t)=-\frac{a^3}{3 H}V'(\hat\phi_s,\phi_1)=-\frac{a^3}{3 H}V''\l(\phi_1\r)\hat\phi(\vk_s,t)\l(1+O\l(\lambda^{\frac{1}{4}}\r)\r)\,,
\ee
so we can express the correlation functions of the momenta through correlation functions of the short fields times some dependence on the long modes without an explicit use of the wave function. 
Specifying to $V(\phi)=\lambda\phi^4/4$, after some straightforward manipulations, we get
\bea
&&\frac{\d}{\d t} P^{(w)}_1(\gamma_1,t)=\l(-a^{-3}\pi_1\frac{\d}{\d \phi_1}  P^{(w)}_1 (\gamma_1,t)\r.+ a^3\lambda\phi_1^3 \frac{\d}{\d \pi_1}P^{(w)}_1 (\gamma_1,t)+\\
&& \l.+\frac{H^3}{4\pi^2}\l(\frac{1}{2}\;\frac{\d^2}{\d\phi_{1}^2}P^{(w)}_1(\gamma_1,t) -a^3\frac{\d^2}{\d\phi_{1}\d\pi_{1}}\l[\frac{\lambda \phi_1^2}{H}\;P^{(w)}_1(\gamma_1,t)\r]+a^6\frac{1}{2}\;\frac{\d^2}{\d\pi_{1}^2}\l[\frac{\lambda^2\phi_1^4}{H^2} P^{(w)}_1(\gamma_1,t)\r]\r)\r)
\nn\\
&&\times\l(1+O(\hbar \eps^3,\eps^2,\lambda^{\frac{1}{2}},\delta)\r)\;.\nn
\eea
To analyze this equation it is convenient to introduce the ``velocity'' variable $v_1=\pi_1 a^{-3}$, in terms of which the equation simplifies a bit:
\bea
\label{1ptleadv}
&&\d_t P_1^{(w)}(\phi_1,v_1,t)= \Bigg[ \frac{\d}{\d v_1}\left[ \left(3 H v_1 + \lambda\phi_1^3\right) P^{(w)}_1(\phi_1,v_1,t)\right]-\frac{\d}{\d\phi_1}(v_1 P^{(w)}_1(\phi_1,v_1,t))+\nn\\ \nn
&&+\frac{H^3}{4\pi^2}\l(\frac{1}{2}\;\frac{\d^2 }{\d\phi_1^2}\left[ P^{(w)}_1(\phi_1,v_1,t)\right]+\frac{1}{2}\;\frac{\d^2 }{\d v_1^2}\left[ \frac{\lambda^2\phi_1^4}{H^2} \;P^{(w)}_1(\phi_1,v_1,t)\right]-\r.\nn\\
&&\l.\frac{\d^2 }{\d \phi_1 \d v_1}\left[ \frac{\lambda \phi_1^2}{H}\; P^{(w)}_1(\phi_1,v_1,t)\right]\r)\Bigg]\l(1+O(\hbar \eps^3,\eps^2,\lambda^{\frac{1}{2}},\delta)\r)\;.
\eea
This is an analog of equation \eqref{1pteqtime}, however, now it describes an evolution on the classical phase space of long modes without assuming any particular exact dependence of the long momentum on the long field. We will see momentarily that \eqref{1pteqtime} follows from \eqref{1ptleadv}.

The equation controlling the leading-order time evolution of the single-point Wigner distribution is now a partial differential equation in three variables. Moreover, its second-derivative part is degenerate, which makes the estimates of the subleading terms slightly more subtle. To proceed, it is useful to perform the moments transform with respect to the velocity variable:
\be
M_{1,n}(\phi_1,t)=\int d v_1 \,v_1^n \, P_1^{(w)}(\phi_1,v_1,t) \,,
\ee
which, for a state close to BD, implies that $M_{1,n}\sim \lambda^{3n/4} P_1$. Of course, $M_{1,0}(\phi_1)$ is simply $P_1(\phi_1)$. Multiplying \eqref{1ptleadv} by powers of $v_1$ and integrating over it, we get the following equations for the moments
\bea
\label{momentsevol}
&&\d_t M_{1,0}(\phi_1,t)=-\frac{\d}{\d \phi_1}M_{1,1}(\phi_1,t)+\frac{H^3}{8\pi^2} \frac{\d^2}{\d \phi_1^2}M_{1,0}(\phi_1,t)+O(\lambda)M_{1,0}(\phi_1,t)\,,\\\nn
&&\d_t M_{1,n}(\phi_1,t)=-3 n H M_{1,n}(\phi_1,t)-n \lambda\phi_1^3  M_{1,n-1}(\phi_1,t)+O\l(\sqrt{\lambda}\r) M_{1,n}(\phi_1,t),\quad n\geq1\,.
\eea
We separated the first equation because its right hand side is of order $O\l(\sqrt{\lambda}\r)$, while the rest of the equations contain in it terms of order $O(1)$.

It is straightforward to find the equilibrium solution of these equations. One finds
\bea
M^{ eq}_{1,n}(\phi_1)=\l(-\frac{\lambda\phi_1^3}{3 H}\r)^{n} M^{ eq}_{1,0}(\phi_1),\qquad
\Gamma_{\phi_1}M^{eq}_{1,0}(\phi_1)=0\,,
\eea
with $\Gamma_\phi$ as in \eqref{Gammaphi}. For the Wigner distribution itself it implies
\bea
\label{Wignereq}
P^{(w),eq}_1(\phi_1,v_1)=P_1^{eq}(\phi_1)\,\delta\l(v_1+\frac{\lambda \phi_1^3}{3 H}\r)\,,
\eea
where $P_1^{eq}$ given in \eqref{1pteqsol}.

\subsubsection{Time dependence of the moments \label{sec:moment-time-dep}}
Let us now analyze the time dependence of the moments and check that they {if all of them are close to the equilibrium distribution \eqref{Wignereq}, they will indeed converge to it.} 
The simplest way is to use an inductive argument. First, 
let us assume that all moments up to $M_{1,n-1}(\phi_1,t)$ {are slowly changing with time. $M_{1,0}$ satisfies this hypothesis given the right hand side of the first equation in (\ref{momentsevol}) {is $O(\sqrt{\lambda})$.} Then we can integrate  \eqref{momentsevol} for the $n$-th moment:
\be
 M_{1,n}(\phi_1,t)=c_n e^{-3 n H  t}-e^{-3n  H t}\int^t dt'\, e^{ 3 n H  t'}n \lambda\phi_1^3\,  M_{1,n-1}(\phi_1,t')\;,
 \ee
 where $c_n$ is an integration constant.
Since, by our assumption, $M_{1,n-1}$ is slowly varying, after times larger than $(n H)^{-1}$, $  M_{1,n}(\phi_1,t)$ will approach $-\lambda\phi_1^3  M_{1,n-1}(\phi_1,t)/(3 H )$ and hence will also become a slowly varying function. Finally, if $M_{1,1}(\phi_1,t)=-\lambda\phi_1^3  M_{1,0}(\phi_1,t)/(3 H )$, then by the first equation in \eqref{momentsevol}, 
\be
\d_t M_{0,n}(\phi_1,t)=\Gamma_{\phi_1}M_{0,n}(\phi_1,t)\,,
\ee
which is nothing but equation \eqref{1pteqtime}.
 Hence, following the results of section~\ref{sec:nonequil1pt}, $M_{0,n}$ is slowly approaching $M_{0,n}^{eq}$ and consequently the equilibrium solution \eqref{Wignereq} is an attractor. Of course this behavior of the Wigner distribution is closely related to the discussion of section \ref{sec:stability} about attractiveness of the BD wave function. Similar arguments can be used to demonstrate the attractor behavior of multi-point Wigner distributions.

\section{Outline of finite $\delta$ corrections\label{app:delta_corrections}}

\subsection{Diffusion term including $\Omega_{\Lambda(t)}$ }
\label{app:smooth-diffusion}
In this appendix we derive the diffusion term including the effect of the smooth window function $\Omega_{\Lambda(t)}$. We have
\bea
&&{\rm Diffus.}=\int {\cal{D}}\phi \; \frac{\d}{\d t}\left( \delta\left[\phi_{\ell }(\vec x)-\int \frac{d^3  k}{(2\pi)^3}\; e^{i\vec k\cdot\vec x}\; \Omega_{\Lambda(t)}(k)\,\phi(\vec k)\right]\; \right)\; P[\phi,t]\\ \nn
&&\quad= \int d^3x_{A} \frac{\delta}{\delta\phi_{\ell}(\vx_1)}\int {\cal{D}}\phi \; \delta\left[\phi_{\ell }(\vec y)-\int \frac{d^3  k}{(2\pi)^3}\; e^{i\vec k\cdot\vec y}\; \Omega_{\Lambda(t)}(k)\,\phi(\vec k)\right]\;\times \\ \nn
&&\qquad\quad\qquad \times \ \left(-\int \frac{d^3  k_A}{(2\pi)^3}\; e^{i\vec k_A\cdot\vec x_A}\; \frac{\d}{\d t}\left(\Omega_{\Lambda(t)}(k_A)\right)\,\phi(\vec k_A)\right)\; P[\phi,t]\ .
\eea
By hypothesis, $\Omega_{\Lambda(t)}$ is equal to 1 for  $k< \Lambda(t)$, vanishes for $k\geq (1+\delta)\Lambda(t)$, and smoothly interpolates between these two values in the intermediate $k$'s.  It is now useful to split the modes in the $\delta$-function as `long' (those with $k\leq \Lambda(t)$, for which the window function is equal to 1), `very short' (those with $k\geq (1+\delta) \Lambda(t)$, and for which the window function is vanishing), and `intermediate' (those with $\Lambda(t)< k< (1+\delta) \Lambda(t)$, and which are the only ones for which $\d \Omega_{\Lambda(t)}/\d t$ has support): 
\bea\label{eq:smooth-diff2}
&&{\rm Diffus.}=   \int d^3x_A\frac{\delta}{\delta\phi_{\ell}(\vx_A)}\int {\cal{D}}\phi \;  \\ \nn
&&\quad \times\  \delta_{k\leq \Lambda(t)}\left[\phi_{\ell }(\vec k)-\phi(\vec k)\right]\; \delta_{k\geq (1+\delta)\Lambda(t)}\left[\phi_{\ell }(\vec k)\right]\; \; \delta_{\Lambda(t)< k< (1+\delta) \Lambda(t)}\left[\phi_{\ell }(\vec k)-\Omega_{\Lambda(t)}(k)\phi(\vec k)\right] \;\times\\ \nn
&& \quad\times\;\left(-\int \frac{d^3  k_A}{(2\pi)^3}\; e^{i\vec k_A\cdot\vec x_A}\; \frac{\d}{\d t}\left(\Omega_{\Lambda(t)}(k_A)\right)\,\phi(\vec k_A)\right)\; P[\phi,t]\ .
\eea
We now Taylor expand the $\delta$-function over the `intermediate' wavenumbers around $\phi_\ell(\vec k)$, in the following way: 
\bea
&&\delta_{\Lambda(t)< k< (1+\delta) \Lambda(t)}\left[\phi_{\ell }( k)-\Omega_{\Lambda(t)}(k)\phi(k)\right]= \\ \nn
 &&\qquad=  \sum_{n=0}^{+\infty}\frac{(-1)^n}{n!}\int d^3x_1\ldots \int d^3x_n\; \frac{\delta}{\delta\phi_{\ell}(\vx_1)}\ldots \frac{\delta}{\delta\phi_{\ell}(\vx_n)}\;
 \delta_{\Lambda(t)< k< (1+\delta) \Lambda(t)}\left[\phi_{\ell }(\vec k)\right]\ \times\\ \nn
 &&\times\  \int_{\Lambda(t)}^{(1+\delta)\Lambda(t)} \frac{d k_1}{(2\pi)^3}\; \ldots \int_{\Lambda(t)}^{(1+\delta)\Lambda(t)} \frac{d k_n}{(2\pi)^3}\;\;e^{i \left(\vec k_1\cdot \vec x_1+\ldots +\vec k_n\cdot \vec x_n\right)}\;\Omega_{\Lambda(t)}(k_1)\phi(\vec k_1) \ldots \Omega_{\Lambda(t)}(k_n)\phi(\vec k_n)\ .
 \eea 
where we used that  $\frac{\delta}{\delta \phi_{\ell}(\vec k_i)}=\frac{1}{(2 \pi)^3}\int d^3x_i\; e^{i  \vec k_i \cdot \vec x_i}\, \frac{\delta}{\delta \phi_{\ell}(\vec x_i)}$. We can now plug back this expression into~(\ref{eq:smooth-diff2}), and obtain
\bea\label{eq:smooth-diff3}\nn
&&{\rm Diffus.}=  \sum_{n=0}^{+\infty}\frac{(-1)^{n+1}}{n!}\int d^3x_A\int d^3x_1\ldots \int d^3x_n\;\frac{\delta}{\delta \phi_{\ell}(\vx_A)}  \frac{\delta}{\delta\phi_{\ell}(\vx_1)}\ldots \frac{\delta}{\delta\phi_{\ell}(\vx_n)}  \\ \nn
&& \times\;\int_{\Lambda(t)}^{(1+\delta)\Lambda(t)} \frac{d^3  k_A}{(2\pi)^3}\,\frac{d^3 k_1}{(2\pi)^3}\; \ldots \frac{d^3 k_n}{(2\pi)^3}\;  \;e^{i \left(\vec k_A\cdot\vec x_A+\vec k_1\cdot \vec x_1+\ldots +\vec k_n\cdot \vec x_n\right)} \frac{\d}{\d t}\left(\Omega_{\Lambda(t)}(k_A)\right)\,\Omega_{\Lambda(t)}(k_1)\ldots \Omega_{\Lambda(t)}(k_n) \\ 
&& \times\ \int {\cal{D}}\phi \; \delta_{k\leq \Lambda(t)}\left[\phi_{\ell }(\vec k)-\phi(\vec k)\right]\; \delta_{k\geq\Lambda(t)}\left[\phi_{\ell }(\vec k)\right] \;P[\phi,t]\;\phi(\vec k_A)\phi(\vec k_1) \ldots \phi(\vec k_n)\ .
\eea
The path integral in the last line is nothing but the expectation value of $\phi(\vec k_A)\phi(\vec k_1) \ldots \phi(\vec k_n)$ for a given value of the long modes $\phi_\ell$. Increasing number of phase space integrals gives increasing powers of $\delta$ by which higher $n$ terms are suppressed. Notice that what, at the beginning of this Appendix, we called the `intermediate' modes are being integrated out as well, and so are part of the `short' modes. We therefore obtain 
\bea\label{eq:smooth-diff4}\nn
&&{\rm Diffus.}=  \sum_{n=0}^{+\infty}\frac{(-1)^{n+1}}{n!}\int d^3x_A\int d^3x_1\ldots \int d^3x_n\;\frac{\delta}{\delta \phi_{\ell}(\vx_A)}  \frac{\delta}{\delta\phi_{\ell}(\vx_1)}\ldots \frac{\delta}{\delta\phi_{\ell}(\vx_n)}  \\ \nn
&& \times\;\int_{\Lambda(t)}^{(1+\delta)\Lambda(t)} \frac{d^3  k_A}{(2\pi)^3}\,\frac{d^3 k_1}{(2\pi)^3}\; \ldots \frac{d^3 k_n}{(2\pi)^3}\;  \;e^{i \left(\vec k_A\cdot\vec x_A+\vec k_1\cdot \vec x_1+\ldots +\vec k_n\cdot \vec x_n\right)} \frac{\d}{\d t}\left(\Omega_{\Lambda(t)}(k_A)\right)\,\Omega_{\Lambda(t)}(k_1)\ldots \Omega_{\Lambda(t)}(k_n) \\ 
&& \times\;P_\ell[\phi_\ell,t]\;\langle\phi(\vec k_A)\phi(\vec k_1) \ldots \phi(\vec k_n)\rangle_{\phi_\ell}\ .
\eea

This is the final expression for the diffusion term including the effect of the window function. With respect to the expression that we gave in~(\ref{eq:diffstep}), the differences are minor. In fact, the $n=0$ and $n=1$ case give respectively the one-derivative and the two-derivative terms of~(\ref{eq:diffstep}).
With respect to~(\ref{eq:diffstep}), we notice the presence of higher-derivative diffusion terms. It is easy to realize that only the one- and two-derivative terms are not suppressed by positive powers of $\delta$. Indeed momentum conservation makes these absent in the two-derivative term. Therefore, with respect to~(\ref{eq:diffstep}), the main difference is the presence of positive $\delta$-corrections that we neglect, { and that we could explicitly compute if we wished}. 

\subsection{Remaining $O(\delta)$ corrections}
Let us now collect the rest of perturbative-in-$\delta$ corrections. First, let us justify the formula \eqref{deltacor}. Using the definition of $P_\Omega$ and decomposing and Taylor expanding the delta-function as in \eqref{eq:smooth-diff2} and \eqref{eq:smooth-diff3} we get
\bea
\label{eq:PomegaPell}
&& P_\Omega[\phi_\ell,t]=\int {\cal{D}}\phi \;\delta\left[\phi_{\ell }(\vec x)-\int\frac{d^3 k}{(2\pi)^3}\; \Omega_{\Lambda(t)}(k)\; e^{i\vec k\cdot\vec x}\; \phi(\vec k)\right]\;  P[\phi,t]= \nn\\
&&\sum_{n=0}^{+\infty}\frac{(-1)^{n}}{n!}\int d^3x_1\ldots \int d^3x_n\;  \frac{\delta}{\delta\phi_{\ell}(\vx_1)}\ldots \frac{\delta}{\delta\phi_{\ell}(\vx_n)}  \nn\\ 
&& \times\;\int_{\Lambda(t)}^{(1+\delta)\Lambda(t)}\,\frac{d^3 k_1}{(2\pi)^3}\; \ldots \frac{d^3 k_n}{(2\pi)^3}\;  \;e^{i \left(\vec k_1\cdot \vec x_1+\ldots +\vec k_n\cdot \vec x_n\right)} \Omega_{\Lambda(t)}(k_1)\ldots \Omega_{\Lambda(t)}(k_n) \nn\\ 
&& \times\ \int {\cal{D}}\phi \; \delta\l[\phi_\ell(\vx)-\int^{\Lambda(t)}\frac{d^3 k}{(2\pi)^3} e^{i\vec k\cdot \vec x}\phi(\vk)\r] \;P[\phi,t]\;\phi(\vec k_1) \ldots \phi(\vec k_n)=\nn\\
&&=\,P_\ell[\phi_\ell,t]+\sum_{n=1}^{+\infty}\frac{(-1)^{n}}{n!}\int d^3x_1\ldots \int d^3x_n\;  \frac{\delta}{\delta\phi_{\ell}(\vx_1)}\ldots \frac{\delta}{\delta\phi_{\ell}(\vx_n)}  \\ \nn
&& \qquad\qquad\qquad\times\   \;P_\ell[\phi_\ell,t]\;\l\langle\Delta\phi(\vx_1) \ldots \Delta\phi(\vx_n)\r\rangle_{\phi_\ell}\ ,
\eea
where in the last line we used the definition of $\Delta\phi$ from \eqref{dgamma} as well as \eqref{eq:expvalue}. We see that apart form the first term that we separated explicitly, all the terms involve the integrals over the phase space, whose volume is proportional to $\delta$, and hence are proportional to powers of $\delta$. Clearly, \eqref{eq:PomegaPell} can be inverted perturbatively in $\delta$ if needed.

Similar manipulations can be used to account for finite-$\delta$ corrections in the drift term in equation \eqref{eq:drift_general}. For this term we get
\bea
\label{eq:driftdelta}
&&{\rm Drift} =-\int {\cal{D}}\phi \; \delta\left[\phi_{\ell }(\vec x)-\int \frac{d^3 k}{(2\pi)^3}\; \Omega_{\Lambda(t)}(k)\; e^{i\vec k\cdot\vec x}\, \phi(\vec k)\right] \int  d^3 x \frac{\delta }{\delta\phi(\vec x)}\l(\Pi[\phi,\vx,t]P[\phi,t]\r)\nn\\
&&=\sum_{n=0}^{+\infty}\frac{(-1)^{n+1}}{n!}\int d^3x_1\ldots \int d^3x_n\;  \frac{\delta}{\delta\phi_{\ell}(\vx_1)}\ldots \frac{\delta}{\delta\phi_{\ell}(\vx_n)} \;   \int {\cal{D}}\phi\; \Delta\phi(\vx_1) \ldots \Delta\phi(\vx_n)\nn \\ 
&& \times\,\delta\left[\phi_{\ell }(\vec x)-\int^{\Lambda(t)} \frac{d^3 k}{(2\pi)^3} e^{i\vec k\cdot\vec x} \, \phi(\vec k)\right] \int  d^3 x \int_0^{\Lambda(t)(1+\delta)} d^3 k'e^{-i\vk \cdot \vx}\frac{\delta }{\delta\phi(\vec k')}\l(\Pi[\phi,\vx,t]P[\phi,t]\r)\nn\\
&&=\sum_{n=0}^{+\infty}\frac{(-1)^{n+1}}{n!}\int d^3 x\int d^3x_1\ldots \int d^3x_n\;  \frac{\delta}{\delta\phi_{\ell}(\vx_1)}\ldots \frac{\delta}{\delta\phi_{\ell}(\vx_n)}\times\\
&&\l( \frac{\delta}{\delta\phi_{\ell}(\vx)}\l(\l \langle \Delta\phi(\vx_1) \ldots \Delta\phi(\vx_n) \Pi[\phi,\vx,t]  \r \rangle_{\phi_\ell}  P_\ell[\phi_\ell,t]\r)-\l\langle  \int_{\Lambda(t)}^{\Lambda(t)(1+\delta)} d^3 k'e^{-i\vk \cdot \vx}\frac{\delta }{\delta\phi(\vec k')}\r.\r.\nn\\
&& \l(\Delta\phi(\vx_1) \ldots \Delta\phi(\vx_n)\r) \,\Pi[\phi,\vx,t]\Bigg\rangle_{\phi_\ell}\,P_\ell[\phi_\ell,t]\Bigg)\nn\,.
\eea
Again we see that all new ({\it i.e.} $n>0$) terms contain phase space integrals {over a thin momentum shell} and hence are proportional to powers of $\delta$. Note, in particular, a new term, in which the integrated by parts variational derivative acts on the product of the intermediate momentum modes.

To summarize, we see that inclusion of finite-$\delta$ effects leads to lengthy, but conceptually simple, expressions. It is thus straightforward to combine \eqref{eq:smooth-diff4}, \eqref{eq:PomegaPell} and \eqref{eq:driftdelta} to produce the equation describing the time evolution of the functional distribution valid to all orders in $\delta$.

\subsection{\label{app:locality-space}Locality in space}

We now show that if we use a smooth window function, the dependence of the correlation functions of the short modes depends locally on the long modes. This was used throughout the paper, starting from the subleading diffusion term computed around (\ref{diffsl}). Here we prove that this assumption is justified for this term, and then assume that it holds also for any other quantities, as it is expected. The way we will be able to prove this fact is that the diffusion depends on the two-point function of the long-wavelength distribution, whose behavior is well described by a decomposition in Eigenfunctions. As it will be clear, to show that the response of the short modes to the long ones is local, we just need to show that that all Eigenfunctions contribute equally. For this reason, we drop in the proof all the irrelevant terms that affect equally all Eigenfunctions.

In deriving (\ref{diffsl}) we simply assumed that $\delta m_s^2$ is a space-time constant. Instead, more, carefully, a short mode with, say, momentum $\vk_2$ is {sourced by} the Fourier mode of $\lambda \phi(x)^3$ with the same momentum.\footnote{One can see this either by studying the equations of motion, or by looking at the real part of the quartic in fields contribution to the wave function.} We thus get for the subleading diffusion term
\bea\nn\label{vphi-difflocal1}
&&{\rm Diffus.}\sim \frac{\d^2}{\d\phi_1^2}\;\int_{\Lambda(t)}^{(1+\delta)\Lambda(t)} \frac{d^3  k_1}{(2\pi)^3}\,\frac{d^3 k_2}{(2\pi)^3}  \; e^{i \left(\vec k_1+\vec k_2\right)\cdot\vec x_1} \frac{\d}{\d t}\left(\Omega_{\Lambda(t)}(\vec k_1)\Omega_{\Lambda(t)}(\vec k_2)\right)\,\times\\ 
&& 
\ \times \int {\cal{D}}\phi_\ell\; \delta\left(\phi_1-\phi_\ell(\vec x_1)\right) \;P_\ell[\phi_\ell,t]\;e^{i\left(\vec k_1 \cdot \vec x_1+\vec k_2\cdot \vec x_2\right)}\langle \phi(\vec k_1)\lambda \phi^3(\vec k_2)\rangle_{\phi_\ell}\, .
\eea
Now we use that the leading contributing term from $\lambda \phi(x)^3$ comes from taking one short mode and two long modes. We write, for the short momenta of order $\Lambda(t)$ that are of interest here: 
\bea
&&\lambda\phi^3(\vec k_2)=\lambda \int d^3  x_A\; e^{-i \vec k_2\cdot\vec x_2 } \phi_\ell(\vx_2)^2 \phi_s(\vec x_2)=\\ \nn
&&\qquad=\lambda \int d^3  x_A\;  \int\frac{d^3  k_A}{(2\pi)^3}\; e^{-i (\vec k_2-\vec k_A)\cdot\vec x_2 } \phi_\ell(x_2)^2 \phi_s(\vec k_A)\ .
\eea
Plugging back in~(\ref{vphi-difflocal1}), we can do the path integral over the long modes at all locations but at $\vec x_2$, to obtain
\bea\nn\label{vphi-difflocal2}
&&{\rm Diffus.}\sim\lambda \frac{\d^2}{\d\phi_1^2}\;\int_{\Lambda(t)}^{(1+\delta)\Lambda(t)} \frac{d^3  k_1}{(2\pi)^3}\,\frac{d^3 k_2}{(2\pi)^3}  \; \frac{\d}{\d t}\left(\Omega_{\Lambda(t)}(\vec k_1)\Omega_{\Lambda(t)}(\vec k_2)\right)\,\times\\ \nn
&& 
\qquad \times\ \int d^3  x_2\;  \frac{d^3  k_A}{(2\pi)^3}\;   \int {d}\phi_2\;P_{2}(\phi_1,t;\phi_2,t;\vec x_1-\vec x_2)\;e^{i\left(\vec k_1 \cdot \vec x_1+\vec k_2\cdot \left(\vec x_1-\vec x_2\right)+\vec k_A\cdot \vec x_2\right)}\phi_2^2\;\langle \phi_s(\vec k_1)\phi_s(\vec k_A)\rangle\sim\\ \nn
&&\quad\sim \lambda \frac{\d^2}{\d\phi_1^2}\;\int_{\Lambda(t)}^{(1+\delta)\Lambda(t)} \frac{d^3  k_1}{(2\pi)^3}\,\frac{d^3 k_2}{(2\pi)^3}  \; \frac{1}{k_1^3} \frac{\d}{\d t}\left(\Omega_{\Lambda(t)}(\vec k_1)\Omega_{\Lambda(t)}(\vec k_2)\right)\;\int {d}\phi_2\; \phi_2^2 \times\\ 
&& 
\qquad \times\   \int d^3  \Delta x\; P_{2}(\phi_1,t;\phi_2,t;\Delta \vx)\;e^{i\left(\vec k_1 +\vec k_2\right)\cdot \vec{\Delta x}} \ . 
\eea
where in the second passage we took the expectation value and defined ${\Delta \vx}=\vec x_1-\vec x_2$.  

We see that we need to compute the Fourier transform of the equal time two-location distribution $P_{2}(\phi_1,t;\phi_2,t;\Delta \vx)$. We will do this simply for the equilibrium distribution, and assume it holds for any state. 
{We expect the main contribution to come} from $|\vk_1+\vk_2|\sim0$ due to approximate momentum conservation.  We therefore have 
\bea
&&\tilde P_2(\phi_1,t;\phi_2,t;\vec k_1+\vec k_2)\equiv \int d^3  \Delta x\; P_{2}(\phi_1,t;\phi_2,t;\Delta \vx)\;e^{i\left(\vec k_1 +\vec k_2\right)\cdot \vec{\Delta x}} \sim\\\nn
&&\qquad\sim\sum_{n=1}^\infty \Phi_n(\phi_1)\Phi_n(\phi_2)  \frac{\lambda_n{(a(t) H)^{-2\lambda_n/H}}}{|\vec k_1+\vec k_2|^{3-2\lambda_n/H}} +  \Phi_0(\phi_1)\Phi_0(\phi_2)\delta^{(3)}(\vk_1+\vk_2) \nn
\eea
{where in the last passage we used \eqref{2ptfourier} but added the $n=0$ contribution.}
Plugging back in~(\ref{vphi-difflocal2}), we have 
\bea\nn\label{vphi-difflocal3}
&&{\rm Diffus.}\sim \lambda \frac{\d^2}{\d\phi_1^2}\; \int {d}\phi_2\;  \phi_2^2 \int_{\Lambda(t)}^{(1+\delta)\Lambda(t)} \frac{d^3  k_1}{(2\pi)^3}\,\frac{d^3 k_2}{(2\pi)^3}  \; \frac{1}{k_1^3} \frac{\d}{\d t}\left(\Omega_{\Lambda(t)}(\vec k_1)\Omega_{\Lambda(t)}(\vec k_2)\right) \times\\ 
&& 
 \times \l(  \sum_{n=1}^{\infty} \Phi_n(\phi_1)\Phi_n(\phi_2)\frac{4\pi^2\lambda_n {(a(t) H)^{-2\lambda_n/H}}}{|\vec k_1+\vec k_2|^{3-2\lambda_n/H}}+\Phi_0(\phi_1)\Phi_0(\phi_2)(2\pi)^3\delta^{(3)}(\vk_1+\vk_2)\r) \ . 
\eea
In the positive real axis, the factor $\frac{\d}{\d t}\left(\Omega_{\Lambda(t)}(\vec k_1)\Omega_{\Lambda(t)}(\vec k_2)\right)$ can be interpreted as a test function for distributions.\footnote{This is correct as long as test functions are more smooth than the distributions.} We can therefore use an identity that is valid for distributions (see for example section 4.6 of chapter I of~\cite{gelfand}~\footnote{Notice that we are using a slight generalization of this formula where $\int_{1}^{+\infty} d^3x$ is replaced by $\int_{\delta\Lambda(t)}^{+\infty} d^3k$, as this is more appropriate for us given that our test functions have support in the interval $[\Lambda(t),(1+\delta)\Lambda(t)]$.}. For the 1-dimensional case, it is easy to specify a choice of window function and verify the results below directly):
\bea\label{eq:galfangexpansion}
&&\frac{\lambda_n}{|\vec k_1+\vec k_2|^{3-2\lambda_n/H}}=(2\pi)\delta^{(3)}(\vec k_1+\vec k_2){(\delta \Lambda(t))^{2\lambda_n/H}}+\\ \nn
&& \qquad+\sum_{i=0}^{\infty} 4\pi 2^i\frac{(-\lambda_n)^{1+i}}{i!} \, {\cal{P}}_{\delta \Lambda(t)}\left(\frac{1}{|\vec k_1+\vec k_2|^{3}}\left(\log(|\vec k_1+\vec k_2|)\right)^i\right)\\ \nn
&&\qquad-4\pi\Theta(\delta\Lambda(t)-|\vec k_1+\vec k_2|) \frac{\lambda_n}{|\vec k_1+\vec k_2|^{3-2\lambda_n/H}} ,
\eea
where $ {\cal{P}}_{\delta \Lambda(t)}(f)$ is defined in terms of distributions, in such a way that, against a test function~$g$, we have
\be
\int dx\; g(x) {\cal{P}}_{\delta \Lambda(t)}(f(x))=\int dx\; f(x)\,\left(g(x)-\left(g(0)-g'(0)x-\ldots\right)\Theta(\delta\Lambda(t)-x)\right)\ ,
\ee
with as many derivatives as needed to make the integral convergent. Let us now consider the various contributions of (\ref{eq:galfangexpansion}). Let us start with the term in $\delta^{(3)}(\vec k_1+\vec k_2)$. If we plug this factor in (\ref{vphi-difflocal3}), we obtain
\bea\nn\label{vphi-difflocal4}
&&{\rm Diffus.}\supset \lambda \frac{\d^2}{\d\phi_1^2}\; \int {d}\phi_2\;  \phi_2^2 \sum_n \Phi_n(\phi_1)\Phi_n(\phi_2) \; \left({\eps}\delta\right)^{\lambda_n}\times\\ 
&& 
\qquad \times\  \int_{\Lambda(t)}^{(1+\delta)\Lambda(t)} \frac{d^3  k_1}{(2\pi)^3} \; \frac{1}{k_1^3} \frac{\d}{\d t}\left(\Omega_{\Lambda(t)}(\vec k_1)\right)\,\Omega_{\Lambda(t)}(\vec k_1)\;  \ . 
\eea
Using that $ \sum_n \Phi_n(\phi_1)\Phi_n(\phi_2)=\delta^{(1)}(\phi_1-\phi_2)  P_1^{eq}(\phi_1)$, and Taylor expanding $\left({\eps}\delta\right)^{\lambda_n}=1+\lambda_n\log(\epsilon\,\delta)+\dots$, we have
\bea\label{vphi-difflocal5}
&&{\rm Diffus.}\supset \lambda \frac{\d^2}{\d\phi_1^2}\; \left(\int {d}\phi_2\; \delta(\phi_2-\phi_1)\right) \phi_1^2 P_1^{eq}(\phi_1) \times\\ \nn
&& 
\qquad \times\  \int_{\Lambda(t)}^{(1+\delta)\Lambda(t)} \frac{d^3  k_1}{(2\pi)^3} \; \frac{1}{k_1^3} \frac{\d}{\d t}\left(\Omega_{\Lambda(t)}(\vec k_1)\right)\,\Omega_{\Lambda(t)}(\vec k_1)\ \times\ \left(1+O(\sqrt{\lambda}\log(\delta\,\epsilon))\right)\;  \ . 
\eea
This is exactly the form of the diffusion term that we guessed in the local-in-space case, up to terms suppressed by $\sqrt{\lambda}\log(\delta\,\epsilon)\ll1$. It is clear that the overall numerical factors are unimportant: it matters only that all Eigenfunctions contribute in the same way. Therefore, in order to complete the proof that the smooth $\Omega_{\Lambda(t)}$ gives approximate locality in space, we need to show that the rest of the contribution from~(\ref{eq:galfangexpansion}) is suppressed.  Let us therefore first see that happens when we plug the term with ${\cal P}$ in (\ref{eq:galfangexpansion}) into~(\ref{vphi-difflocal3}). We have
\bea\nn\label{vphi-difflocal6}
&&{\rm Diffus.}\sim \lambda \frac{\d^2}{\d\phi_1^2}\; \int {d}\phi_2\;  \phi_2^2 \sum_n \Phi_n(\phi_1)\Phi_n(\phi_2) \times\\ \nn
&& 
\qquad \times\  \int_{\Lambda(t)}^{(1+\delta)\Lambda(t)} \frac{d^3  k_1}{(2\pi)^3}\,\frac{d^3 k_2}{(2\pi)^3}  \; \frac{1}{k_1^3} \frac{\d}{\d t}\left(\Omega_{\Lambda(t)}(\vec k_1)\,\Omega_{\Lambda(t)}(\vec k_2)\right)\;\times \\
&&\qquad\times\ \sum_{i=0}^{\infty} \frac{\lambda_n^{1+i}}{i!} \, {\cal{P}}\left(\frac{1}{|\vec k_1+\vec k_2|^{3}}\left(\log(|\vec k_1+\vec k_2|/(a H))\right)^i\right)\ .
\eea
Let us estimate the integral in the last two lines for a generic $i$. We have, by using the definition of ${\cal{P}}$ and the fact that the integral is barely divergent as $\vec k_1+\vec k_2\to 0$:
\bea
 &&{\rm Int}_i=\frac{1}{i!}\int_{\Lambda(t)}^{(1+\delta)\Lambda(t)} \frac{d^3  k_1}{(2\pi)^3}\,\frac{d^3 k_2}{(2\pi)^3}  \; \frac{1}{k_1^3}  \lambda_n^{1+i} \, \left(\frac{1}{|\vec k_1+\vec k_2|^{3}}\left(\log(|\vec k_1+\vec k_2|/aH)\right)^i\right)  \;\times \\ \nn
&&\times\ \left(\frac{\d}{\d t}\left(\Omega_{\Lambda(t)}(\vec k_1)\,\Omega_{\Lambda(t)}(\vec k_2)\right)- \frac{\d}{\d t}\left(\Omega_{\Lambda(t)}(\vec k_1)\,\Omega_{\Lambda(t)}(\vec k_2)\right)\Big|_{\vec k_1+\vec k_2=0}\Theta(\delta\Lambda(t)-|\vec k_1+\vec k_2|)\right)\ .
\eea
Let us now estimate the factor containing the window function. The derivative in time scales as $1/\delta$ (as can be understood by noticing that upon integration over space, that goes as $\delta$, we get an unsuppressed contribution, or equivalently because for $\delta\to 0$, $\Omega_{\Lambda(t)}$ goes to a unit-step function.). Since the derivative in time comes through the dependence over $k/\Lambda(t)$, also the first derivative in $k$ goes as $1/\delta$, times some dimensionfull factors  of order $|\vec k_1+\vec k_2|$. Therefore, within order one numbers, we have
\bea\nn
 &&|{\rm Int}|\lesssim\int_{\Lambda(t)}^{(1+\delta)\Lambda(t)} \frac{d^3  k_1}{(2\pi)^3}\,\frac{d^3 k_2}{(2\pi)^3}  \; \frac{1}{k_1^3} \frac{|\vec k_1+\vec k_2|}{\delta^2}\; \lambda_n^{1+i} \, \left(\frac{1}{|\vec k_1+\vec k_2|^{3}}\left(\log(|\vec k_1+\vec k_2|/aH)\right)^i\right) \\ 
&& \qquad\lesssim \lambda_n^{1+i}  \left(\log\delta\right)^i\ .
\eea
Plugging back into (\ref{vphi-difflocal6}), we find that the contribution to ${\rm Diffus.}$ of the term in ${\cal{P}}$ of~(\ref{eq:galfangexpansion}) is of order
\bea\nn\label{vphi-difflocal7}
&&{\rm Diffus.}\lesssim \sum_{n,i=0}^\infty\lambda\, \lambda_n^{1+i} \left(\log\delta\right)^i \frac{\d^2}{\d\phi_1^2}\; \int {d}\phi_2\;  \phi_2^2 \sum_n \Phi_n(\phi_1)\Phi_n(\phi_2) \\ 
&&\qquad\qquad\sim   \sum_{i=0}^\infty \lambda^{3/2+i/2} \left(\log\delta\right)^i \ ,
\eea
where in the second step we have used that $\lambda_n\sim \sqrt{\lambda}$ and $\phi_\ell\sim 1/\lambda^{1/4}$. We therefore find that this contribution is suppressed with respect to the leading local-in-space term by expressions of the form $\sqrt{\lambda}(1+\sqrt{\lambda}\log\delta+\ldots$.  At last, the last term of~(\ref{eq:galfangexpansion}) { contributes to the diffusion coefficient} simply as $\delta^2\frac{1}{\delta}\lambda_n\sim \delta\sqrt{\lambda}$, where the factor of $\delta^2$ comes from the measure of integration, and the $1/\delta$ from the time-derivative of the window function.

Overall, as expected, we find that the  smoothness of the window function implies that the short modes depend locally on the long ones, up to computable corrections of order $\sqrt{\lambda}$ and $\sqrt{\lambda}\log(\delta\,\epsilon)$. This analysis shows why it was necessary to keep $\delta$ finite, or, more precisely, $\frac{e^{-\frac{1}{\sqrt{\lambda}}}}{\epsilon}\ll \delta$, as was anticipated in section \ref{sec:functionequation}. Otherwise the fact that all Eigenfunctions contribute the same at the leading order would not be true and, consequently, the corrections to short expectation values would not be simply proportional to the values of long fields at the same spacial points. {We discuss this effect in details in Appendix \ref{app:sharp-diffusion}.}

 We can also sketch the functional form of the further subleading correction in $\sqrt{\lambda}$. We use the fact that the $n$ dependence comes only through the factor of $\lambda_n$, to write, for the equilibrium distribution, that
\be
\sum_n \lambda_n \Phi_n(\phi_1)  \Phi_n(\phi_2)=-\Gamma_{\phi_1} \sum_n  \Phi_n(\phi_1)  \Phi_n(\phi_2)=-\Gamma_{\phi_1}  \delta^{(1)}(\phi_1-\phi_2)  P^{eq}_1(\phi_1)\ ,
\ee
so that the subleading contribution to ${\rm Diffus.}$, ${\rm Diffus.}_{{\rm \, subl.}}$, goes as
\bea\label{vphi-difflocal8}
&&{\rm Diffus.}_{ {\rm \, subl.}}= \beta_1 \frac{\lambda \log(\delta \eps)}{3H^2}\, \frac{\d^2}{\d\phi_1^2}\;\left(  \Gamma_{\phi_1}\,\phi_1^2\, P^{eq}_1(\phi_1)\right)\left\langle  \phi(\vk_s,t) \phi(-\vk_s,t)\r\rangle' \ ,
\eea
with $\beta_1$ an order one number.

\section{Fourier transform of the two-point distribution\label{app:2-point-Fourier}}
In this Appendix we compute some integrals appearing in the Fourier transform of the two-location distributions.

Let us start with the Fourier transform of the leading two point function given in \eqref{P2k}. We need to compute
\be
\int d^3 x\, e^{-i \vk \cdot \vx} x^{-p}= 4 \pi\int_0^{\infty} d x \,\frac{x \sin(k x)}{k}x^{-p}=4 \pi\Gamma(2-p)\sin\l(\frac{ \pi p}{2}\r)k^{-3+p}\,,
\ee
where we assumed that the integral is convergent which is true for $1<p<3$. We are interested, however, in $p=2\lambda_n/H\ll1$, for which the integral is divergent. The correct procedure is to analytically continue in $p$, after which we can Taylor expand for small $\lambda_n$ (see, for example, \cite{gelfand} for a careful treatment of Fourier transforms of distributions). We thus get
\be
4 \pi\Gamma(2-2\lambda_n/H)\sin\l(\frac{ \pi \lambda_n}{H}\r)k^{-3+2\lambda_n/H}=\frac{4\pi ^2 \lambda_n}{H}k^{-3+2\lambda_n/H}+O(\sqrt{\lambda})\,.
\ee

We next switch to the integrals relevant for the subleading two-location distribution~\eqref{threeints}. Their evaluation is straightforward but somewhat long. We thus focus on the most relevant part, which comes from the integral over the intermediate region, namely
\be
\label{midint}
\bar {\cal I}_{II}=2\lambda_n \int_{x_i}^{x_f}d^3 x   \int_{t_i(x)}^t d\, t' \l[j_0\l(\Lambda(t') x\r)-1\r] e^{-i \vk \cdot \vx}\,.
\ee
We remind the reader that
\be
 \Lambda(t)=\eps a(t) H,\quad x_i=\Lambda(t)^{-1} e^{-\Delta}, \quad x_f=\Lambda(t)^{-1} e^{\Delta}, \quad t_i(x)=(-\log(\eps H x)-\Delta)/H\,.
\ee
We are interested in the regime $k\sim \Lambda(t)$ and $\Delta\to \infty$, however, as we discussed, $\Delta \sqrt{\lambda}$ should still be small in order for sudden perturbation theory to be under control.
First, let us take the time integral in \eqref{midint}. After taking the large $\Delta$ limit we get the following expression:
\bea
&&{\cal I}_t= \int_{t_i(x)}^t d\, t' \l[j_0\l(\Lambda(t') x\r)-1\r] =\frac{1}{H}\l(1-\gamma_{\rm E}+{\rm{Ci}}(x\Lambda(t))-\log( x \Lambda(t))-\frac{\sin(x \Lambda(t))}{x \Lambda(t)}\r)\nn\\
&&\qquad\qquad\qquad\qquad\qquad\qquad\qquad\qquad\qquad\qquad\qquad\qquad +\,O(e^{-2 \Delta})\,.
\eea
By expanding this expression at large $x$ one can readily see that the integral \eqref{midint} has contributions that grow with $\Delta$. All these contributions, however, cancel against corresponding growing contributions from ${\cal I}_{I}$, ${\cal I}_{III}$ as well as the part of ${\cal I}_{II}$ not included in $\bar{ \cal I}_{II}$. The relevant finite contribution, after the angular integrals in it are taken, reads
\be
 {\cal I}_{I}+ {\cal I}_{II}+{\cal I}_{III}=8 \pi\frac{\lambda_n}{H}\l(\frac{\pi}{2 k^3}+ \int_{x_i}^{x_f} d x \l({\rm Ci}(\Lambda x) -\frac{\sin( \Lambda x)}{ \Lambda x}\r) \frac{x \sin(k x)}{k}\r) + O(e^{-\Delta},\lambda)\,.
\ee
Note that this integral is convergent in the $x_f\to \infty$ limit, however, it's derivative with respect to $k$ is not, consequently it can give $\theta$-function-like singularity that we are expecting. Indeed, since the boundary terms are of $O(e^{-\Delta})$, we can integrate by parts, and, by defining a dimensionless integration variable $y=\Lambda x$, we get a simpler-looking integral which can be readily evaluated, again dropping some terms of order $O(e^{-\Delta})$:
\bea
\label{threeintsanswer}
 && \tilde A_{nm}(k)={\cal I}_{I}+ {\cal I}_{II}+{\cal I}_{III}=\\ \nn
 &&\quad= \frac{8 \pi\lambda_n}{H k^3}\l(\frac{\pi}{2}+ \int_{e^{-\Delta}}^{e^{\Delta}}d y\,\frac{\sin y}{y^2} \l(\frac{k y}{\Lambda} \cos\frac{ k y}{\Lambda}-\sin\frac{k y}{\Lambda} \r)\r)+ O(e^{-\Delta},\lambda)=\nn\\\nn
 &&\quad= \frac{4 \pi\lambda_n}{H k^3}\l(\pi+{\rm Si}\l(\l(1-\frac{k}{\Lambda}\r)e^{\Delta}\r)-{\rm Si}\l(\l(1+\frac{k}{\Lambda}\r)e^{\Delta}\r)\r)+ O(e^{-\Delta},{\lambda})\,.
\eea
The value of the sine integral for large absolute values of the argument depends on the sign of the argument, namely ${\rm Si}(\pm\infty)=\pm \pi/2$. Consequently, for $k>\Lambda$ and not too close to $\Lambda$ we get zero, while for $k<\Lambda$ { (and not too close to $\Lambda$)}, we get $\frac{4 \pi^2\lambda_n}{H k^3}$, as stated in \eqref{Annk}. The detailed analysis we just did shows us that the width of the $\theta$-function is given by $k/\Lambda-1\sim e^{-\Delta}\gg e^{-1/\sqrt{\lambda}}$. So even though it can be made more narrow than any power of $\lambda$, it cannot be made arbitrarily narrow. 

We add the following comment. The smearing of the $\theta$-function that we just found is of order $e^{-\Delta}$. Since $\Delta$ is an artificial parameter, any dependence on it will cancel order by order in perturbation theory. It is expected that such a smearing will ultimately take the size {of the only physical} parameter of the theory, which is $e^{-1/\sqrt{\lambda}}$.

As we mentioned earlier, in section~\ref{sec:twolocations}, in this range of momenta, the contribution of the late-time perturbation to the two-point function contributes at the same order as the subleading sudden perturbation theory terms. Let us turn to computing this contribution, which is given by
\bea
&&P_2^{\rm late}(\phi_1,t;\phi_2,t;\vx)=-\int_{t_f(x)}^t dt 'j_0\l(\eps a(t') H x\r) H'(\phi_1,\phi_2)\; P_2^{\rm eq}(\phi_1,t';\phi_2,t';\vx)\,
\eea
where for $P_2^{\rm eq}$ it suffices to take the leading distribution \eqref{2loclead}, the interaction Hamiltonian $H'$ is defined in (\ref{eq:Hprdef}), and $t>t_f$. After taking the time integral, the part which is relevant in the  large $\Delta$ limit reads
\bea
\label{Platemodes}
&&P_2^{\rm late}(\phi_1,t;\phi_2,t;\vx)=\frac{\cos(\Lambda(t) x)}{H \Lambda(t)^2 x^2} H' \sum_n \Phi_n(\phi_1)\Phi_n(\phi_2) \l(\frac{x}{x_f}\r)^{-2 \lambda_n/H}\l(1+O\l(\sqrt{\lambda},e^{-\Delta}\r)\r)\,.\nn\\
\eea
Since $x>x_f=\exp{\Delta}/\Lambda(t)$ in coordinate space this expression is exponentially small in $\Delta$ and hence can be ignored at the leading order. However, for $k=\Lambda(t)$ its Fourier transform is divergent, at least for $\lambda_n=0$, so we might expect a significant contribution  in momentum space for $k\sim\Lambda(t)$. Let us first study the $n=0$ term.
\bea
&&\tilde P_2^{\rm late}(\phi_1,t;\phi_2,t;\vk)\supset \int_{x_f}^{\infty}d^3 x\, e^{-i\vk\cdot\vx}\frac{\cos(\Lambda(t) x) }{{H}\Lambda(t)^2 x^2} H'  \Phi_0(\phi_1)\Phi_0(\phi_2)\\\nn
&&\supset 2 \pi  \int_{x_f}^{\infty}d x \; \frac{\sin\l((k-\Lambda(t)) x \r)}{H \Lambda(t)^2 k x} H'  \Phi_0(\phi_1)\Phi_0(\phi_2)=\\\nn
&&=\frac{2 \pi}{H \Lambda(t)^{ 2} { k}}\l[\frac{\pi}{2}{\rm Sign}(k-\Lambda(t))-{\rm Si}\l(\l(\frac{k}{\Lambda(t)}-1\r)e^{\Delta}\r)\r]  H'  \Phi_0(\phi_1)\Phi_0(\phi_2)\l(1+O\l(e^{-\Delta}\r)\r)\,.
\eea
Let us note several features of this result. First, as we anticipated, it is of order $\sqrt{\lambda}$ (since $H'\sim\sqrt{\lambda} H$), that is, it is of the same order as \eqref{threeintsanswer}, in the region $|k/\Lambda-1|\sim e^{-\Delta}$. It becomes otherwise very small outside of this interval. Second, it is nonzero for $k/\Lambda-1=0_+$, namely
\bea\label{eq:P2late}
\tilde P_2^{\rm late}(\phi_1,t;\phi_2,t;k=\Lambda(t)_+)=-\frac{H^2}{4 \Lambda(t)^3}\frac{\d^2}{\d \phi_1\d\phi_2}\Phi_0(\phi_1)\Phi_0(\phi_2)\,.
\eea
This feature of the distribution of long modes plays an important role for the computation of the tadpole in the limit of the sharp window function, see discussion in Appendix \ref{app:sharp-diffusion}.

Finally, let us discuss briefly the contribution of nonzero Eigenvalues to \eqref{Platemodes}. Each of them contributes as
\bea\label{eq:P2non-local-nonzeroeigen}
&&\tilde P_2^{\rm late}(\phi_1,t;\phi_2,t;\vk)\supset2 \pi  \int_{x_f}^{\infty}d x \; \frac{\sin\l((k-\Lambda) x \r)}{H \Lambda^2 k x}  \l(\frac{x}{x_f}\r)^{-2 \lambda_n/H} H'  \Phi_n(\phi_1)\Phi_n(\phi_2)\,.\nn\\
\eea
This integral is convergent for any $k$ and for $k=\Lambda$ the result vanishes up to terms of order $e^{-\Delta}$. However, for other $k$'s in the range $|k/\Lambda-1|\sim e^{-\Delta}$ it gives a non-zero contribution which is also comparable to~\eqref{threeintsanswer}. Consequently, if one is interested in the detailed functional profile of the distribution in this range of $k$'s, one needs to include this contribution as well. 
For us this will not be important since the contribution of this thin momentum shell to all physical observables or components of our equations is suppressed.

\section{Distribution in the case of  sharp window function \label{app:sharp-diffusion}}

As we anticipated in the main text, in the case where the window function is sharp, $\Omega_{\Lambda(t)}(k)=\Theta(\Lambda(t)-k)$, the dependence on the long-modes of the correlation functions of the short modes, which are the building blocks of the Fokker-Planck-like equation for the long modes, is not local in space. Here in this section we show how the calculation carries forward in this situation, and we check how we obtain the same result for $P^{eq}_1$ at subleading order in $\sqrt{\lambda}$. In passing, we will explicitly show how the tadpole of $\phi$ is negligible at the order we work on if we use a smooth window function.

Let us {first study the tadpole term in the case of the single-point distribution}. At leading order, we can compute the expectation value by Taylor expanding the wave function in (\ref{eq:wavefuncquartic}) by pulling downstairs the quartic term, to obtain 
\bea\label{eq:tadpole1}\nn
&&{\left\langle\dot\Delta \phi(\vx_1)\right\rangle_{\phi_1}\;} P_1(\phi_1,t)= \int {\cal{D}}\phi_\ell \; \delta^{(1)}(\phi_1-\phi_\ell(\vec x_1)) P_\ell[\phi_\ell,t]\ \int^{(1+\delta)\Lambda(t)}_{\Lambda(t)} \frac{d^3k}{(2\pi)^3} \;\d_t\Omega_{\Lambda(t)}(k)\; e^{i\vec k\cdot\vec x_1} \ \times \\ \nn
&&\quad\times\  \frac{\lambda}{H^4}  \int  \frac{d^3k_1}{(2\pi)^3}  \frac{d^3k_2}{(2\pi)^3} \frac{d^3k_3}{(2\pi)^3} \frac{d^3k_4}{(2\pi)^3}  \;(2\pi)^3\delta^{(3)}\left(\sum_{i=1}^4\vec k_i\right)\; \ \times\\
&& \quad\times \ k_{\Sigma^3} \; f_\psi\left(\frac{k_2}{k_1},\frac{k_3}{k_1},\frac{k_4}{k_1}\right)\;  \langle\phi(\vec k_1)\phi(\vec k_2)\phi(\vec k_3)\phi(\vec k_4)\; \phi(\vec k)\rangle\ ,
 \eea    
{ where $ f_\psi\left(\frac{k_2}{k_1},\frac{k_3}{k_1},\frac{k_4}{k_1}\right)$ is a unitless function of the ratio of the wavenumbers.  }
 The leading order in $\sqrt{\lambda}$ contribution comes from contracting $\phi(\vec k)$ with $\phi(\vec k_4)$ at free-theory level, and considering all the remaining $\phi(\vec k_1)\phi(\vec k_2)\phi(\vec k_3)$ as long wavelength fields.  Because of the window function, the mode $k$ is sharply peaked at  $\Lambda(t)$. After the contraction, $k_4$ will also be forced to be close to $\Lambda(t)$. The $\delta$-function of the quartic vertex will therefore force $|\sum_{i=1,2,3}\vec  k_i|\simeq\Lambda(t)$. At leading order in $\sqrt{\lambda}$, the leading contribution comes from taking one of the long modes to be close to $\Lambda(t)$,  and the other two as long as possible. In this kinematical regime, the expression for $f_\psi$ greatly simplifies. Accounting for combinatorics, we obtain
 \bea\label{eq:tadpole2}\nn
&&\left\langle\dot\Delta \phi(\vx_1)\right\rangle_{\phi_1}\; P_1(\phi_1,t)= \int {\cal{D}}\phi_\ell \; \delta^{(1)}(\phi_1-\phi_\ell(\vec x_1)) P_\ell[\phi_\ell,t]\ \; \int^{(1+\delta)\Lambda(t)}_{\Lambda(t)} \frac{d^3k}{(2\pi)^3} \;\d_t\Omega_{\Lambda(t)}(k) \; e^{i\vec k\cdot\vec x_1} \ \times \\ \nn
&&\quad\times\ \frac{4}{3} \frac{\lambda}{H^4} \left(\log(\epsilon/2)-\psi(3/2)\right) k^3 \int  \frac{d^3k_1}{(2\pi)^3}  \;(2\pi)^3\delta^{(3)}(\vec k+\vec k_1)\;  \phi_\ell^3({\vec k_1}) \frac{H^2}{2 k^3}=\\ \nn
&&=\int {\cal{D}}\phi_\ell \; \delta^{(1)}(\phi_1-\phi_\ell(\vec x_1)) P_\ell[\phi_\ell,t]\ \; \int^{(1+\delta)\Lambda(t)}_{\Lambda(t)} \frac{d^3k}{(2\pi)^3} \;\d_t\Omega_{\Lambda(t)}(k) \; e^{i\vec k\cdot\vec x_1} \ \times \\ \nn
&&\quad\times\ \frac{2}{3}  \frac{\lambda}{H^2} \left(\log(\epsilon/2)-\psi(3/2)\right)   \phi_\ell^3(\vec k) \ .
 \eea    
 Now, as usual, we can write the long field at wavenumber $\vec k$ as the Fourier transform of the field at an auxiliary location, and then perform the integral  over all the locations that do not appear in the expression, obtaining the probability distribution of the field at equal times at two locations:
 \bea\label{eq:tadpole3}\nn
&&\left\langle\dot\Delta \phi(\vx_1)\right\rangle_{\phi_1}\; P_1(\phi_1,t)= \frac{2}{3}  \frac{\lambda}{H^2} \left(\log(\epsilon/2)-\psi(3/2)\right)  \int^{(1+\delta)\Lambda(t)}_{\Lambda(t)} \frac{d^3k}{(2\pi)^3} \;\d_t\Omega_{\Lambda(t)}(k) \int d\phi_2\;  \phi_2^3\;  \ \times \\ \nn
&&\qquad\qquad\qquad\qquad\times\  \int d^3x_{12}\; P_2{(\phi_1,t;\phi_2,t;\vx_{12})}\ \;  \; e^{i\vec k\cdot\vec x_{12}} = \\ \nn
&&\qquad\quad=\frac{2}{3}  \frac{\lambda}{H^2} \left(\log(\epsilon/2)-\psi(3/2)\right)  \int^{(1+\delta)\Lambda(t)}_{\Lambda(t)} \frac{d^3k}{(2\pi)^3} \;\d_t\Omega_{\Lambda(t)}(k) \int d\phi_2\;  \phi_2^3\;    P_2(\phi_1,t;\phi_2,t;\vx_{12})\ ,
 \eea    
  where we used the fact that, by translation invariance, $P_2$ depends only on the relative separations $\vec x_{12}=\vec x_2-\vec x_1$. 
 
  We are therefore led to consider the Fourier transform of the equal-time two location distribution, at momenta $k\simeq \Lambda(t)$. We will specialize here on the equilibrium distribution, though the result is expected to hold in general.  As we saw in the former appendix, this is rather delicate. There are two contribution. The first is given by using the expression in~(\ref{Annk}), where instead of the smeared $\theta$-function we  use directly the expression in~(\ref{threeintsanswer}). Since   the window function is sharp, we can take expression (\ref{threeintsanswer}) at $k=\Lambda(t)_+$. This is equivalent to taking the expression in~(\ref{Annk}) with the $\theta$-function evaluated to $1/2$. Then there is a second contribution, coming from $P_2^{\rm late}$ of (\ref{eq:P2late}). We obtain
 \bea\label{eq:tadpole4}\nn
&&\left\langle\dot\Delta \phi(\vx_1)\right\rangle_{\phi_1} P_1^{eq}(\phi_1)= \frac{2}{3}  \frac{\lambda}{H^2} \left(\log(\epsilon/2)-\psi(3/2)\right)  \int^{(1+\delta)\Lambda(t)}_{\Lambda(t)} \frac{d^3k}{(2\pi)^3} \;\d_t\Omega_{\Lambda(t)}(k) \int d\phi_2\;  \phi_2^3\;  \ \times \\ 
&&\left(\sum_n \frac{2\pi^2}{k^3}\left(\frac{\lambda_n}{H}\right) \Phi_n(\phi_1)\Phi_n(\phi_2)-\frac{H^2}{4 \Lambda(t)^3}\frac{\d^2}{\d\phi_1\d\phi_2} \Phi_0(\phi_1)\Phi_0(\phi_2)\right)\ .
 \eea  
 We can now use that 
 \be
 \sum_n\lambda_n \Phi_n(\phi_1)\Phi_n(\phi_2)=-\Gamma_{\phi_1} \sum_n\Phi_n(\phi_1)\Phi_n(\phi_2)=-\Gamma_{\phi_1} \left(\delta^{(1)}(\phi_1-\phi_2) P^{eq}_1(\phi_1)\right)\ , 
 \ee
 to write, after performing the $\int d\phi_2$: 
   \bea\label{eq:tadpole5}\nn
&&\left\langle\dot\Delta \phi(\vx_1)\right\rangle_{\phi_1}\;P_1^{eq}(\phi_1)= \frac{2}{3}  \frac{\lambda}{H^2} \left(\log(\epsilon/2)-\psi(3/2)\right)  \int^{(1+\delta)\Lambda(t)}_{\Lambda(t)} \frac{d^3k}{(2\pi)^3} \;\d_t\Omega_{\Lambda(t)}(k)   \ \times \\ 
&&\qquad\times\ \left(-\frac{1}{H}\frac{2\pi^2}{k^3} \left[\Gamma_{\phi_1},\phi_1^3\right]P_1^{eq}(\phi_1)+\frac{3}{4}\frac{H^2}{\Lambda(t)^3}\langle\phi_2^2\rangle \frac{\d}{\d\phi_1}P^{eq}_1(\phi_1)\right)\ ,
 \eea 
 where  we used that $\Gamma_{\phi_1}P^{eq}_1(\phi_1)=0$. Using now that, in the limit $\delta\to 0$, $\d_t\Omega_{\Lambda(t)}=\delta^{(1)}(\Lambda(t)-k) H \Lambda(t)$, we can perform the last $k$-integral, to finally obtain the following expression for the tadpole in the case  of a sharp window function
     \bea\label{eq:tadpole6}\nn
&&\left\langle\dot\Delta \phi(\vx_1)\right\rangle_{\phi_1}\;P_1^{eq}(\phi_1,t)= \frac{2}{3}  \frac{\lambda}{H^2} \left(\log(\epsilon/2)-\psi(3/2)\right) )   \ \times \\ 
&&\qquad \times \ \left(- \left[\Gamma_{\phi_1},\phi_1^3\right]P_1^{eq}(\phi_1)+\frac{1}{2\pi^2}\frac{3}{4}H^3\langle\phi_2^2\rangle \frac{\d}{\d\phi_1}P^{eq}_1(\phi_1)\right)\ .
 \eea

 We now derive the subleading form of the diffusion term. {We start from~(\ref{vphi-difflocal3}), as, until that point, we did not specify the sharpness of the window function, but now we will take the limit $\delta\to0$ at the end of the calculation. Let us first establish that the contribution from the Eigenvalues in $n>0$ goes to zero in this limit. In fact, when the derivative acts on $\Omega_{\Lambda(t)}(k_1)$, we obtain a $\delta^{(1)}(k_1-\Lambda(t))$ that fixes $k_1=\Lambda(t)$. Then, the integral in $k_2$ is convergent, and will be bounded by $\delta^{\lambda_{n}}$. This goes to zero as $\delta\to 0$. The contribution of the zero mode is instead different, as the $\delta^{(3)}(\vec k_1+\vec k_2)$ removes the integration over $k_2$. This is the only contributing term in this limit, and reads, using that $\Phi_0(\phi_1)=P^{eq}_1(\phi_1)$:}
\bea\label{diffnonlocal}
&&{\rm Diffus.}\supset\lim_{\delta\to 0} \lambda \frac{\d^2}{\d\phi_1^2}\; \left(\int {d}\phi_2\;  \phi_2^2 P_1^{eq}(\phi_2) \right)\times\\ \nn
&& \quad \times\  P_1^{eq}(\phi_1) \int_{\Lambda(t)}^{(1+\delta)\Lambda(t)} \frac{d^3  k_1}{(2\pi)^3} \; \frac{1}{k_1^3} \frac{\d}{\d t}\left(\Omega_{\Lambda(t)}(\vec k_1)\right)\,\Omega_{\Lambda(t)}(\vec k_1)\;  \\ \nn
&&=\lim_{\delta\to 0} \lambda { \langle\phi_2^2\rangle\,}\frac{\d^2}{\d\phi_1^2}\; P_1^{eq}(\phi_1)\   \int_{\Lambda(t)}^{(1+\delta)\Lambda(t)} \frac{d^3  k_1}{(2\pi)^3} \; \frac{1}{k_1^3} \frac{\d}{\d t}\left(\Omega_{\Lambda(t)}(\vec k_1)\right)\,\Omega_{\Lambda(t)}(\vec k_1)\;  \ . 
\eea
where $\langle\phi_2^2\rangle$ is just the expectation value of the square of the long field at one location. We therefore see that in the case of the sharp window function, the factors of $\phi_1^2$ in the diffusion get replaced by~$\langle\phi_2^2\rangle$.  We therefore obtain, in the sharp-window function case, the following equation for the one-location {equilibrium} probability distribution in replacement of (\ref{1pteqsublead}):
\be
\label{1pteqsubleadnonlocal}
\l[-\frac{\d}{\d\phi_1} \left\langle\dot\Delta \phi(\vx_1)\right\rangle_{\phi_1}+\frac{H^3}{8\pi^2}\frac{\d^2}{\d\phi_1^2}\l(1+a \langle\phi_2^2\rangle\r)+\frac{\d}{\d \phi_1}\l(\frac{\lambda\phi_1^3}{3 H}+b\phi_1+c\phi_1^5\r)\r]P_1^{eq}(\phi_1)\l(1+O(\lambda)\r)=0\,,
\ee
where $a,b,c$ are the same as in (\ref{abc}).
One can check that the same solution (\ref{eq:subleading_sol_1loc}) that we found using the broad window function satisfies this equation. This confirms, by facing the most extreme case, that our results are independent of the window function.

We have just seen that in the case of the sharp window function the tadpole of $\dot\Delta\phi$ was non-negligible at subleading order. This calculation allows us to show that in the case of the broad window function such a contribution is negligible and the tadpole contributes at one more order in $\sqrt{\lambda}$. Neglecting irrelevant factors, in the kinematical regime for the long modes that we considered, the size of the tadpole is controlled relevantly by the following phase-space integral:
\bea
\left\langle\dot\Delta \phi(\vx_1)\right\rangle_{\phi_1}P_1^{eq}(\phi_1)\supset\lambda \int d\phi_2 \; \phi_2^3\int_{\Lambda(t)}^{(1+\delta)\Lambda(t)} \frac{dk}{k}\; \frac{\d}{\d t}\Omega_{\Lambda(t)}(k)\; \tilde P^{eq}_2(\phi_1,t;\phi_2,t;k)\ .
\eea
In the case of the sharp window function, the width of the window was much smaller than the variation scale of $\tilde P_2(\phi_1,t;\phi_2,t;k)$, and therefore we could simply approximate the $k$-dependence of $\tilde P_2$ with {its value at $k=\Lambda(t)_+$}. Instead, in the case of a broad window function, the variation of $P_2$ while we integrate over the wide shell of order $\Lambda \delta$ selected by the window function cannot be neglected, and indeed one can take the window function as approximately constant in that interval, with size $1/\delta$. In detail, for $k\simeq\Lambda(t)_+$, the relevant expression for $\tilde P_2$ is given by eq.~(\ref{threeintsanswer}),  (\ref{eq:P2late}), and~(\ref{eq:P2non-local-nonzeroeigen}). In all cases, $\tilde P_2$ is unsuppressed only on a thin shell of order $k-\Lambda\sim +\Lambda e^{-\Delta}$. Let us analyze just the contribution from~(\ref{threeintsanswer}) for definiteness, as the others gives the same parametric result. We have
\bea\nn\label{eq;tadpoleborad1}
&&{\l\langle\dot\Delta\phi\r\rangle_{\phi_1,\,\rm broad\; window} }P_1^{eq}(\phi_1)\supset\lambda \sum_n \int d\phi_2\; \phi_2^3\;\int_{\Lambda(t)}^{\Lambda(t)(1+e^{-\Delta})} \frac{dk}{k}\; \frac{\d}{\d t}\Omega_{\Lambda(t)}(k)\; \frac{\lambda_n}{k^{3}} \ \Phi_{n}(\phi_1)\Phi_n(\phi_2)\\ 
&&\quad \simeq \frac{\lambda}{\delta} \sum_n \int d\phi_2 \;\phi_2^3\,\lambda_n \log\left(1+e^{-\Delta}\right) \ \Phi_{n}(\phi_1)\Phi_n(\phi_2)\simeq - \frac{\lambda e^{-\Delta}}{\delta} \left[\Gamma_{\phi_1},\phi_1^3\right] \; P_1^{eq}(\phi_1)\ .
\eea
where in the last passage we Taylor expanded $\log(1+e^{-\Delta})$, and we have also written, as usual, $\sum_n \lambda_n \Phi_n(\phi_1)\Phi_n(\phi_2)=-\Gamma_{\phi_1} \delta(\phi_1-\phi_2)P_1^{eq}(\phi_1)$. Let us now estimate the size of this term. The parameter $\Delta$ is the one that gives us control over sudden perturbation theory, and the dependence on it (as for the one on $\epsilon$ and $\Delta$), will cancel order by order in perturbation theory. We expect therefore that the widening of the $\theta-$function in $\tilde P_2$ will be saturated by $e^{-\Delta}\sim e^{-1/\sqrt{\lambda}}$, which is the scale associated to the problem (our perturbative calculation, in order to achieve control, indeed gives a suboptimal estimate). Now, we see that the size of the tadpole in the case of the broad window function is suppressed with respect to the case of a sharp window by a factor of $e^{-\frac{1}{\sqrt{\lambda}}-\log\delta}=e^{-\frac{1}{\sqrt{\lambda}}\left(1+\sqrt{\lambda}\log{\delta}\right)}$. Using the  bound on $\delta$ imposed by locality that we obtained below~(\ref{vphi-difflocal5}), $\sqrt{\lambda}\log\delta\ll1$, we see that the contribution of this tadpole term is suppressed by $e^{-\frac{1}{\sqrt{\lambda}}}$ with respect to the case of a sharp window function. Notice that if we saturate the bound on $\delta$ from locality (neglecting the factor of $\epsilon$), and we take $\delta\sim e^{-1/\sqrt{\lambda}}$, then the suppression of the tadpole from the widening of the window function disappears. This is expected because this is indeed when the window function becomes too sharp and one loses locality.  

Finally, we point out  that $\langle\dot \Delta\phi\rangle_{\rm broad\; window}$ is in reality only suppressed by a factor of $\sqrt{\lambda}$ with respect to the case of a sharp window, as we could have considered a kinematical regime for the same perturbative diagram where we take two long modes to be close to $\Lambda(t)$ rather than just one. We would have paid {an extra} factor of $\sqrt{\lambda}$, but the phase-space integrals would have been unsuppressed.
 
\bibliography{reference_leo}
\bibliographystyle{jhep}

\end{document}